\documentclass[twocolumn,superscriptaddress,prb,aps,preprintnumbers,nofootinbib]{revtex4-1}
\usepackage{amsmath,amssymb,bm,graphicx,color,gensymb,bbold,appendix,hyperref}

\graphicspath{{./figures/}}

\bibpunct{[}{]}{,}{n}{}{}

\begin{document}

\title{Anisotropic-Exchange Magnets on a Triangular Lattice:\\
Spin Waves, Accidental Degeneracies,  and Dual Spin Liquids}

\author{P. A. Maksimov}
\affiliation{Department of Physics and Astronomy, University of California, Irvine, California 92697, USA}
\author{Zhenyue Zhu}
\affiliation{Department of Physics and Astronomy, University of California, Irvine, California 92697, USA}
\author{Steven R. White}
\affiliation{Department of Physics and Astronomy, University of California, Irvine, California 92697, USA}
\author{A. L. Chernyshev}
\affiliation{Department of Physics and Astronomy, University of California, Irvine, California 92697, USA}
\date{\today}
\begin{abstract}
We present an extensive overview of the phase diagram, spin-wave excitations, and finite-temperature transitions 
of the anisotropic-exchange magnets on an ideal nearest-neighbor triangular lattice. 
We investigate transitions between five principal classical phases of the corresponding model: ferromagnetic,
N\'{e}el, its dual, and the two stripe phases. Transitions are identified by the spin-wave instabilities and 
by the Luttinger-Tisza approach, and we highlight the benefits of the former while outlining the shortcomings of the 
latter. Some of the transitions are direct and 
others occur via intermediate phases with more complicated forms of ordering. 
The spin-wave spectrum in the  N\'{e}el phase is obtained and is shown to be non-reciprocal, 
$\varepsilon_{\alpha,{\bf k}}\!\neq\!\varepsilon_{\alpha,-{\bf k}}$,
in the presence of anisotropic bond-dependent interactions. 
In a portion of the N\'{e}el phase, we find spin-wave instabilities to a long-range spiral-like state. 
This transition boundary is similar to that of the spin-liquid phase of the 
$S\!=\!1/2$ model, discovered in our prior work, suggesting a possible connection between the two.
Further, in the stripe phases, quantum fluctuations are mostly negligible, leaving the ordered  moment nearly saturated 
even for the $S\!=\!1/2$ case. However, for a two-dimensional (2D) surface of the full 3D parameter space, the 
spin-wave spectrum in one of the stripe phases exhibits an enigmatic accidental degeneracy manifested by  
pseudo-Goldstone modes. As a result,  despite the nearly classical ground state, the ordering transition temperature 
in a wide region of the phase diagram is significantly suppressed from the mean-field expectation. 
We identify this accidental degeneracy 
as due to an exact correspondence to an extended 
Kitaev-Heisenberg model with emergent symmetries that naturally lead to the pseudo-Goldstone modes. 
There are previously studied dualities within the Kitaev-Heisenberg model on the triangular lattice 
that are exposed here in a wider parameter space. One important implication of this correspondence
for the $S\!=\!1/2$ case is the existence of a region of the spin-liquid phase
that is dual to the spin-liquid phase discovered recently by us.
We complement our studies by the density-matrix renormalization group of the $S\!=\!1/2$ model
to confirm some of the duality relations and to verify the existence of the dual spin-liquid phase.  
\end{abstract}
\maketitle

\section{Introduction}
\vskip -0.07cm

Ever since the seminal works by Wannier \cite{wannier} and Anderson \cite{spinliquid}, a 
motif of spins on a triangular-lattice network epitomizes the idea of   geometric frustration that can give rise to 
non-magnetic spin-liquid states \cite{SavaryBalentsSL16,Balents10}. 
A variety of materials and models with the triangular,  kagom\'e,  and pyrochlore  
lattices have provided a natural playground for  geometric frustration and realized various exotic and 
quantum-disordered states \cite{spinliquid,Balents10,Kagome_review,Pyrochlore_review,SavaryBalentsSL16,ZhouSLreview}.

More recently, magnetic materials with anisotropic spin-spin interactions, which arise from  spin-orbit coupling
in their magnetic ions, have offered a different path to achieve similar goals.
A strong mixing of spin  and orbital degrees of freedom leads to the bond-dependent anisotropic-exchange interactions
providing an alternative mechanism for  frustration  
\cite{Witczak, RauKee_review}. 
A  particular case is that of the so-called compass model \cite{Nussinov} on the tri-coordinated honeycomb 
lattice with each spin component interacting selectively along only one of the  bonds via an Ising-like interaction.
In a celebrated work  \cite{Kitaev},  Kitaev showed that it has a spin-liquid ground state 
with fractionalized excitations, a finding that set off a significant research effort.
In real materials, however, desired terms occur along with the other  diagonal and off-diagonal components of 
the anisotropic-exchange matrix that are allowed by the lattice symmetry 
\cite{Winter_review,JK09,JK10,Chaloupka, Rau18}. 
These terms have proven to be detrimental to the Kitaev spin liquid and so far have prevented its definite 
realization \cite{Winter_review}. 

Combining the geometric frustration of the lattice with the spin-orbit-induced anisotropic exchanges 
is a potentially very fruitful and less explored area.
The recently synthesized rare-earth compound YbMgGaO$_4$ (YMGO) has offered an example of 
such a synergy, with the pseudo-spin $S\!=\!1/2$ states of the strongly spin-orbit-coupled $f$-shells of the 
magnetic Yb$^{3+}$ ions  arranged in nearly perfect triangular-lattice layers \cite{SciRep, Chen1}.
It has been initially marketed as a spin-liquid candidate, given the lack of ordering and broad features in its 
dynamical response \cite{SciRep, Chen1,MM,muons,Yuesheng17a,Chen2,Zhao_neutron_field17,Ruegg,Chen4,Chen5}.
However, this initial optimism has faded considerably due to experimental evidence and theoretical arguments 
in favor of the intrinsic disorder causing a ``mimicry'' of a spin liquid 
\cite{MM,MM2,Yuesheng17,kappa,Wen_freeze18,us,Balents18,Kimchi}. 
Nevertheless, the problem of the triangular-lattice anisotropic-exchange magnets has attracted considerable interest 
\cite{Kimchi,multiQ,Chen3,Balents17,Balents18,us,topography,multiQ,Wang17,Chen_field,ChenTm}
and remains the focus of much research as a wider family of materials become available 
\cite{Cava16,newYb,newTm1,newTm2,Yb_chalc}. 

In our first study, Ref.~\cite{us}, we  argued that the experimental range of  parameters of 
the model that should describe a disorder-free YMGO does not support a spin-liquid state,
in agreement with exact-diagonalization  \cite{Wang17} and   
variational Monte Carlo  studies \cite{Balents17}.
In our more recent work, Ref.~\cite{topography}, we  provided a detailed exploration of the phase diagram of 
the most generic nearest-neighbor triangular-lattice model 
in order to find out whether anisotropic-exchange interactions on this lattice can potentially 
create much desired exotic states. We have used the density-matrix renormalization group (DMRG) for the $S\!=\!1/2$ model
and discovered a spin-liquid region in its 3D phase diagram \cite{topography}. 
We have also established a close similarity and 
a direct isomorphism of this newly found spin-liquid phase to the much studied 
spin liquid of the fully isotropic Heisenberg $J_1$--$J_2$ model on the triangular lattice
\cite{ZhuWhite,Sheng1,Sheng2,Bishop,Kaneko,Iqbal}. 
We have also pointed out that the spin liquid with open spinon Fermi surface 
\cite{Chen2,Chen5,Zhao_neutron_field17,Chen_field}  
is not realized in the phase diagram of the model. This is also in accord with Ref.~\cite{Balents17}.


In the present study, we expand our previous work in several directions.
First, we provide a quasiclassical description of the five principal  magnetically-ordered  single-${\bf Q}$
phases that span the 3D phase diagram of the nearest-neighbor anisotropic-exchange model 
on the triangular lattice. These phases are ferromagnetic, 120${\degree}$ N\'{e}el, dual 120${\degree}$, 
and two different stripe states. 
For four of them, we find explicit expressions for their
spin-wave spectra. To the best of our knowledge, the spin-wave 
spectrum of the 120${\degree}$ N\'{e}el phase with anisotropic terms 
has not been discussed previously. 
We demonstrate that it is, generally, non-reciprocal in this case, 
$\varepsilon_{\alpha,{\bf k}}\!\neq\!\varepsilon_{\alpha,-{\bf k}}$.

Next, we analyze the transition boundaries between these principal  phases as given by the 
instabilities of their magnon spectra. We find that such an approach closely and reliably 
reproduces phase boundaries obtained by a numerical optimization of classical energy in large clusters
of spins \cite{multiQ}, offering obvious advantages over this technique. In agreement with prior studies 
\cite{Chen3,multiQ,Balents18}, we also find consistent discrepancies of the phase boundaries  
provided by the Luttinger-Tisza method \cite{lt_original} and discuss a potential reason for that.

For the  120${\degree}$ phase close to the Heisenberg limit, we find an instability toward 
a long-range spiral state that is similar to the $Z_2$ vortex state found in the 
triangular-lattice Kitaev-Heisenberg model \cite{Trebst_tr,Ioannis}. 
With the correspondence to that model discussed below in more detail,
we note that the identified transition boundary is similar to the boundary of the spin-liquid
phase advocated in our previous work \cite{topography} for the quantum $S\!=\!1/2$ case, 
suggesting a possible relation between the two.

In the present  study, we also explore the ferromagnetic and stripe parts of the phase diagram in order to check
whether the regions dominated by anisotropic interactions can lead to 
strongly frustrated and highly degenerate states. The on-site magnetization is nearly classical 
and quantum fluctuations are negligible for most of these regions  even in the quantum $S\!=\!1/2$ limit. 
Enigmatically, however, the ordering N\'{e}el temperature
calculated from the spin-wave spectrum  in one of the stripe phases is suppressed in the vicinity of a  
surface of parameters in the 3D parameter space. This suppression originates from the gapless 
pseudo-Goldstone spin-wave modes, which occur due to an accidental degeneracy, 
with the Mermin-Wagner theorem dictating $T_N\!=\!0$ for a two-dimensional (2D) system.
Although quantum fluctuations do induce a gap in the pseudo-Goldstone spectrum via an order-by-disorder effect 
\cite{Rau_obd},
the ordering temperature remains suppressed in that region compared to the mean-field expectations. 
Thus, while the system is almost classical, large values of the factor $f\!=\!T_\text{MF}/T_N$, which is used to 
identify a proximity to a quantum-disordered state \cite{Chen3,Ruegg}, can be highly misleading, questioning 
it as a useful measure in such cases.

Crucially, this surface of accidental degeneracy in the anisotropic-exchange model 
is identified as corresponding to an extended Kitaev-Heisenberg model. 
This latter model possesses emergent symmetries that naturally lead to the pseudo-Goldstone modes
in the quasiclassical limit, thus explaining the enigmatic trends described above. There are also additional 
symmetry transformations within that model, known as Klein dualities \cite{Khal_ProgSupp,Kimchi14}, 
that allow one to make deeper connections  between different parts of the parameter space. 
 
Using these insights, we have performed DMRG studies of the quantum $S\!=\!1/2$ anisotropic-exchange 
model in previously unexplored parts of the phase diagram \cite{topography}. We have 
validated quasiclassical phase boundaries discussed above and verified previous studies of the Kitaev-Heisenberg model 
\cite{Trebst_tr,Ioannis,Tohyama,Avella,Rau_tr} that are exposed here in a wider parameter space. 
We demonstrate that the so-called nematic phase \cite{Trebst_tr,Ioannis,Tohyama,Rau_tr} corresponds to the 
boundary between two stripe phases and does not represent a separate state in the quantum limit. 

We emphasize that the region that we have previously identified as a spin-liquid phase 
in Ref.~\cite{topography} includes a sector of the line that corresponds to the Kitaev-Heisenberg model. 
This implies that previous numerical studies of the $S\!=\!1/2$ Kitaev-Heisenberg  model
\cite{Trebst_tr,Tohyama} must have overlooked the spin-liquid phase,
either due to smallness of their clusters \cite{Trebst_tr} or due to periodic boundary conditions \cite{Tohyama} 
that are unfavorable for DMRG. 

However, the most important implication of the correspondence to the Kitaev-Heisenberg model is that
it necessitates an existence of another spin liquid, the one that is Klein dual to the spin liquid found in Ref.~\cite{topography}.
In our present DMRG study, we confirm the existence of this dual spin-liquid phase. Interestingly, 
for the exchange matrix written in crystallographic axes, the dual spin liquid occurs in the region 
dominated by anisotropic terms.
We  use the structure factor $S({\bf q})$ to argue that 
the dual spin liquid can be seen as a result of a ``melting'' of the dual 120${\degree}$  phase,
just as the spin liquid of Ref.~\cite{topography} is a molten 120${\degree}$ phase, with both phases 
maintaining the shapes of the structure factor similar to that of their parent ordered states.
The confirmation of the dual spin liquid strengthens our case for both of them.

The paper is structured as follows. 
Sec.~\ref{sec_classical_pd} presents the model and simplified classical phase diagram.
Sec.~\ref{sec_lswt} shows spin-wave spectra of the key phases. 
In Sec.~\ref{sec_swt_inst} phase boundaries are discussed. 
Sec.~\ref{sec_avgs_tn} discusses finite-temperature transitions.
In Sec.~\ref{sec_cubic_axes} a transformation to the extended Kitaev-Heisenberg model
is given. Sec.~\ref{sec_dmrg} presents the DMRG results. We conclude in Sec.~\ref{sec_conc}, and the Appendices 
contain further details.

\section{Model and Classical phases}
\label{sec_classical_pd}

\subsection{Model}

In systems with spin-orbit coupling, magnetic degrees of freedom are entangled with the orbital orientations that are 
tied to the lattice due to crystal fields \cite{Witczak}. 
Because of that,  Hamiltonians of the low-energy effective pseudo-spins  involve 
bond-dependent interactions that obey only discrete symmetries of the underlying lattice, thus explicitly breaking 
spin-rotational symmetries \cite{Chen3}. 

The most general nearest-neighbor 
spin-orbit-induced anisotropic-exchange Hamiltonian, applicable to a variety of systems, can be written as \cite{Rau18}
\begin{align}
\hat{\cal H}=\sum_{\langle ij\rangle} \mathbf{S}^{\rm T}_i \hat{\bm J}_{ij} \mathbf{S}_j
\label{eq_Hij}
\end{align}
where ${\bf S}_i^{\rm T}\!=\!\left(S_i^{x},S_i^{y},S_i^{z}\right)$, $\langle ij\rangle$ denotes nearest-neighbor sites, 
and  $\hat{\bm J}_{ij}$ is a $3\!\times\! 3$ exchange matrix that depends on the  bond orientation.
Since the spin-rotational symmetries are, generally, absent, constraints on the matrix elements of $\hat{\bm J}_{ij}$
come solely from the space group symmetry of the lattice.

The effect of these constraints on the   Hamiltonian (\ref{eq_Hij}) 
for the triangular-lattice materials, such as YbMgGaO$_4$ and others, has been thoroughly 
discussed in Refs.~\cite{Chen3, Balents17,us,Chen5,ChenTm}.
Here we would like to provide a brief and intuitive derivation of the main results.

\begin{figure}
\includegraphics[width=0.99\linewidth]{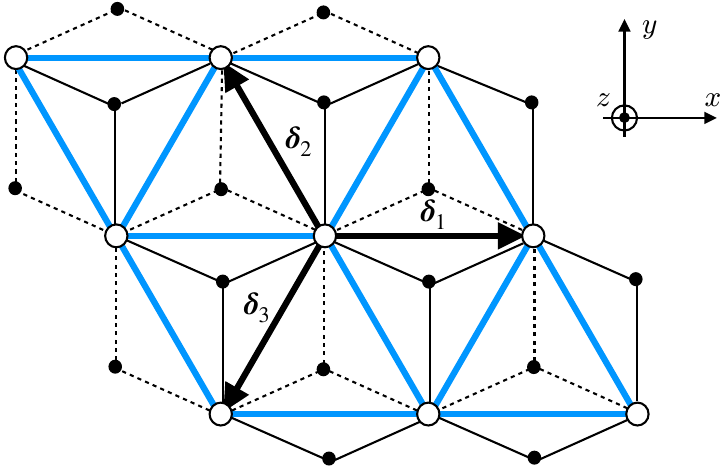}
\caption{A sketch of the triangular-lattice layer of magnetic ions (empty circles) embedded in the octahedra 
of ligands (black dots) with the primitive vectors. Thick (blue) bonds are between magnetic ions and 
ion-ligand bonds are the thin solid (dashed) lines for above (below) the plane. }
\label{fig_0}
\end{figure}

Consider the Hamiltonian (\ref{eq_Hij})  on the 
${\bm \delta}_1$ bond, see Fig.~\ref{fig_0}, with the $x$ axis parallel to it
\begin{eqnarray}
\hat{\cal H}_{ij}={\bf S}^{\rm T}_i \left( \begin{array}{ccc} 
J_{xx} & J_{xy} & J_{xz} \\ 
J_{yx} & J_{yy} & J_{yz} \\
J_{zx} & J_{zy} & J_{zz} 
\end{array}\right) {\bf S}_j,
\label{eq_H12}
\end{eqnarray}
As can be seen from Fig.~\ref{fig_0}, the symmetries of the lattice  are the $C_3$ rotation around the $z$ axis, 
$C_2$ rotation around each bond, site inversion symmetry $\mathcal{I}$, and two translations, 
$\mathcal{T}_1$ and $\mathcal{T}_2$ along ${\bm \delta}_1$ and ${\bm \delta}_2$, respectively \cite{Chen3}. 
These symmetries eliminate most of the elements of the exchange matrix. First, the 180${\degree}$ 
rotation around the ${\bm \delta}_1$ bond changes $y\rightarrow -y$ and $z\rightarrow -z$, 
but should leave the two-site form \eqref{eq_H12} invariant, leaving
us with
\begin{eqnarray}
\hat{\cal H}_{ij}={\bf S}^{\rm T}_i \hat{\bm J}_1 {\bf S}_j={\bf S}^{\rm T}_i \left( \begin{array}{ccc} 
J_{xx} & 0 & 0 \\ 
0 & J_{yy} & J_{yz} \\
0 & J_{zy} & J_{zz} 
\end{array}\right) {\bf S}_j.
\label{eq_H12a}
\end{eqnarray}
Then, inversion with respect to the bond center, which is a combination of the site inversion and 
$\mathcal{T}_1$ translation, and change $i\!\leftrightarrow\! j$ should also  
leave \eqref{eq_H12a} invariant, allowing only the symmetric off-diagonal term, 
$J_{zy}\!=\!J_{yz}$. Renaming it as $J_{zy}\!=\!J_{z\pm}$, and rewriting the diagonal terms using $XXZ$-like parametrization
$J_{zz}\!=\!\Delta  J$, with $J\!=\!(J_{xx}+J_{yy})/2$ and $J_{\pm\pm}\!=\!(J_{xx}-J_{yy})/4$
yields the two-site Hamiltonian for ${\bm \delta}_1$ in a ``spin-ice'' form \cite{Ross11}
\begin{eqnarray}
&&\hat{\cal H}_{ij}=J\Big(\Delta S^{z}_i S^{z}_j+S^{x}_i S^{x}_j+S^{y}_i S^{y}_j\Big)
\label{eq_H_1b}\\
&&+2J_{\pm\pm}\Big(S^{x}_i S^{x}_j-S^{y}_i S^{y}_j\Big)
+J_{z\pm}\Big(S^{z}_i S^{y}_j+S^{y}_i S^{z}_j\Big).
\nonumber
\end{eqnarray}

For the other bonds, using the $C_3$ invariance with the $z$ axis 
to transform (\ref{eq_H12a}) to the ${\bm \delta}_\alpha$ bond in Fig.~\ref{fig_0}, 
changes the $\hat{\bm J}_1$ matrix in  Eq.~(\ref{eq_H12a}) to 
$\hat{\bm J}_\alpha\!=\!\hat{\mathbf{R}}^{-1}_{\alpha} \hat{\bm J}_1 \hat{\mathbf{R}}_{\alpha}$ where
\begin{equation}
\hat{\mathbf{R}}_\alpha=\left( \begin{array}{ccc} 
\cos\tilde{\varphi}_\alpha & \sin\tilde{\varphi}_\alpha &0\\ 
-\sin\tilde{\varphi}_\alpha & \cos\tilde{\varphi}_\alpha&0\\
0&0&1
\end{array}\right),
\end{equation}
is the rotation matrix, 
or, explicitly 
\begin{eqnarray}
\hat{\bm J}_\alpha = \left( \begin{array}{ccc} 
J+ 2J_{\pm\pm}\tilde{c}_\alpha & -2J_{\pm\pm}\tilde{s}_\alpha &-J_{z\pm} \tilde{s}_\alpha\\ 
-2J_{\pm\pm}\tilde{s}_\alpha & J- 2J_{\pm\pm}\tilde{c}_\alpha & J_{z\pm} \tilde{c}_\alpha\\
-J_{z\pm} \tilde{s}_\alpha & J_{z\pm} \tilde{c}_\alpha & \Delta J
\end{array}\right),
\label{Jij}
\end{eqnarray}
where the abbreviations are 
$\tilde{c}_\alpha\!=\!\cos\tilde{\varphi}_\alpha$ and $\tilde{s}_\alpha\!=\!\sin\tilde{\varphi}_\alpha$.

Altogether, the most general Hamiltonian \eqref{eq_Hij} on the triangular lattice becomes 
\begin{align}
\label{HJpm}
{\cal H}=&\sum_{\langle ij\rangle}\Big\{
J \Big(S^{x}_i S^{x}_j+S^{y}_i S^{y}_j+\Delta S^{z}_i S^{z}_j\Big)\\
+&2 J_{\pm \pm} \Big[ \Big( S^x_i S^x_j - S^y_i S^y_j \Big) \tilde{c}_\alpha 
-\Big( S^x_i S^y_j+S^y_i S^x_j\Big)\tilde{s}_\alpha \Big]\nonumber\\
 +&J_{z\pm}\Big[ \Big( S^y_i S^z_j +S^z_i S^y_j \Big) \tilde{c}_\alpha 
 -\Big( S^x_i S^z_j+S^z_i S^x_j\Big)\tilde{s}_\alpha \Big]\Big\},\nonumber
\end{align}
where $\tilde{c}(\tilde{s})_\alpha\!=\!\cos(\sin)\tilde{\varphi}_\alpha$ as above, the bond angles $\tilde{\varphi}_\alpha$
are that of the primitive vectors ${\bm \delta}_\alpha$ with the $x$ axis, $\tilde{\varphi}_\alpha\!=\!\{0,2\pi/3,-2\pi/3\}$, 
and the spin operators are in crystallographic axes that are tied to the lattice, see Fig.~\ref{fig_0}.

The Hamiltonian (\ref{HJpm}) is naturally divided in the bond-independent $XXZ$ 
part and the bond-dependent anisotropic $J_{\pm \pm}$ and $J_{z\pm}$ terms, 
also referred to as the pseudo-dipolar terms \cite{Balents17},
which generally break continuous spin-rotational symmetries down to discrete ones.   

\begin{figure*}
\includegraphics[width=1.0\linewidth]{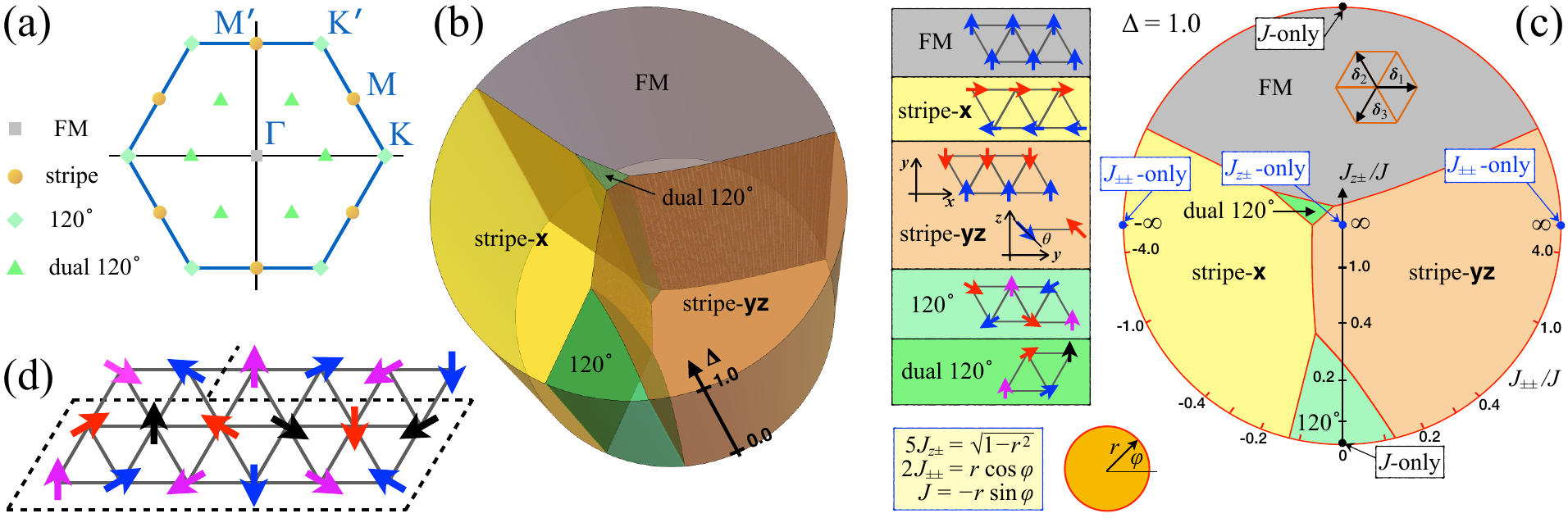}
\caption{(a) Brillouin zone of the triangular
lattice with  the ordering vectors for each phase. (b)
The 3D classical phase diagram of the model (\ref{HJpm}) for the single-$\mathbf{Q}$  states, 
see text for their description. The vertical axis is  $0\!\leq\!\Delta\!\leq\!1$. 
(c) The 2D cut of (b) at $\Delta\!=\!1$ with a sketch of spin structures 
and parametrization of the radial and angular coordinates.  (d) 
A detailed sketch of the dual 120${\degree}$ state. It is, generally, noncoplanar, see Sec.~\ref{sec_cubic_axes}, and 
consists of 12 sublattices with the unit cell indicated.}
\label{fig_pd_classical}
\end{figure*}

\subsection{Classical phase diagram}

Since there are four parameters in the model \eqref{HJpm}, its parameter space is three dimensional,
with the fourth degree of freedom setting the energy scale. 
We apply one  physical constraint on it by assuming the easy-plane type of the $XXZ$ anisotropy, $0\!\leq\!\Delta\!\leq\!1$, 
as this is natural for a variety of  layered systems of interest \cite{Chen1, MM}.
As was  pointed out in Ref.~\cite{Chen3}, one needs to consider only positive $J_{z\pm}$ since the 
global $\pi$ rotation around the $z$ axis that should leave the Hamiltonian invariant is equivalent to 
changing $J_{z\pm}\!\rightarrow \!-J_{z\pm}$.
With these two constraints, one can map the entire 3D parameter space on a cylinder,
with the vertical axis represented by the $XXZ$ anisotropy $\Delta$, $J_{z\pm}$ as the radial, and 
$J_{\pm\pm}/J$ as the polar variables, so that each horizontal cut represents 
an entire 2D phase space of the model (\ref{HJpm}) for a fixed $\Delta$. 

In  Figures~\ref{fig_pd_classical}(b) and (c) we use a parametrization 
\begin{eqnarray}
\label{eq_s_param}
\left(J,2J_{\pm\pm},5J_{z\pm}\right)=\left(-r\sin\varphi,r\cos\varphi,\sqrt{1-r^2}\right),
\end{eqnarray}
such that $\sqrt{J^2+(2J_{\pm\pm})^2+(5J_{z\pm})^2}\!=\!1$, with the choice of numerical coefficients made
to exaggerate the region where all parameters are of the same order, 
$J_{\pm\pm}, J_{z\pm}\!\alt\!J$. 
The $XY$ and the Heisenberg limits of the $XXZ$ part of the model, $\Delta\!=\!0$ and $\Delta\!=\!1$, correspond 
to the bottom and the top of the cylinder, respectively. The  2D phase diagram of the latter is 
shown in more detail in Fig.~\ref{fig_pd_classical}(c).

By the energy minimization for the {\it commensurate single-$\mathbf{Q}$} states,
there are five ordered phases shown in the classical phase diagram
in Figs.~\ref{fig_pd_classical}(b) and (c) with spin arrangements shown 
in Fig.~\ref{fig_pd_classical}(c) and the ordering vectors in the Brillouin zone in Fig.~\ref{fig_pd_classical}(a). 
Two of the phases  are favored by the $XXZ$ part 
of the model (\ref{HJpm}), the ferromagnetic phase with $\mathbf{Q}\!=\!\Gamma$ and the 
$120{\degree}$ phase with  $\mathbf{Q}_{120{\degree}}\!=\!K$,  
for $J\!<\!0$ and $J\!>\!0$, respectively.

Although not so obvious, the two stripe phases are favored by the $J_{\pm\pm}$ and $J_{z\pm}$ 
bond-dependent terms that are selecting the states that satisfy them fully on one of the bonds 
and partially on the others \cite{topography}.
While the stripes have the same ordering vector, $\mathbf{Q}\!=\!M$ or equivalent, 
they differ by the mutual orientation of spins and bonds. In the stripe-${\bf x}$ phase, 
favored only by the $J_{\pm\pm}\!<\!0$ term,
spins are in plane and along one of the bonds.
In the stripe-${\bf yz}$ phase, spins are perpendicular to one of the bonds and are also tilted out of 
plane, taking advantage of both $J_{\pm\pm}$ and $J_{z\pm}$  terms,
see Fig.~\ref{fig_pd_classical}(c).  

The remaining small region is the dual 120${\degree}$ phase with ordering vector 
$\mathbf{Q}_{\rm d120{\degree}}\!=\!K/2$ or equivalent.
While the reason for this terminology and the logic behind this state will be made clear in Sec.~\ref{sec_cubic_axes}, 
it is related to the conventional 120${\degree}$ order via the so-called  Klein duality transformation \cite{Khal_ProgSupp}.  
The dual 120${\degree}$ state is a 12-sublattice state, which is a combination 
of four counterrotating 120${\degree}$ structures, shown in different colors in Fig.~\ref{fig_pd_classical}(d). 

The classical per-site energies  of these phases [in units of $S(S+1)$] are as follows
\begin{align}
& E_\text{FM}=3J, \ \ \ E_{120{\degree}}=-\frac{3}{2}J, \ \  \ E_\text{\rm stripe-{\bf x}}=-J+4J_{\pm\pm},\nonumber\\
& E_\text{\rm stripe-{\bf yz}}=-\widetilde{J}_{c}-\Delta J-\sqrt{4J_{z\pm}^2+\widetilde{J}_{c}^2},
\label{eq_Ecl_all}\\
& E_\text{d120{\degree}}=\frac12\left(\widetilde{J}_{c}+\Delta J-\sqrt{4J_{z\pm}^2+\widetilde{J}_{c}^2}\right),\nonumber
\end{align}
where we abbreviated $\widetilde{J}_{c}\!=\!\left[J(1-\Delta)+4J_{\pm\pm}\right]/2$
and the (negative) out-of-plane tilt angle of the stripe-${\bf yz}$ state is found by energy minimization as:
$\tan 2 \theta\!=\!-2 J_{z\pm}/\widetilde{J}_{c}$.

Our discussion reproduces results of previous studies of the single-$\mathbf{Q}$ ordered ground states 
of the model (\ref{HJpm})  that have identified the stripe and 120${\degree}$ states for $J\!>\!0$
\cite{Chen3, Balents17,Balents18, topography}; see, in particular, Ref.~\cite{Wang17}. We also extend
the same approach to the entire available parameter space.
However, as was first pointed out in Ref.~\cite{multiQ} using numerical energy optimization in large clusters,  
more complicated multi-$\mathbf{Q}$ ordered structures become ground states of the classical model 
near the phase boundaries 
of the stripe and 120${\degree}$ regions. We confirm these findings by studying instabilities in the spin-wave spectra,---%
see Sec.~\ref{sec_swt_inst},---and also identify a different instability within the 120${\degree}$ phase for a range 
of $\Delta$ near the Heisenberg limit of the $XXZ$ term in (\ref{HJpm}).
This  instability is toward a different multi-$\mathbf{Q}$ state with a 
long-range spiral-like distortion of the 120${\degree}$  order
that is similar to the $Z_2$ vortex state discussed previously for the triangular-lattice Kitaev-Heisenberg (or $K$--$J$) model 
\cite{Trebst_tr,Ioannis}. Our discussion of the correspondence of the model (\ref{HJpm}) to the $K$--$J$ model 
in Sec.~\ref{sec_cubic_axes} extends this earlier finding to a broader range of parameters.

Since the continuous spin-rotational symmetries in model (\ref{HJpm}) are broken, one generally expects 
gapped spin excitations in all ordered phases. This is indeed true for the stripe-${\bf x}$ and 
for most of the stripe-${\bf yz}$ parts of the phase diagram, where in the latter part, a peculiar accidental degeneracy
exists along a 2D surface of parameters  that is related to duality relations discussed in Sec.~\ref{sec_duality}.
However, the ferromagnetic (FM) and the 120${\degree}$ states of the classical model 
exhibit an accidental degeneracy everywhere in their 3D regions of stability, also referred to as an emergent symmetry in 
Ref.~\cite{Balents17}. This degeneracy means that their spectra are gapless and 
the orientation of their spin configurations is not fixed within the lattice plane for 
$\Delta\!<\!1$, or at all for $\Delta\!=\!1$, 
offering examples of the emergent $U(1)$ and $O(3)$ symmetries, respectively.
However, since the model \eqref{HJpm} breaks rotational symmetry, 
it means that quantum and thermal fluctuations will select a preferred direction and gap out the spectrum. 
We discuss the outcomes of such a  quantum order-by-disorder effect in these phases in Sec.~\ref{sec_lswt}. 

It is worth noting that most of the phase diagram in Fig.~\ref{fig_pd_classical} is occupied by the states with 
quantum fluctuations that remain insignificant even for the quantum $S\!=\!1/2$ limit, such as stripes and FM states.
Thus, by and large, strong anisotropic terms on the triangular lattice do not seem to result in a massive degeneracy
of the classical states that would indicate possible exotic phases, contrary to some early expectations \cite{Chen1}.

\section{Magnon spectra}
\label{sec_lswt}

Although  the linear spin-wave spectra and transverse dynamical 
structure factors for the ordered single-${\bf Q}$ spin structures can be obtained numerically, see Ref.~\cite{Toth}
and Supplemental Material of Ref.~\cite{Rau_obd}, 
their analytical forms  can be tremendously useful and informative.
In this Section, we present the linear spin-wave theory (LSWT) for  four out of five ordered single-${\bf Q}$ 
phases shown in Fig.~\ref{fig_pd_classical} and discussed in Sec.~\ref{sec_classical_pd} above:
the stripe-${\bf x}$,  stripe-${\bf yz}$, 120${\degree}$ N\'{e}el, and ferromagnetic states. 
We do not consider the spectrum of the dual 120${\degree}$ state because of its complicated 12-sublattice structure.
Some of our results for the stripe phases have been been discussed previously either in  limiting cases \cite{us} 
or with minimal details \cite{Chen3}. We would like to emphasize that the spectrum of the 120${\degree}$ phase 
in the presence of the bond-dependent anisotropic terms has not been calculated previously. The same is true for the
ferromagnetic phase, which is, however, much simpler.

The spin-wave expansion requires a rotation of the axes on each site from the laboratory reference frame 
$\{x,y,z\}$, in which the Hamiltonian is typically written, to a {\it local} reference frame $\{\tilde{x},\tilde{y},\tilde{z}\}$
with the $\tilde{z}$ along the spin's quantization axis given by the classical energy minimization 
for a spin configuration
\begin{align}
\mathbf{S}_i =\hat{\bf R}_i \, \widetilde{\mathbf{S}}_i  \, ,
\label{eq_Srot}
\end{align}
where $\widetilde{\mathbf{S}}_i$ is the spin vector in the local reference frame at the site $i$ and $\hat{\bf R}_i$
is the rotation matrix for that site.
Thus, the spin Hamiltonian in (\ref{eq_Hij}) can be rewritten as 
\begin{align}
\hat{\cal H}=\sum_{\langle ij\rangle} \mathbf{S}_i^{\rm T} \hat{\bm J}_{ij} \mathbf{S}_j=
\sum_{\langle ij\rangle} \widetilde{\mathbf{S}}_i^{\rm T} 
\widetilde{\bm J}_{ij} \widetilde{\mathbf{S}}_j\, ,
\label{eq_Hij_loc}
\end{align}
where the ``rotated'' exchange matrix is
\begin{align}
\widetilde{\bm J}_{ij} =\hat{\bf R}_i^{\rm T}\hat{\bm J}_{ij}\hat{\bf R}_j\, .
\label{eq_Jij_rot}
\end{align}
This procedure is followed by a standard bosonization of spin operators via the
Holstein-Primakoff transformation, $\widetilde{S}_i^z\!=\!S-a_i^\dag a_i$, $\widetilde{S}^+\!\approx\!a_i\sqrt{2S}$,
and a subsequent diagonalization of the harmonic part of the Hamiltonian.

In the following, we will use a  two-stage rotation from the 
crystallographic to local reference frame, $\hat{\bf R}\!=\!\hat{\bf R}_{\varphi}\cdot \hat{\bf R}_{\theta}$,
with the first rotation in the  $x$-$y$ plane by an angle $\varphi$ 
\begin{eqnarray}
\hat{\bf R}_{\varphi}=
\left( \begin{array}{ccc} 
\cos \varphi & -\sin\varphi & 0\\ 
\sin\varphi & \cos\varphi & 0\\
0 & 0 & 1 
\end{array}\right),
\label{Rphi}
\end{eqnarray}
and the second  rotation in the $x$-$z$ plane
\begin{eqnarray}
\hat{\bf R}_{\theta}=\left( \begin{array}{ccc} \sin \theta& 0 & \cos \theta \\ 
0 & 1 & 0\\
-\cos\theta & 0 & \sin\theta \end{array}\right).
\label{Rth}
\end{eqnarray}
Thus, the spin transformation  is given by
\begin{eqnarray}
\hat{\bf R}_i=\left( \begin{array}{ccc} \sin \theta_i \cos \varphi_i & -\sin\varphi_i & \cos \theta_i\cos\varphi_i \\ 
\sin\theta_i\sin\varphi_i & \cos\varphi_i & \cos\theta_i\sin\varphi_i\\
-\cos\theta_i & 0 & \sin\theta_i \end{array}\right),
\label{eq_frametransform_gen}
\end{eqnarray}
where $\theta_i$ are the out-of-plane canting angles of the spins. 
While not unique, this approach is physically intuitive and is motivated by the  in-plane
orderings.   

A particularly simple and useful example that is used below 
is that of the states that are coplanar with the  lattice plane,
such as stripe-${\bf x}$ and 120${\degree}$ N\'{e}el states, for which all $\theta_i\!=\!0$ and 
the rotation matrix (\ref{eq_frametransform_gen}) simplifies to 
\begin{eqnarray}
\hat{\bf R}_i=\left( \begin{array}{ccc} 0& -\sin\varphi_i & \cos\varphi_i \\ 
0 & \cos\varphi_i & \sin\varphi_i\\
-1 & 0 & 0 \end{array}\right).
\label{eq_coplanar}
\end{eqnarray}
After this rotation, the Hamiltonian (\ref{HJpm}) that yields the 
 LSWT  becomes
\begin{align}
\mathcal{H}=&\sum_{\langle ij\rangle} \Big\{ J\Big[ \Delta \widetilde{S}^x_i \widetilde{S}^x_j +
\cos\left(\varphi_i -\varphi_j \right)\Big(\widetilde{S}^z_i \widetilde{S}^z_j +
\widetilde{S}^y_i \widetilde{S}^y_j \Big) \Big] \nonumber\\
&\label{Hcoplanar}
 \quad +2J_{\pm\pm} \cos \left( \tilde{\varphi}_\alpha+\varphi_i+\varphi_j\right)
   \Big( \widetilde{S}^z_i \widetilde{S}^z_j -\widetilde{S}^y_i \widetilde{S}^y_j\Big) \\
-&J_{z\pm}\Big[\cos \left( \tilde{\varphi}_\alpha-\varphi_j \right)\widetilde{S}^x_i \widetilde{S}^y_j +
\cos \left( \tilde{\varphi}_\alpha-\varphi_i \right)\widetilde{S}^y_i \widetilde{S}^x_j \Big]\Big\}\,,\nonumber
\end{align}
where $\varphi_{i(j)}$ are the angles of the spins with the $x$ axis in the laboratory frame
and we have omitted the  terms that contribute only to the anharmonic order of the SWT.

Choosing the ordering vector according to the classical energy minimization dictates the number of 
sublattices within the magnetic unit cell, $n_s$. Upon introducing the corresponding Holstein-Primakoff boson
species and  after the Fourier transform the LSWT Hamiltonian reads
\begin{align}
\mathcal{H}_2 =\frac{1}{2}\sum_{\bf k}\hat{\bf x}_{\bf k}^\dagger 
\hat{\bf H}_{\bf k}\hat{\bf x}_{\bf k}^{\phantom{\dagger}}.
\end{align}
where $\hat{\bf x}^\dag_{\bf k}=\left( a^\dag_{\bf k}, b^\dag_{\bf k},\dots, 
a^{\phantom \dag}_{\bf -k},b^{\phantom \dag}_{\bf -k},\dots\right)$ is a vector of length $2 n_s$ 
and $\hat{\bf H}_{\bf k}$ is a $2 n_s\times 2 n_s$ matrix
\begin{eqnarray}
\label{LSWTmatrix}
\hat{\bf H}_{\bf k}=S 
\left( \begin{array}{cc} 
\hat{\bf A}^{\phantom \dagger}_{\bf k} &  \hat{\bf B}^{\phantom \dagger}_{\bf k}\\[0.5ex] 
\hat{\bf B}^\dagger_{\bf k}  & \hat{\bf A}^*_{\bf -k}
\end{array}\right),
\end{eqnarray}
where we isolated the factor  $S$. Then the eigenvalues of $\hat{\bf g} \hat{\bf H}_{\bf k}$,
$\{\varepsilon_{1{\bf k}},  \varepsilon_{2{\bf k}}, \dots,  -\varepsilon_{1-{\bf k}},  -\varepsilon_{2-{\bf k}},
\dots\}$, give magnon energies. Here $\hat{\bf g}$ is a
diagonal matrix $[1,1,\dots,-1,-1,\dots]$, see Ref.~\cite{Colpa}.

\subsection{Stripe-x phase}

\begin{figure}
\includegraphics[width=0.99\linewidth]{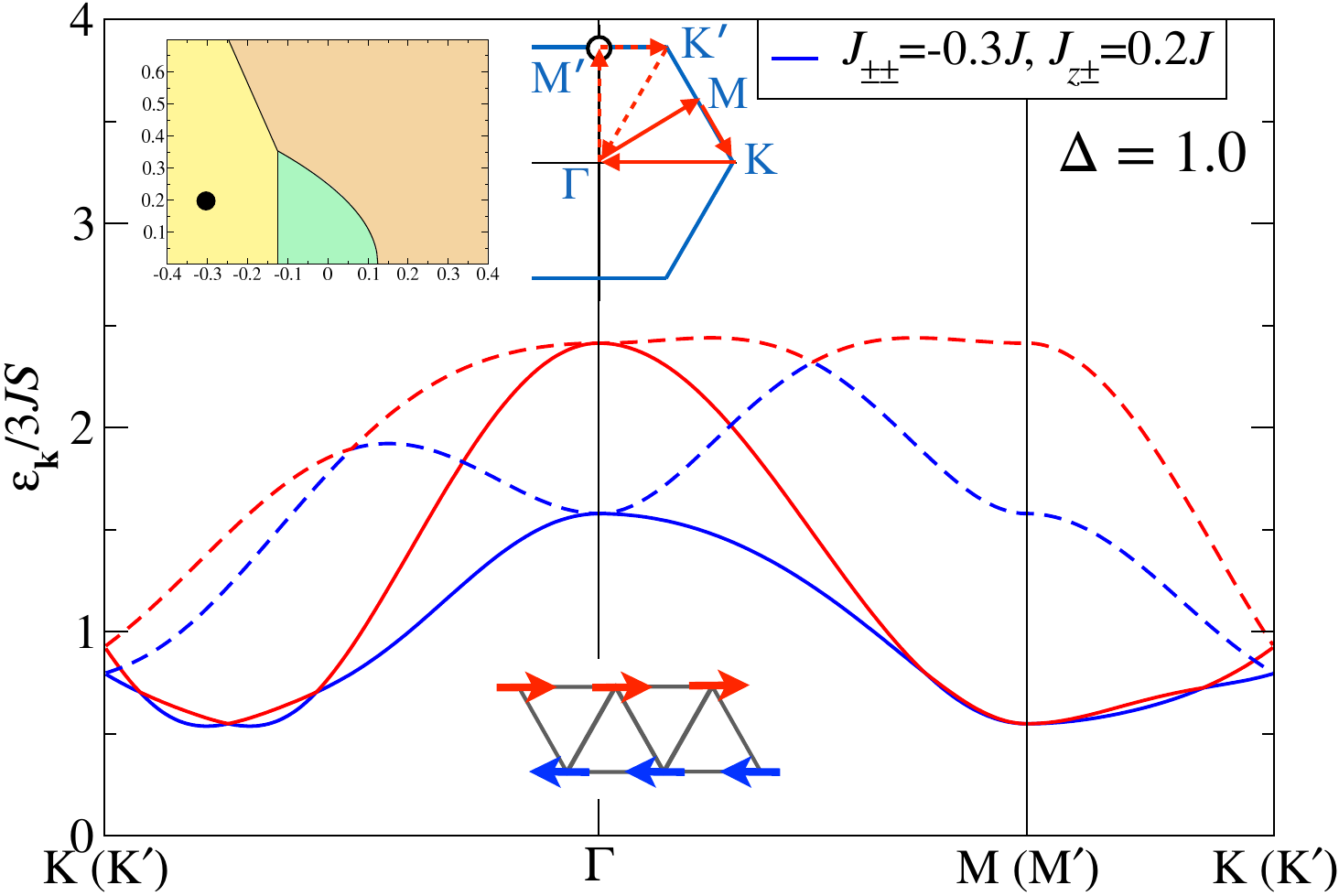}
\caption{Magnon energies $\varepsilon_{1,2{\bf k}}$ (upper and lower curves) from Eq.~(\ref{E12stripe}) 
for $J\!>\!0$, $\Delta\!=\!1.0$, $J_{\pm\pm}\!=\!-0.3J$, and $J_{z\pm}\!=\!0.2J$. 
Solid (dashed) lines are along the $\Gamma MK\Gamma$ ($\Gamma M^\prime K^\prime\Gamma$) paths;  
the ordering vector at $M^\prime$ is indicated in the inset. 
Left inset: the 2D $J_{z\pm}$--$J_{\pm\pm}$ phase diagram for $\Delta\!=\!1.0$
with the point indicating the chosen set of parameters.}
\label{fig_lswtxyz}
\end{figure}

For the stripe-${\bf x}$ phase, the ordering vector is one of the M-points and spins are split into two sublattices
with spins along one of the bonds, $\varphi_i\!=\!\varphi_0+({\bf Q}{\bf r}_i)$. 
Choosing the ordering vector ${\bf Q}\!=\!M^\prime\!=\!(0,2\pi/\sqrt{3})$ 
for convenience, dictates $\varphi_A\!=\!0$ and $\varphi_B\!=\!\pi$ for the $A$ and $B$ sublattices, 
see Fig.~\ref{fig_pd_classical}.
After some straightforward algebra with the Hamiltonian in (\ref{Hcoplanar}), separating the bonds into intra- and
inter-sublattice ones, yields the $4\times 4$ LSWT matrix in (\ref{LSWTmatrix}) with the $2\times 2$ matrices
$\hat{\bf A}^{\phantom \dag}_{\bf k}$ and $\hat{\bf B}^{\phantom \dag}_{\bf k}$  
\begin{eqnarray}
\label{ABstripex}
\hat{\bf A}^{\phantom \dag}_{\bf k}=
\left( \begin{array}{cc} 
A_{\bf k} &  B_{\bf k}\\
B^*_{\bf k}  & A_{\bf k}
\end{array}\right), \quad \quad 
\hat{\bf B}^{\phantom \dag}_{\bf k}=
\left( \begin{array}{cc} 
D_{\bf k} &  C_{\bf k}\\
C_{\bf k}  & D^*_{\bf k}
\end{array}\right),
\end{eqnarray}
and  the elements of the matrices given by
\begin{align}
A_{\bf k}&=2\big(J-4J_{\pm\pm}\big)+
   \big(J(1+\Delta)-2J_{\pm\pm}\big)\cos {\bf k}\bm{\delta}_1\, ,\nonumber\\
B_{\bf k}&=\big(J \left(\Delta-1\right) -J_{\pm \pm}+iJ_{z\pm}\big)
\big( \cos {\bf k}\bm{\delta}_2 +\cos {\bf k}\bm{\delta}_3\big),\nonumber\\
\label{ABCDstripex}
C_{\bf k}&=\big(J\left(1+\Delta\right)+J_{\pm \pm}\big)
\big(\cos {\bf k}\bm{\delta}_2 +\cos {\bf k}\bm{\delta}_3\big),\\
D_{\bf k}&=\big(J\left(\Delta -1\right)+2J_{\pm\pm}-2iJ_{z\pm}\big) \cos {\bf k}\bm{\delta}_1\,.\nonumber
\end{align}
In this case, the eigenvalues of the Hamiltonian matrix \eqref{LSWTmatrix} can be found analytically 
by diagonalizing $\big(\hat{\bf g}\hat{\bf H}_{\bf k}\big)^2$ instead of $\hat{\bf g}\hat{\bf H}_{\bf k}$, giving 
two magnon branches
\begin{align}
\label{E12stripe}
\varepsilon_{1,2{\bf k}}^2&=S^2\bigg(A_{\bf k}^2+|B_{\bf k}|^2-C_{\bf k}^2-|D_{\bf k}|^2\\\nonumber
&\pm\sqrt{4|A_{\bf k} B_{\bf k}- C_{\bf k} D_{\bf k}|^2-|B_{\bf k} D^*_{\bf k}-B_{\bf k}^{*}D_{\bf k} |^2}
\bigg).
\end{align}
Figure \ref{fig_lswtxyz} gives an example of the two magnon branches for a representative 
point within the stripe-${\bf x}$ phase along the two paths in the Brillouin zone shown in the inset
with the ordering vector at $M^\prime$ also indicated.
The left inset shows a sketch of the $J_{z\pm}$--$J_{\pm\pm}$ phase diagram for $\Delta\!=\!1.0$
in Cartesian coordinates with the point indicating the chosen set of parameters. 
As expected, the spectrum is fully gapped due to the symmetry-breaking terms.

\subsection{Stripe-yz phase}

In the stripe-${\bf yz}$ phase, the ordering vector and the number of sublattices are 
the same as in the stripe-${\bf x}$ case, but the spins 
are tilted away from the laboratory frame, see Fig.~\ref{fig_pd_classical}(c). 
Choosing again  ${\bf Q}\!=\!M^\prime$ fixes all $\varphi_i\!=\!\pi/2$ while the out-of-plane angles 
are $\theta_i\!=\!\theta+({\bf Q}{\bf r}_i)$, leading to $\theta_A\!=\!\theta$ and $\theta_B\!=\!\theta+\pi$, 
where $\theta\!<\!0$ according to the classical energy minimization in Sec.~\ref{sec_classical_pd}~B. Thus, 
the matrix of rotation to local reference frames (\ref{eq_frametransform_gen}) becomes 
\begin{eqnarray}
\hat{\bf R}_i=\left( \begin{array}{ccc} 0 & -1 & 0 \\ 
\sin\theta_i & 0 & \cos\theta_i \\
-\cos\theta_i & 0 & \sin\theta_i \end{array}\right),
\label{eq_rotation_yz}
\end{eqnarray}
which yields a somewhat lengthly ``rotated'' Hamiltonian  Eq.~(\ref{eq_Hij_loc}), shown in Appendix~\ref{app_stryz}. 
The subsequent spin-wave expansion yields the LSWT matrix (\ref{LSWTmatrix}) 
with the structure identical to the stripe-${\bf x}$ case, Eq.~(\ref{ABstripex}) above. 
We delegate explicit expressions for $A_{\bf k}, B_{\bf k}, C_{\bf k}$, and  $D_{\bf k}$ for the stripe-${\bf yz}$ phase
to  Appendix~\ref{app_stryz} for brevity. Needless to say, the magnon energies are calculated 
using the same expressions (\ref{E12stripe}).

\begin{figure}
\includegraphics[width=0.99\linewidth]{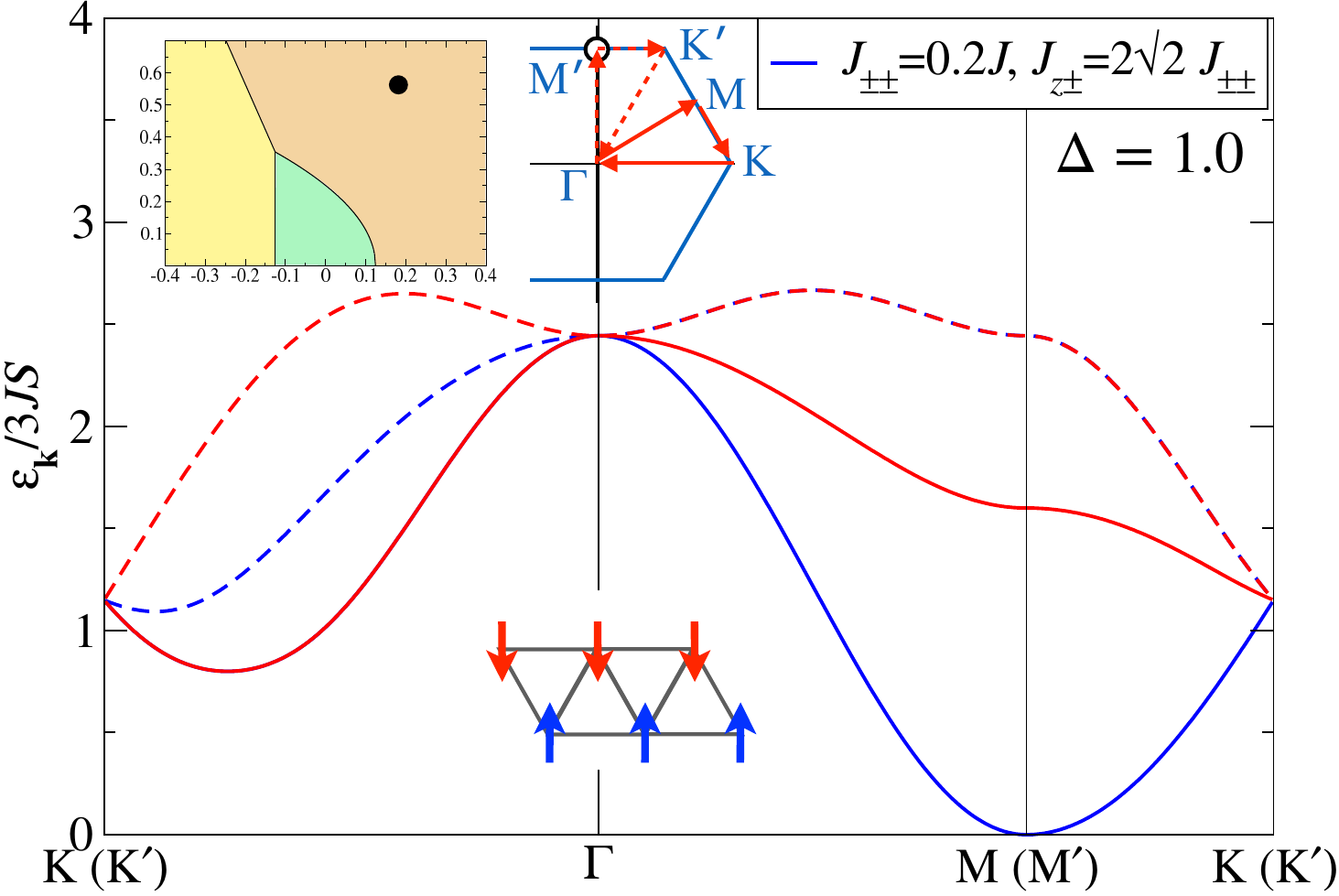}
\vskip -0.1cm
\caption{Same as Fig.~\ref{fig_lswtxyz} for $J\!>\!0$,  $\Delta\!=\!1.0$, 
$J_{\pm\pm}\!=\!0.2J$, and $J_{z\pm}\!=\!2\sqrt{2}J_{\pm\pm}$ 
within the stripe-${\bf yz}$ phase. See text for the discussion of the accidental degeneracy at 
the $M$ point.}
\label{fig_lswtxyz2}
\vskip -0.3cm
\end{figure}

Figure \ref{fig_lswtxyz2} shows magnon energies for a point within the stripe-${\bf yz}$ phase, along the same two paths 
in the Brillouin zone as in Fig.~\ref{fig_lswtxyz},
and for the same ordering vector ${\bf Q}\!=\!M^\prime$.
Although one expects the spectrum to be fully gapped due to the symmetry-breaking terms as in the stripe-${\bf x}$ case,
there is a gapless mode at the $M$ point, which is not the ordering vector, and the 
spectrum is fully gapped at $M^\prime$. 
The gap vanishes because of the choice of parameters in Fig.~\ref{fig_lswtxyz2}, $\Delta\!=\!1.0$, 
$J_{\pm\pm}\!=\!0.2J$, and $J_{z\pm}\!=\!2\sqrt{2}J_{\pm\pm}$, that belong to a 2D surface of parameters 
\begin{align}
J_{z\pm}=\big( 4J_{\pm\pm}+J(1-\Delta)\big)/\sqrt{2},
\label{eq_yzzero}
\end{align}
which yields an accidental degeneracy and a pseudo-Goldstone (gapless) mode.
For $\Delta\!=\!1$, Eq.~(\ref{eq_yzzero}) gives a line of points defined by $J_{z\pm}\!=\!2\sqrt{2} J_{\pm\pm}$ and the 
out-of-plane tilt of spins in this case is given by $\tan\theta\!=-\!1/\sqrt{2}$.
At this stage, the condition  (\ref{eq_yzzero}) can be seen as a serendipitous discovery  
that is sufficient to ensure an accidental degeneracy of the magnon band. 
For the LSWT matrix elements in (\ref{ABstripex}), Eq. (\ref{eq_yzzero}) follows from requiring $D_M\!=\!0$, 
which  simultaneously makes $A_M\!=\!|B_M|$, see Appendix~\ref{app_stryz} for their explicit expressions.

While this accidental degeneracy looks similar to the case of the $J_1$--$J_2$ model on a triangular lattice \cite{Chubukov}, 
the symmetry associated with it is hidden. 
We identify this 2D surface of accidental degeneracies in the parameter space
 as corresponding to an extended Kitaev-Heisenberg model, which 
possesses emergent symmetries that naturally lead to the pseudo-Goldstone modes
in the quasiclassical limit; see Sec.~\ref{sec_duality}.
For a generic choice of parameters within the 3D region of the stripe-${\bf yz}$ phase, 
the accidental degeneracy is not present, but
the gap remains small in a large portion of the phase diagram. In Appendix~\ref{app_stryz}, we provide two additional plots of  
magnon energies for the stripe-${\bf yz}$ phase to substantiate this point.

We also point out that the presence of this gapless or nearly gapless mode does not affect quantum fluctuations, 
which remain small even in the quantum $S\!=\!1/2$ limit throughout the stripe-${\bf yz}$ phase.
However, the ordering N\'{e}el temperature is necessarily suppressed in a vicinity of the 2D   
surface in Eq.~(\ref{eq_yzzero}) due to the Mermin-Wagner theorem. We discuss this dichotomy in Sec.~\ref{sec_avgs_tn}.

\vspace{-0.15cm}

\subsection{120${\degree}$ phase}

For the 120${\degree}$ phase, the ordering vector is one of the $K$-points, see Fig.~\ref{fig_pd_classical}(a), 
and spins form a three-sublattice structure.
For instance, choosing the ordering vector ${\bf Q}\!=\!K^\prime\!=\!(4\pi/3,0)$ and fixing 
the angle on the $A$ sublattices to $\varphi_A\!$, defines the other angles via 
$\varphi_i\!=\!\varphi_0+({\bf Q}{\bf r}_i)$. Thus,
$\varphi_B\!=\!\varphi_A+(\bm{\delta}_1{\bf Q})\!=\!\varphi_A-2\pi/3$ and 
$\varphi_C\!=\!\varphi_A-(\bm{\delta}_1{\bf Q})\!=\!\varphi_A+2\pi/3$ as expected for the 120${\degree}$ pattern.
Using these phases in the Hamiltonian  (\ref{Hcoplanar}), after some tedious but straightforward algebra, one
obtains the $3\times 3$ $\hat{\bf A}_{\bf k}$ and $\hat{\bf B}_{\bf k}$ 
LSWT matrices in (\ref{LSWTmatrix})
\begin{align}
\hat{\bf A}_{\bf k}&=3J \left(\hat{\bf \mathbb{1}}+\frac14\left(2\Delta-1 \right)\hat{\bf M}_1 \right)
-3J_{\pm\pm} \hat{\bf M}_2-i\frac{3J_{z\pm}}{2}\hat{\bf M}_3,\nonumber\\
\label{AB120}
\hat{\bf B}_{\bf k}&=\frac{3J}{4}\left(1+2\Delta\right) \hat{\bf M}_1 
+3J_{\pm\pm} \hat{\bf M}_2-i\frac{3J_{z\pm}}{2}\hat{\bf M}_4.
\end{align}
Since the nearest-neighbor interactions couple only different sublattices, the resultant $\hat{\bf M}_i$
matrices are all traceless and are built from the hopping amplitudes, which are either bond independent, 
as for $\hat{\bf M}_1$ in the $XXZ$ term
\begin{eqnarray}
\hat{\bf M}_1=\left( \begin{array}{ccc} 
0 & \gamma & \gamma^*\\ 
\gamma^* & 0 & \gamma\\
\gamma & \gamma^* & 0 
\end{array}\right), 
\quad \gamma=\frac{1}{3}\sum_{\alpha}  e^{i{\bf k} \bm{\delta}_\alpha},
\end{eqnarray}
or bond dependent, as  for the ones originating from the  anisotropic $J_{\pm\pm}$ 
and $J_{z\pm}$ terms 
\begin{eqnarray}
&&\hat{\bf M}_2=
 \left( \begin{array}{ccc} 
0 & \gamma_{AB} & \gamma^*_{AC}\\ 
\gamma^*_{AB} & 0 & \gamma_{BC}\\
\gamma_{AC} & \gamma^*_{BC} & 0 
\end{array}\right),\\
&&\hat{\bf M}_3=
\left( \begin{array}{ccc} 
0 & \overline{\gamma}_{AB} & \overline{\gamma}_{AC}^*\\ 
-\overline{\gamma}_{AB}^* & 0 & \overline{\gamma}_{BC}\\
-\overline{\gamma}_{AC} & -\overline{\gamma}_{BC}^* & 0 
\end{array}\right),\\
&&\hat{\bf M}_4=
\left( \begin{array}{ccc} 
0 & \widetilde{\gamma}_{AB} & \widetilde{\gamma}_{AC}^*\\ 
\widetilde{\gamma}_{AB}^* & 0 & \widetilde{\gamma}_{BC}\\
\widetilde{\gamma}_{AC} & \widetilde{\gamma}_{BC}^* & 0 
\end{array}\right),
\end{eqnarray}
where
\begin{eqnarray}
&&\gamma_{RS}=\frac{1}{3}\sum_{\alpha}  e^{i{\bf k} \bm{\delta}_\alpha} 
\cos \left( \varphi_R +\varphi_S +\tilde{\varphi}_\alpha \right),\\
&&\overline{\gamma}_{RS}=\frac{1}{3}\sum_{\alpha}  e^{i{\bf k} \bm{\delta}_\alpha} 
\big(\cos  \left(\varphi_R-\tilde{\varphi}_\alpha\right) -\cos\left(\varphi_S-\tilde{\varphi}_\alpha\right)  \big),\nonumber\\
&&\widetilde{\gamma}_{RS}=\frac{1}{3}\sum_{\alpha}  e^{i{\bf k} \bm{\delta}_\alpha} 
\big(\cos  \left(\varphi_R-\tilde{\varphi}_\alpha\right) +\cos\left(\varphi_S-\tilde{\varphi}_\alpha\right) \big), \ \ \ \ \
\nonumber
\end{eqnarray}
and sums over $\alpha$ involve only three primitive vectors $\bm{\delta}_\alpha$.

\begin{figure}
\includegraphics[width=0.99\linewidth]{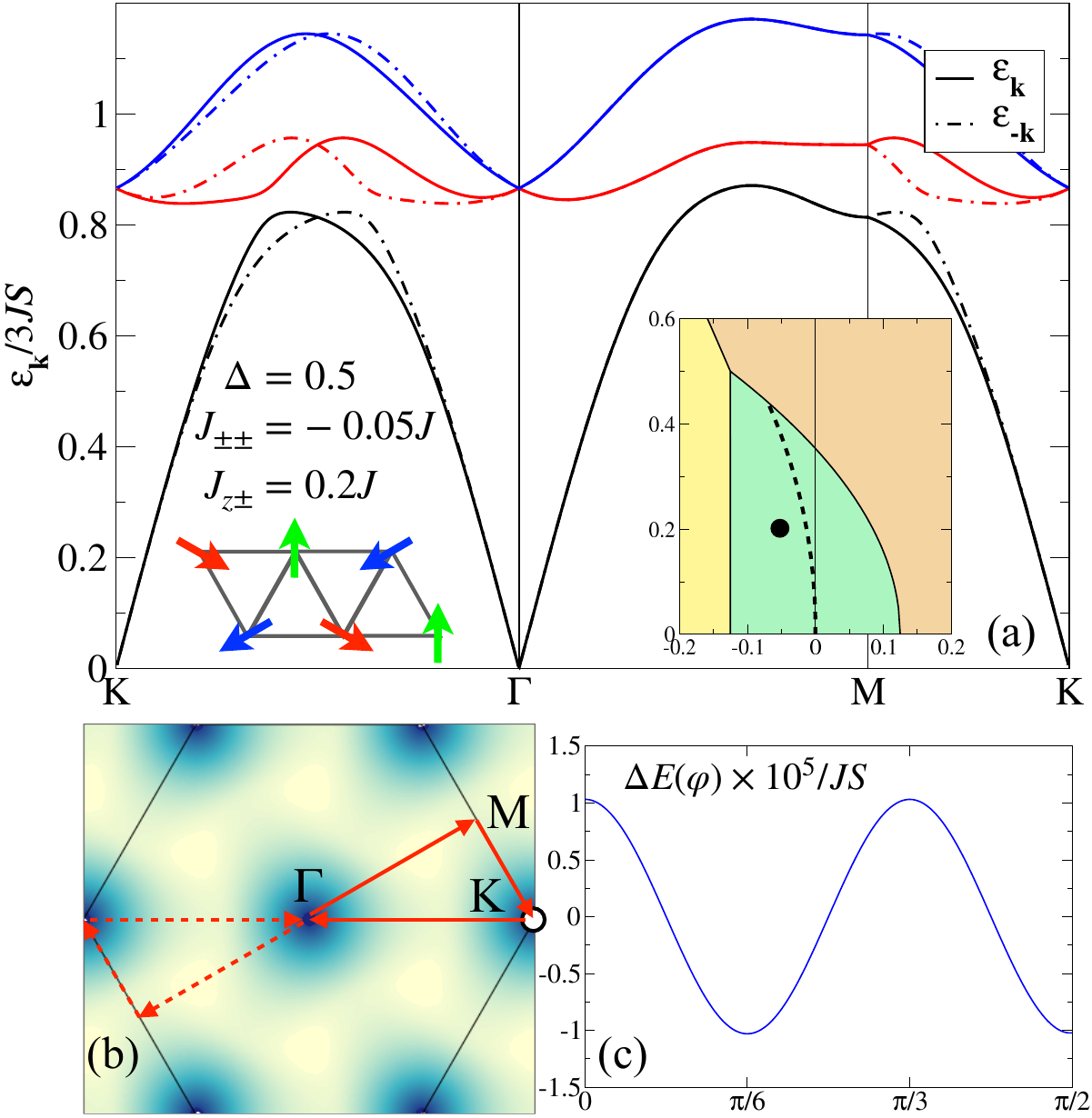}
\caption{(a) Magnon energies for a point in the 120${\degree}$ phase 
along two reciprocal ${\bf k}$-contours (solid and dashed) in (b), with the ordering vector indicated. 
(b)  Intensity plot of the lower branch. (c) The $\varphi$-dependent part of the zero-point energy vs $\varphi$ 
that selects the structure sketched in (a). The spectrum retains  $C_3$, but not  ${\cal I}$ symmetry, see text.}
\label{fig_120}
\end{figure}

With the $\hat{\bf A}_{\bf k}$ and $\hat{\bf B}_{\bf k}$ matrices (\ref{AB120}) written out explicitly, the $6\!\times\!6$
LSWT Hamiltonian (\ref{LSWTmatrix}) has to be diagonalized numerically.
Figure \ref{fig_120}(a) shows the resultant magnon spectrum for a representative point in the 
120${\degree}$ phase for $J_{\pm\pm}\!<\!0$ indicated in the insets along the two paths in the Brillouin zone shown 
in Fig.~\ref{fig_120}(b), which also shows an intensity map of the lowest magnon branch. 
Note that due to the three-sublattice structure, the magnetic Brillouin zone is one-third of the full one and
the spectrum in Fig.~\ref{fig_120}(a) is symmetric with the middle of the $\Gamma-K$ line, so  the $K$ and
$\Gamma$ points are equivalent. Thus the spectrum possesses only one Goldstone mode for $\Delta\!<\!1$. 

While one expects the  spectrum to be gapless because of the emergent $U(1)$ symmetry of the classical model 
discussed in Sec.~\ref{sec_classical_pd}~B, the unusual feature in Figs.~\ref{fig_120}(a) and \ref{fig_120}(b) is the 
nonreciprocal character of the magnon dispersions, $\varepsilon_{\alpha,{\bf k}}\!\neq\!\varepsilon_{\alpha,-{\bf k}}$ 
\cite{nonreciprocal,umbrella}, which is due to the $J_{z\pm}$ term that breaks 
the inversion symmetry of the LSWT Hamiltonian  
in the  120${\degree}$ state. While 
$\hat{\bf H}_{\bf -k}\!\neq\!\hat{\bf H}_{\bf k}$ does not automatically imply the nonreciprocity, 
the $J_{z\pm}$ term also makes $\hat{\bf H}_{\bf -k}^*\!\neq\!\hat{\bf H}_{\bf k}$, which together 
seem to be a sufficient condition. 
As one can see in Fig.~\ref{fig_120}(b), the $2\pi/3$-rotation symmetry of the spectrum is preserved, 
but the inversion is not.
An additional figure showing the same behavior for $J_{\pm\pm}\!>\!0$ is given in Appendix~\ref{app_stryz}.

As discussed in Sec.~\ref{sec_classical_pd}~B,  anisotropic terms do not contribute to classical energy 
of the 120${\degree}$ state, yielding an accidental $U(1)$ degeneracy for $\Delta\!<\!1$. 
Since the magnon spectra  depend on the angles of spins with  bonds, this degeneracy will 
be lifted by zero-point fluctuations via the order-by-disorder mechanism \cite{CCL,Henley,Rau_obd} 
that should select a preferred spin direction and open a gap in the spectrum.
In the 120${\degree}$ structure, fixing an angle of one spin fixes the rest, so 
the quantum energy correction is given by 
$\delta E(\varphi)\!=\!-3JS/2+\sum_{\alpha,{\bf k}} \varepsilon_{\alpha,{\bf k}}(\varphi)/6$, 
where $\varepsilon_{\alpha,{\bf k}}(\varphi)$ are the magnon energies that depend on the angle $\varphi$ of a spin
in one of the sublattices with the $x$ axis, and the sum is over the full Brillouin zone. 

In Fig.~\ref{fig_120}(c) we show this quantum correction with its average value 
subtracted, $\Delta E(\varphi)\!=\!\delta E(\varphi)-\langle\delta E(\varphi)\rangle$ vs $\varphi$.
Thus, for the given parameters,  fluctuations pin spins to the bond directions in the manner
shown in the sketch of the 120${\degree}$ configuration in Fig.~\ref{fig_120}(a). That is,
each spin is perpendicular to one of the bonds and bisects an angle of the triangle,
corresponding to the choice $\varphi\!=\!\pi/6+\pi n /3$. 
This choice is for the parameters to the left of the dashed line in the inset in Fig.~\ref{fig_120}(a).
To the right of it,  a state with spins along the bonds is chosen with  
$\varphi\!=\!\pi n /3$, see Appendix~\ref{app_stryz}. Curiously, the fact that the energy minimum must switch 
implies that the $U(1)$ symmetry is retained on a 2D surface within the 120${\degree}$ 
phase.

\subsection{Ferromagnetic phase}

For the ferromagnetic phase, the ordering vector is at ${\bf Q}\!=\!\Gamma\!=\!(0,0)$
and the spin state can be described with a single-sublattice picture, i.e., all  $\varphi_i\!=\!\varphi$ and
$\theta_i\!=\!\theta$. As is mentioned in Sec.~\ref{sec_classical_pd}, for $\Delta\!<\!1$ the $XXZ$ anisotropy 
reduces the symmetry to $U(1)$ and makes  $\theta\!=\!0$. For $\Delta\!=\!1$, the classical 
energy is insensitive to the global direction of the ordered moment and the symmetry is $O(3)$.
Having this latter case in mind, the matrix of rotation to local reference frames (\ref{eq_frametransform_gen})
retains its general form and we list the relevant terms of the rotated exchange matrix  in Appendix~\ref{app_stryz}.
 
The spin-wave expansion yields a simple Hamiltonian   
\begin{align}
\mathcal{H}=3S\sum_{\bf k} \Big(A_{\bf k} a^\dagger_{\bf k} a^{\phantom \dagger}_{\bf k}-
\frac{1}{2} \Big(B_{\bf k} a^{\dagger}_{\bf k} a^{\dagger}_{\bf -k}+\text{H.c.}\Big)\Big),
\end{align}
with somewhat involved expressions for $A_{\bf k}$ and $B_{\bf k}$
\begin{align}
&A_{\bf k}=-2J\big(\cos^2\theta+ \Delta\sin^2\theta\big)
+J\big(2+\left(1-\Delta\right)\cos^2\theta\big)\gamma_{\bf k}\nonumber \\
\label{AB_FM}
&\quad\quad -2J_{\pm\pm}\cos^2 \theta\,\gamma_{F,\bf k}-J_{z\pm}\sin 2 \theta\,\overline{\gamma}_{F,\bf k}\, ,  \\
&B_{\bf k}=J\left(1-\Delta\right)\cos^2\theta\,\gamma_{\bf k}
-2J_{\pm\pm}\left(1+\sin^2\theta\right)\gamma_{F,\bf k}\nonumber\\ 
&+J_{z\pm}\sin 2 \theta\,\overline{\gamma}_{F,\bf k}
+2i\Big(2J_{\pm\pm}\sin\theta\,\widetilde{\gamma}_{F,\bf k}+J_{z\pm}\cos \theta\,\hat{\gamma}_{F,\bf k}\Big)\, ,\nonumber
\end{align}
where, using notations $c_{\bar{\varphi}}\!=\!\cos\bar{\varphi}$ and $s_{\bar{\varphi}}\!=\!\sin\bar{\varphi}$,
\begin{eqnarray}
\label{gammaFM}
&&\gamma_{\bf k}=\frac{1}{3}\sum_{\alpha} c_\alpha , \\
&&\gamma_{F,\bf k}=\frac{1}{3}\sum_{\alpha}  c_\alpha
c_{\varphi_\alpha+2\varphi}, \ \ \ \widetilde{\gamma}_{F,\bf k}=\frac{1}{3}\sum_{\alpha}  c_\alpha
s_{\varphi_\alpha+2\varphi},\nonumber\\
&&\overline{\gamma}_{F,\bf k}=\frac{1}{3}\sum_{\alpha}  c_\alpha 
s_{\varphi-\varphi_\alpha}, \ \ \ 
\hat{\gamma}_{F,\bf k}=\frac{1}{3}\sum_{\alpha}  c_\alpha
c_{\varphi-\varphi_\alpha},\nonumber
\end{eqnarray}
where $c_\alpha\!=\!\cos {\bf k}\bm{\delta}_\alpha$ and sums over $\alpha$ involve only 
three primitive vectors $\bm{\delta}_\alpha$ as before. 

The magnon spectrum is given by
\begin{align}
\varepsilon_{\bf k}=3S\sqrt{A_{\bf k}^2-\left| B_{\bf k} \right|^2},
\end{align}
and expressions for $A_{\bf k}$ and $B_{\bf k}$ 
(\ref{AB_FM}) simplify considerably for $\Delta\!<\!1$ as it forces a coplanar state with $\theta\!=\!0$.

\begin{figure}
\includegraphics[width=0.99\linewidth]{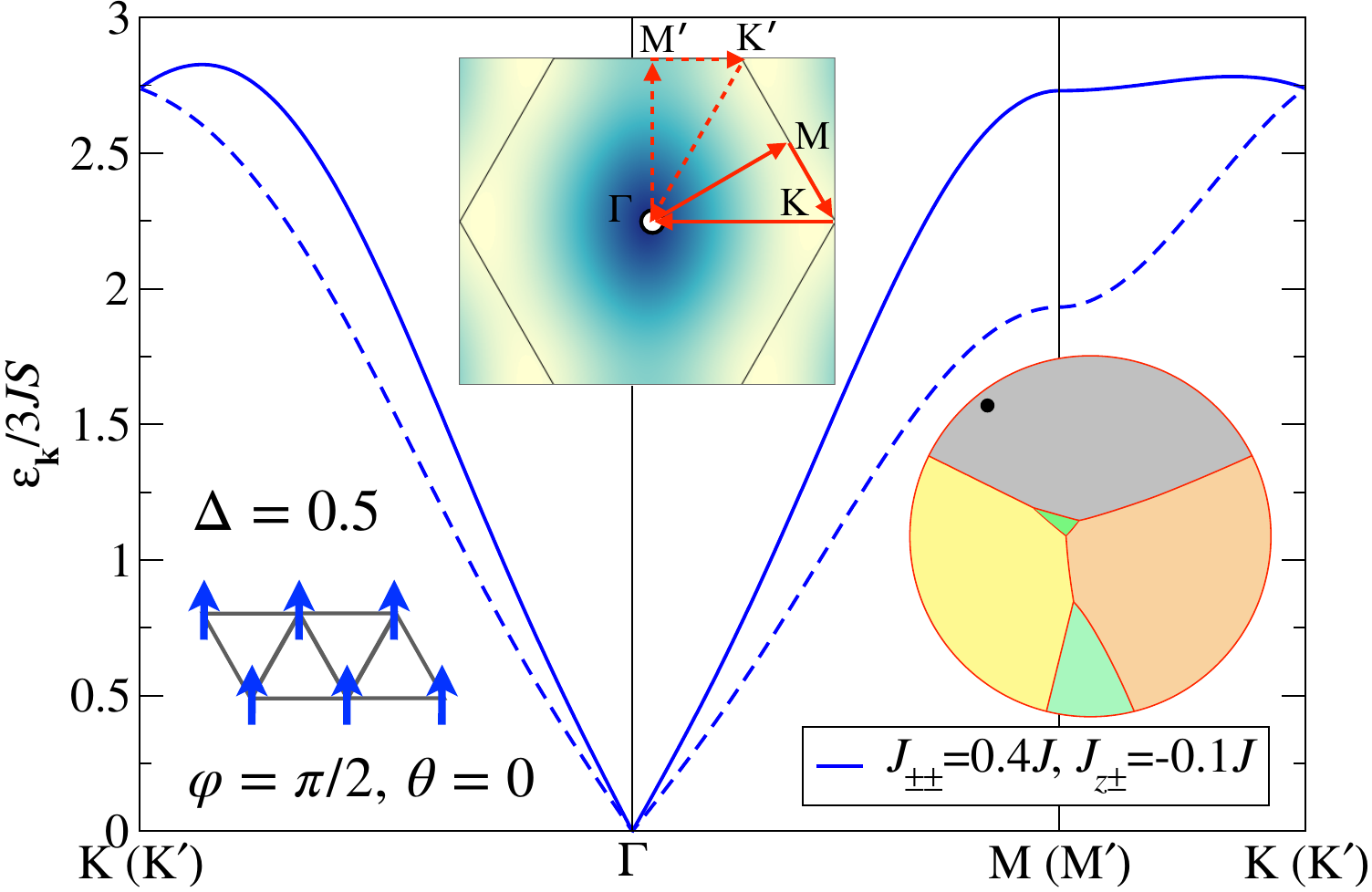}
\caption{Magnon energies along two contours shown in the inset  
for a representative point in the FM phase:  $J\!<\!0$,  $\Delta\!=\!0.5$, $J_{\pm\pm}\!=\!0.4J$, and 
$J_{z\pm}\!=\!-0.1J$. Inset shows the intensity plot of the magnon dispersion.
A sketch depicts the orientation selected  by the order-by-disorder mechanism.}
\label{fig_fm}
\vskip -0.3cm
\end{figure}

Figure~\ref{fig_fm} shows the magnon spectrum along the two paths in the Brillouin zone 
for a representative point in the FM phase as indicated in the insets. Note that $J\!<\!0$.
Similar to the 120${\degree}$ phase, the entire 3D region of the FM phase  has an accidental $U(1)$ degeneracy 
of the classical model for $\Delta\!<\!1$ and $O(3)$ for $\Delta\!=\!1$ as discussed in Sec.~\ref{sec_classical_pd}~B, 
hence the gapless spectrum. 
One can  show from Eqs.~(\ref{AB_FM}) and (\ref{gammaFM}) that 
the Goldstone mode should be linear in ${\bf k}$ for all $\Delta\!<\!1$ and quadratic for $\Delta\!=\!1$; 
see Appendix~\ref{app_stryz} for another plot of the magnon spectrum demonstrating the latter result.

Similar to the stripe states, once the direction of the ordered moment is chosen, the $C_3$ symmetry is broken, 
while inversion remains; see Fig.~\ref{fig_fm}.
The direction of the ordered moment in the FM state is selected via   order by disorder, 
  as for the 120${\degree}$ state. The same type of analysis for the parameters in Fig.~\ref{fig_fm} pins spins 
in the manner shown in the sketch of the ordered structure, i.e., 
perpendicular to one of the bonds with $\varphi\!=\!\pi/2+\pi n /3$.
We find that for $\Delta\!<\!1$, the preferred orientation of the ordered moment 
in the FM phase of the phase diagram switches from the perpendicular ($\varphi\!=\!\pi/2$)
to the parallel ($\varphi\!=\!0$) orientation in a region roughly near $J_{\pm\pm}\!\approx\!0$, 
although the details of such a switch can be more complicated.
In analogy with the 120${\degree}$ phase, this behavior 
also implies a surface where $U(1)$ symmetry is retained.

For $\Delta=1$, the quantum selection of the direction of the 
ordered moment is more complicated because of the $O(3)$ degeneracy. 
However,  in  case of parameters that fall on the line corresponding 
to the $K$--$J$ model (see Sec.~\ref{sec_cubic_axes}),
order by disorder selects the so-called cubic axes as a set of preferred directions, in agreement 
with Ref.~\cite{Ioannis}.

\vspace{-0.15cm}

\section{instabilities of magnon spectra}
\label{sec_swt_inst}

In the preceding Sections, principal single-${\bf Q}$ ordered phases of the model (\ref{HJpm})  
have been identified and their spectra of excitations have been found. Here we advocate a straightforward 
and fruitful approach that provides further insights into the phase diagram of the model via an investigation 
of the stability boundaries of the magnon spectra.
Generally, a magnon spectrum of a state is defined in a corresponding region of the phase diagram.
As a function of the model parameters, a magnon branch may soften and become imaginary, 
or have $\varepsilon_{\bf k}^2\!<\!0$ at one or at a set of ${\bf k}$ points, indicating a transition to a 
different state, which is referred to as magnon instability.   

The well-known examples of such a behavior are magnon-softening transitions in the $J_1$--$J_2$ model  
on the square and triangular lattices and in some of its extensions \cite{Chubukov,Starykh,Roscilde}, in which  
magnon instability occurs exactly at the classical phase boundary. 
There are also other examples, such as the $XXZ$ version of the same model on the triangular lattice  \cite{Ivanov},
in which the magnon stability region extends beyond the boundary of its classical phase. 
Such an overlap of the magnon stability regions of the neighboring phases  suggests a 
first-order transition between them \cite{Chubukov91}. Finally,  
magnon instability may occur before the boundaries of its classical phase are reached,  indicating an 
intermediate phase.

\vspace{-0.15cm}

\subsection{Stripe phases}

First, we explore magnon instabilities in the stripe phases. 
Ref.~\cite{multiQ} used a numerical optimization of energy in large clusters of classical spins and 
was first to demonstrate that the single-${\bf Q}$ classical phase diagram of the model (\ref{HJpm}) 
of the type shown in Fig.~\ref{fig_pd_classical} may be incomplete. 
It was shown that the more complicated  multi-$\mathbf{Q}$ states, which can be described 
as modulated stripe phases with incommensurate ordering vectors, create a layer of intermediate states
between the stripe and  120${\degree}$ phases.

\begin{figure}
\includegraphics[width=0.99\linewidth]{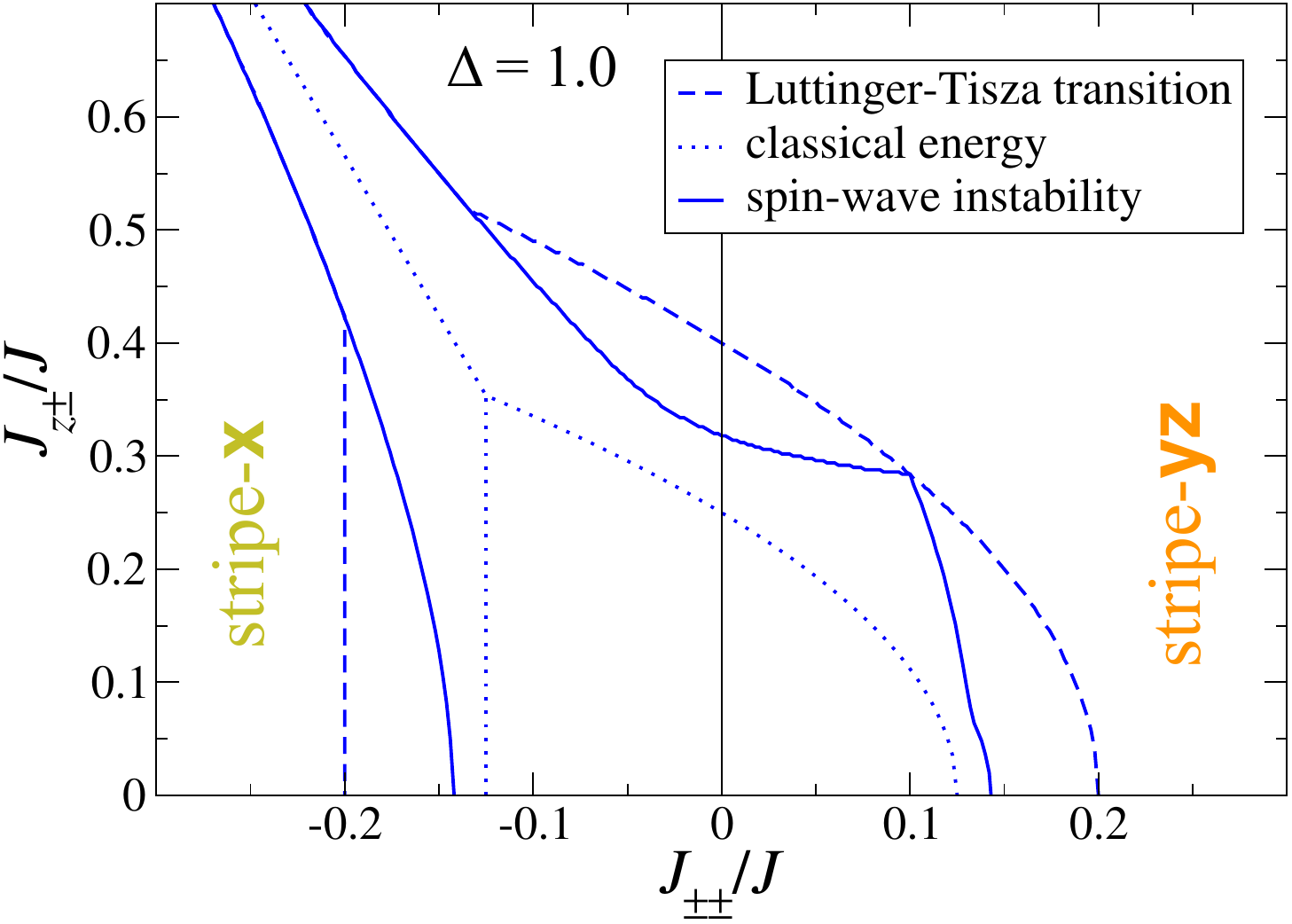}
\caption{The $J_{z\pm}$--$J_{\pm\pm}$ phase diagram of the model (\ref{HJpm}) for $\Delta\!=\!1$
with  transition boundaries from Fig.~\ref{fig_pd_classical} (dotted lines), from magnon 
instabilities in the stripe-${\bf x}$ and stripe-${\bf yz}$ phases (solid lines), and 
 by the LT approach (dashed lines), see text.}
\label{fig_inst_lt}
\vskip -0.3cm
\end{figure}

\begin{figure}
\includegraphics[width=0.99\linewidth]{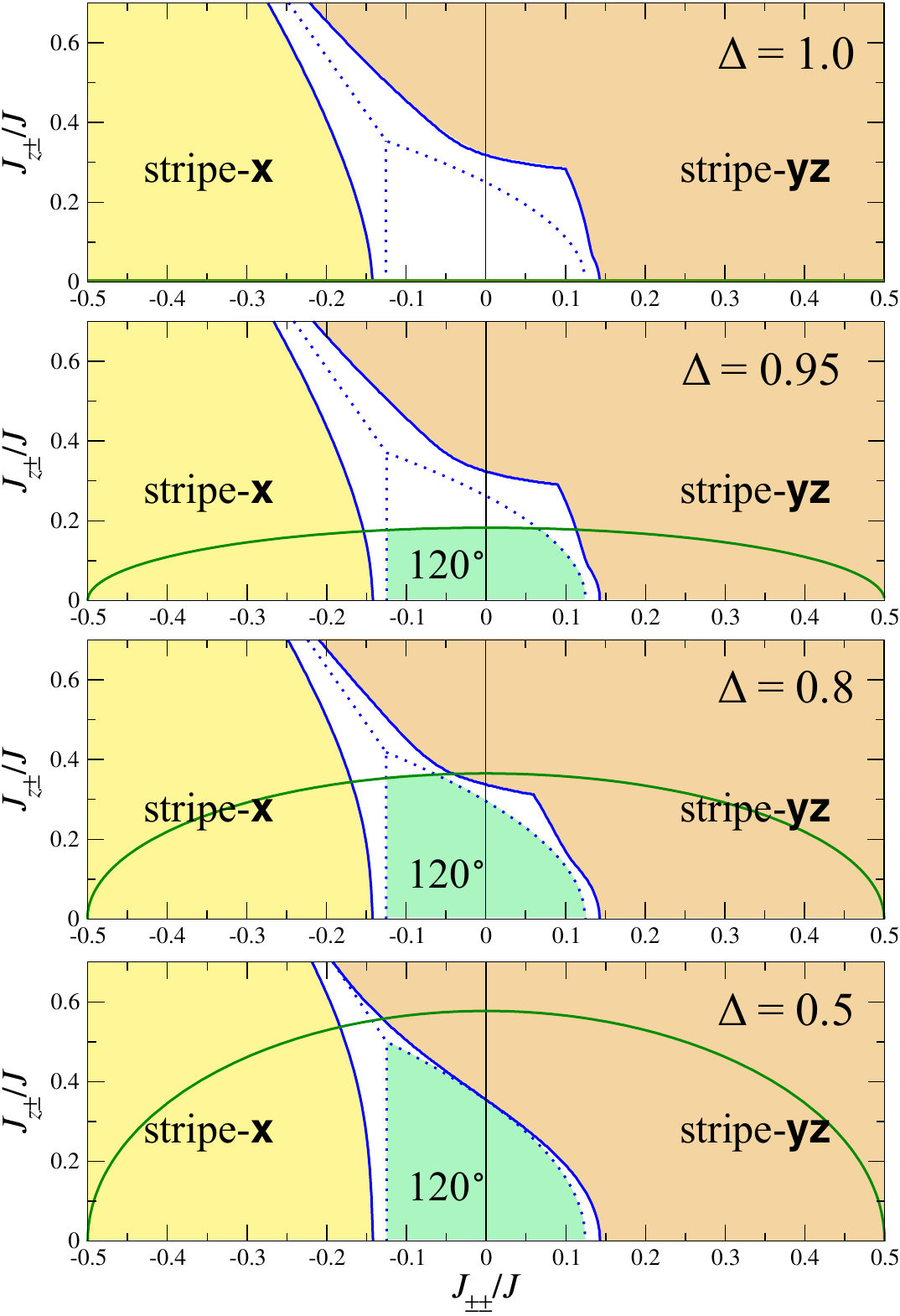}
\caption{Same as in Fig.~\ref{fig_inst_lt} for several $\Delta$;
dotted lines are  transitions in Fig.~\ref{fig_pd_classical}, solid lines are magnon 
instabilities of the stripe and 120$\degree$ phases (half-ellipse shape). 
Filled areas are stable single-${\bf Q}$ phases and blank regions are multi-${\bf Q}$ states.}
\label{fig_inst_120}
\vskip -0.3cm
\end{figure}

We support this finding using magnon instabilities and explore it in a wider region of   phase space.
Our Figs.~\ref{fig_inst_lt} and \ref{fig_inst_120} show the 2D $J_{z\pm}$--$J_{\pm\pm}$ 
phase diagrams for several values of $\Delta$ with magnon instability boundaries obtained from 
Eq.~(\ref{E12stripe}) for the stripe-${\bf x}$ and stripe-${\bf yz}$ phases. 
While not shown, magnon instability boundaries for the choice of $\Delta\!\approx\!0.56$ are virtually indistinguishable
from the phase boundaries of the multi-$\mathbf{Q}$ regions identified numerically in Ref.~\cite{multiQ}.
                
Moreover, the ${\bf k}$-points of the observed magnon instabilities also coincide with the ones 
identified in Ref.~\cite{multiQ} as the new ordering vectors.
For instance, for most of the boundary region with the 120${\degree}$ phase, 
the magnon instability occurs at the incommensurate vector $\mathbf{Q}$ located along the line between the   
$\Gamma$ and $K$ points in the Brillouin zone, in agreement with Ref.~\cite{multiQ}. 
Needless to say, our method of identifying multi-$\mathbf{Q}$ transition boundaries 
offers obvious advantages over the numerical approach, as  it  requires only the knowledge of the 
magnon spectrum of the single-$\mathbf{Q}$ states. 

In  Fig.~\ref{fig_inst_lt}, which displays magnon instability lines (solid lines) from the stripe phases for $\Delta\!=\!1$ 
that occur before the classical boundaries (dotted lines) are reached, we also compare them to the results
of the Luttinger-Tisza (LT) method \cite{lt_original}, shown by the dashed lines. 
The LT method, with the so-called ``weak'' constraint, essentially amounts to  
finding a minimum of the lowest eigenvalue of the Fourier transform of the exchange matrix (\ref{Jij})
$J^{\alpha \beta}(\mathbf{q})$ in the ${\bf q}$-space to find a single-$\mathbf{Q}$ ground state. 
It has been used previously to study the phase diagram of the model \eqref{HJpm} \cite{Chen3,multiQ,Balents18}
and was instrumental in identifying stripe structures as the ground states of the model.
However, it was noted that the LT method sometimes fails and finds an incommensurate state even 
when the true ground state is a commensurate single-${\bf Q}$ state \cite{Chen3,Balents18}. 
In fact, these problems are not specific 
to the model \eqref{HJpm} and have been known and understood as coming from 
the ``weak'' nature of the constraint \cite{ZaZh}.

Figure~\ref{fig_inst_lt} illustrates these consistent discrepancies of the LT method 
with the results of the magnon instability approach for the phase boundaries of the stripe phases.
Curiously, the two methods agree for one point $\{J_{\pm\pm},J_{z\pm}\}\!\approx\!\{0.1,0.3\}$,
which is, actually, a point that belongs to the line corresponding 
to the $K$--$J$ model, see Sec.~\ref{sec_cubic_axes}, so one can suspect an emergent $U(1)$/$O(3)$ symmetry along this line
as the reason for restoring the validity of the LT approach. The other range where LT agrees with our method is the 
range of $J_{z\pm}\!\agt\!0.5$ of the stripe-to-stripe boundary. The reason for that is not totally clear.
We also note that the LT method breaks down within the 120$\degree$ phase for any $\Delta$
at any non-zero $J_{\pm\pm}$, predicting incommensurate states \cite{Balents18}.
Altogether,  significant care must be exercised when using it.

Lastly, we point out that the energy gain created by the modulation of the stripe states in 
the multi-${\bf Q}$ structures was shown to be tiny, $\sim\!10^{-3}J$ \cite{multiQ}. 
This smallness may explain why no multi-${\bf Q}$ states were detected by DMRG \cite{topography}
for the quantum $S\!=\!1/2$ case. 

\subsection{120$\degree$ phase}

While the border regions of the stripe-${\bf x}$ and stripe-${\bf yz}$ phases with the 120${\degree}$ 
phase get replaced by the multi-${\bf Q}$ states, magnon instabilities within the 120${\degree}$ phase 
also indicate  intriguing behavior.
First, the 120${\degree}$ magnons defined by the algebra in Sec.~\ref{sec_lswt}~C are stable with respect to 
the anisotropic $J_{\pm\pm}$ term far beyond the classical boundaries of the 120${\degree}$ phase regardless of 
the value of the $XXZ$ anisotropy $\Delta$, with stability boundaries shown in Fig.~\ref{fig_inst_120} by the 
half-ellipse lines. This fact strongly suggests a first-order transition from the 120${\degree}$ to the stripe phases 
if quantum fluctuations are included.   

Second,  the spectrum stability with respect to the anisotropic $J_{z\pm}$ term is more drastic. In fact, 
while the $XXZ$ anisotropy provides a finite range of stability to magnons in the 120${\degree}$ phase, 
magnons in the  $\Delta\!=\!1$ limit  are unstable to any finite value of $J_{z\pm}$, 
see Fig.~\ref{fig_inst_120}. This instability is toward a different multi-${\bf Q}$ long-range spiral state, 
which corresponds to the spectrum softening at three symmetric ${\bf k}$ points in the immediate vicinity of the 
$\Gamma$ and $K$ points. This new state is very similar to the so-called $Z_2$ vortex state that was discussed 
intensely in the triangular-lattice Kitaev-Heisenberg model \cite{Trebst_tr,Ioannis}. In Sec.~\ref{sec_cubic_axes}, 
we discuss the correspondence of our model (\ref{HJpm}) to that model along the line in the $\Delta\!=\!1$ plane.
Our present consideration shows that a similar state exists in a significantly wider parameter space.

We note that both trends are commensurate with our previous DMRG results for the quantum $S\!=\!1/2$ case 
in Ref.~\cite{topography}. Namely, DMRG observes only a direct transition from the 120${\degree}$ to the stripe 
phases vs $J_{\pm\pm}$ and the transition is likely first order \cite{topography}.
The behavior of the transition boundary of the spin-liquid phase found in Ref.~\cite{topography}
and the overall shape of the phase space occupied by it is similar to the boundary of the long-range spiral state 
discussed above. The main difference is that the footprint of the spin liquid in the $\Delta\!=\!1$ plane is smaller
and an ordered 120${\degree}$-like phase is stabilized for $J_{z\pm}\!\alt\!0.27$.
Both the spiral ($Z_2$ vortex) and the spin-liquid phases shrink and disappear upon reducing $\Delta$. 
This similarity may argue for a possible relation between the two, suggesting the spin-liquid state of Ref.~\cite{topography}
to be a molten $Z_2$ vortex state rather than a molten 120${\degree}$ state. However, to add a word of caution, 
we have found no indication of any sizable shift of the peaks in the spin-liquid structure factor from the 
commensurate ordering vector of the 120${\degree}$ state, nor have we seen any traces of the chirality in the 
spin-liquid wave-function that is non-zero in the $Z_2$ vortex state, see Ref.~\cite{topography}.

\vspace{-0.15cm}

\section{Quantum and thermal fluctuations in the stripe phases}
\label{sec_avgs_tn}

In this Section, we verify whether strong anisotropies in the model (\ref{HJpm}) can result in strong quantum 
or thermal fluctuations in the stripe phases using  a quasiclassical approach.

\vspace{-0.25cm}

\subsection{Quantum fluctuations}

\begin{figure*}
\includegraphics[width=0.9\linewidth]{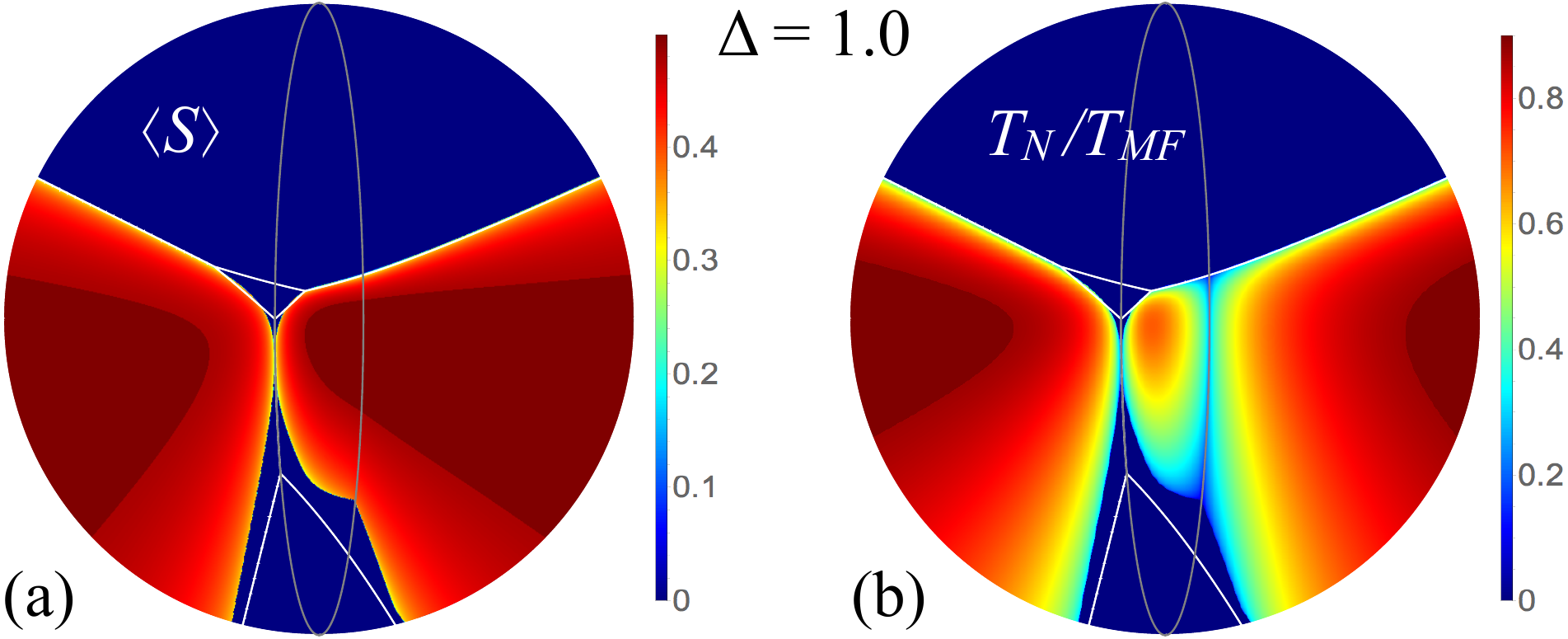}
\caption{Intensity maps of the (a) $T\!=\!0$ on-site ordered moment $\langle S \rangle$, Eq.~\eqref{eq_avgs}, and
(b) ratio $T_N/T_{\rm MF}$,  Eqs.~\eqref{eq_tn} and \eqref{eq_tcmf}, in the stripe phases for $S\!=\!1/2$. 
Parametrization is as in Fig.~\ref{fig_pd_classical}, $\Delta\!=\!1$. 
Ellipse shows $J_{z\pm}\!=\!2\sqrt{2}|J_{\pm\pm}|$ line, Eq.~\eqref{eq_yzzero}, see text.}
\label{fig_tn}
\vskip -0.3cm
\end{figure*}

The on-site ordered moment for $T\!=\!0$ within the LSWT takes the standard form
\begin{align}
\label{eq_avgs}
\langle S \rangle =S-\frac{1}{N}\sum_{\mu,\mathbf{k}}v_{\mu\mathbf{k}}^2,
\end{align}
where $N$ is the total number of lattice sites, the ${\bf k}$-sum is over the full Brillouin zone of the lattice,
$\mu\!=\!1,2$ numerates magnon branches in the stripe phase, $v_{\mu\mathbf{k}}$ are the Bogolyubov 
parameters of the transformation that diagonalizes the Hamiltonian in Eq.~(\ref{LSWTmatrix}), and we have 
used the symmetry of the sublattices in the stripe phases.

In the stripe phases, given the symmetry between   sublattices,
normalization of the Bogolyubov parameters is $u_{\mu{\bf k}}^2-v_{\mu{\bf k}}^2\!=\!1/2$
and their squares can be found analytically following Ref.~\cite{Plakida}
\begin{align}
u_{\mu{\bf k}}^2=
\frac{M_{11}\left[ \varepsilon_{\mu{\bf k}} \right]}
{{\displaystyle \prod_{\nu\neq \mu} \left( \varepsilon_{\mu{\bf k}} - \varepsilon_{\nu{\bf k}} \right)}},
\label{eq_uk}
\end{align}
where $M_{11}[\lambda]$ is the first minor of $\lambda\hat{\bf I}-{\bf \hat{g}\hat{H}_k}$ 
with the LSWT Hamiltonian matrix (\ref{LSWTmatrix}), the product in the denominator is over 
three out of four eigenvalues 
$\{\varepsilon_{1{\bf k}},  \varepsilon_{2{\bf k}},  -\varepsilon_{1-{\bf k}},  -\varepsilon_{2-{\bf k}}\}$
found in Sec.~\ref{sec_lswt}~A, and the explicit expression for  $M_{11}[\lambda]$ given by
\begin{align}
M_{11}[\lambda]&=\left(A_{\bf k}+\lambda\right)
\left(\lambda^2+C_{\bf k}^2+|D_{\bf k}|^2-A_{\bf k}^2\right)\nonumber\\
&+|B_{\bf k}|^2\left(A_{\bf k}-\lambda\right)-C_{\bf k} \left(D^{*}_{\bf k}B_{\bf k}+B_{\bf k}^{*}D_{\bf k}\right),
\end{align}
with the matrix elements for the stripe-${\bf x}$ phase given in Eq.~(\ref{ABCDstripex}) and 
for the stripe-${\bf yz}$ phase in Appendix~\ref{app_stryz}. 

The intensity map of $\langle S \rangle$ obtained from (\ref{eq_avgs}) 
for $S\!=\!1/2$ and $\Delta\!=\!1$ is shown in Fig.~\ref{fig_tn}(a)
throughout the stripe phases. One can see that the on-site magnetization is nearly classical 
and quantum fluctuations are really negligible for these regions   
all the way to the transition boundaries even in the quantum $S\!=\!1/2$ limit. 
While this observation may seem natural for the strongly gapped stripe-${\bf x}$ phase \cite{us}, it is somewhat
less obvious in the stripe-${\bf yz}$ phase because its spectrum has low-lying pseudo-Goldstone modes 
owing to an accidental degeneracy, see Sec.~\ref{sec_lswt}~B. 
As will be argued in Sec.~\ref{sec_duality}, this is because the model in this region is related to a ferromagnetic
state, which translates into a non-divergent contribution of the gapless region to the fluctuations in \eqref{eq_avgs}.

We also note that we do not show the results for the ordered moment 
$\langle S \rangle$ from the 120${\degree}$ and ferromagnetic 
states. In the former state,  the spectrum is unstable toward the spiral-like state for $\Delta\!=\!1$,
as discussed in Sec.~\ref{sec_swt_inst}~B above. In the latter state, the order-by-disorder selection
of the ordered moment direction is complicated for $\Delta\!=\!1$ and has to be done numerically, see 
Sec.~\ref{sec_lswt}~D. 

These problems are avoided for smaller $\Delta$ and we present such calculations
in our Fig.~\ref{fig_TN_d05}(a) for $\Delta\!=\!0.5$. Here the order-by-disorder selection 
in both 120${\degree}$ and FM phases is straightforward, 
see Sec.~\ref{sec_lswt}~C and D, and for the 120${\degree}$ state the formalism for calculating 
$\langle S \rangle$ is the same as in Eqs.~(\ref{eq_avgs})  and (\ref{eq_uk}).
Note that the empty regions in Fig.~\ref{fig_TN_d05}(a) are 
from the dual 120${\degree}$ state and the multi-${\bf Q}$ regions discussed in 
Sec.~\ref{sec_swt_inst}~A, for which it is simply challenging to perform similar calculations. 
There is, however, hardly any doubt that they are equally well ordered as the rest of the phase diagram. 

Thus, we reiterate, once again, that most of the phase diagram of the model (\ref{HJpm}) is occupied by the states 
with insignificant quantum fluctuations even for  $S\!=\!1/2$.

\subsection{Thermal fluctuations}

It is common in  studies of  frustrated magnets and their models to characterize 
them with the ``frustration ratio''
$f=\theta_\text{CW}/T_N$ \cite{Ramirez,Chen3},  the ratio of the Curie-Weiss temperature $\theta_\text{CW}$
to the actual ordering temperature $T_N$, as a measure of a proximity to potentially exotic states. 

A more refined alternative to $\theta_\text{CW}$ in  anisotropic models 
is the mean-field transition temperature \cite{Gingras} 
\begin{align}
T_{\rm MF}=-\frac{S(S+1)}{3 k_B} \ \lambda_{\rm min} (\mathbf{Q}),
\label{eq_tcmf}
\end{align}
where $\lambda_{\rm min} (\mathbf{Q})$ is the lowest eigenvalue of the Fourier transform of the 
exchange matrix (\ref{Jij}), 
$\hat{\bm J}_{ij}$, at the ordering vector. For the stripe orders, 
$\mathbf{Q}\!=\!M^\prime$ and $\lambda_{\rm min} (\mathbf{Q})$ 
is simply the classical energy per unit cell from Eq.~\eqref{eq_Ecl_all}:
$\lambda_{\bf x}\!=\!2E_{\rm stripe-{\bf x}}$ for the stripe-${\bf x}$  phase and 
$\lambda_{\bf yz}\!=\!2E_{\rm stripe-{\bf yz}}$ for  the stripe-${\bf yz}$ phase.

\begin{figure*}
\includegraphics[width=0.9\linewidth]{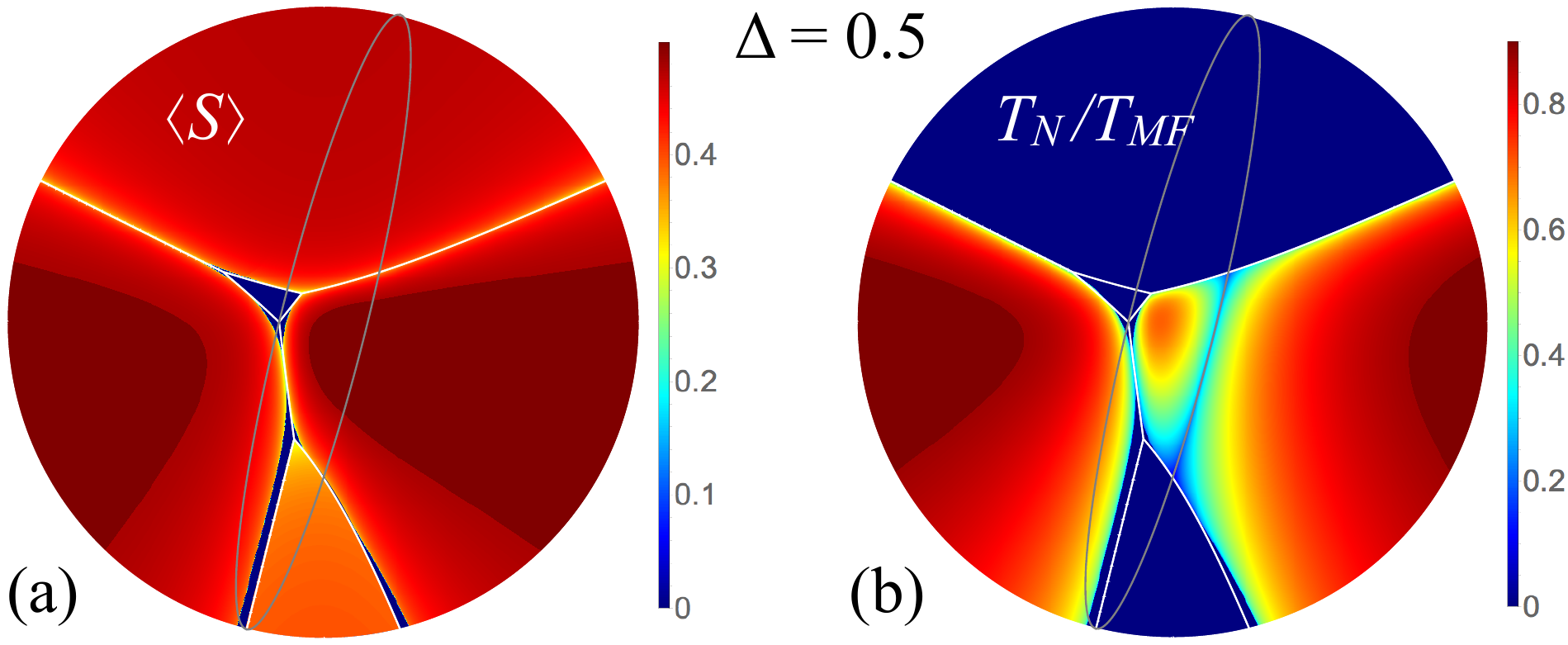}
\caption{Same as in Fig.~\ref{fig_tn} for $\Delta\!=\!0.5$ with $\langle S \rangle$ for the 120${\degree}$  
and FM phases in  (a).
Ellipse shows the line from Eq.~\eqref{eq_yzzero}.}
\label{fig_TN_d05}
\end{figure*}

The transition temperature can be found from the vanishing point of the ordered moment within the LSWT   
\begin{align}
\langle S \rangle_T =\langle S \rangle-\frac{1}{N}\sum_{\mu,\mathbf{k}} 
\left( u_{\mu\mathbf{k}}^2 +v_{\mu\mathbf{k}}^2 \right) n\left(\varepsilon_{\mu\mathbf{k}} \right) ,
\label{eq_avgsT}
\end{align}
where $n\left(\varepsilon_{\mu\mathbf{k}} \right)$ is the magnon occupation number. However,
this approach is known to overestimate  $T_N$ \cite{Katanin} as it leads to $\langle S \rangle_T\!\propto\!(T_N-T)$ 
for $T\!\rightarrow\!T_N$, not to a power law. 
Instead, we use  the self-consistent RPA method of Refs.~\cite{Tyablikov,Plakida}, 
in which one introduces the $T$-dependence in the magnon spectrum as
$\widetilde{\varepsilon}_\mathbf{k}\!=\!2\langle S \rangle_T\varepsilon_\mathbf{k}$.
Then the spin Green's function is given by \cite{Tyablikov}
\begin{align}
\langle S^{-}_\mathbf{k} S^+_\mathbf{-k} \rangle_\omega=2\langle S \rangle_T\sum_{\mu} 
\left(\frac{u_{\mu\mathbf{k}}^2}{\omega - \widetilde{\varepsilon}_{\mu\mathbf{k}}}-
\frac{v_{\mu\mathbf{k}}^2}{\omega +\widetilde{\varepsilon}_{\mu\mathbf{k}}}\right),
\end{align}
and Eq.~\eqref{eq_avgsT} is replaced by a self-consistent condition 
\begin{eqnarray}
\label{eq_app_rpa_s}
\langle S \rangle_T =\frac{1}{2}-
\frac{2\langle S\rangle_T}{N}  \sum_{\mu,\mathbf{k}} 
\Big( u_{\mu\mathbf{k}}^2 n\left(\widetilde{\varepsilon}_{\mu\mathbf{k}} \right)-
v_{\mu\mathbf{k}}^2 n\left(-\widetilde{\varepsilon}_{\mu\mathbf{k}} \right) \Big). \ \ \ \ \ \ 
\end{eqnarray}
At $T\!\rightarrow\!T_N$, the ordered moment $\langle S \rangle_T\!\rightarrow\! 0$ 
and (\ref{eq_app_rpa_s}) yields the ordering  temperature 
\begin{align}
\frac{1}{T_N}=\frac{2}{N}\sum_{\mu,\mathbf{k}} \frac{u_{\mu\mathbf{k}}^2+
v_{\mu\mathbf{k}}^2}{\varepsilon_{\mu\mathbf{k}}}.
\label{eq_tn}
\end{align}
Note that this approach is only valid for  $S\!=\!1/2$, with expressions being more complicated 
 for larger spins \cite{Tyablikov}.

The results of the calculations of $f^{-1}\!=\!T_N/T_{\rm MF}$ are shown in Figs.~\ref{fig_tn}(b) and \ref{fig_TN_d05}(b) 
for $\Delta\!=\!1$ and $\Delta\!=\!0.5$, respectively. 
For the stripe-${\bf x}$ phase, the ordering  temperature is close to the mean-field temperature except near 
the transition boundaries as expected. 
The results for the stripe-${\bf yz}$ phase are more intriguing. There is clearly a line of parameters where $T_N$ 
is exactly zero in both figures. This line is given by Eq.~\eqref{eq_yzzero},
which reduces to $J_{z\pm}\!=\!2\sqrt{2} | J_{\pm\pm}|$ for $\Delta\!=\!1$ in Fig.~\ref{fig_tn}(b). It corresponds 
to the condition for the accidental degeneracy and for the pseudo-Goldstone mode in the magnon spectrum,
see Sec.~\ref{sec_lswt}~B.
On a technical level, it is easy to understand why  the transition temperature vanishes. 
In Eq.~\eqref{eq_tn},  $u_\mathbf{k},~v_\mathbf{k}\!\sim\!\text{const}$ and the magnon energy 
$\varepsilon_\mathbf{k} \!\sim\! {\bf k}^2$ near  the accidental degeneracy point, 
leading to a logarithmic divergence.
Physically, this divergence 
is a manifestation of the Mermin-Wagner theorem  that forbids ordering in 2D in the presence of a
continuous symmetry. 

Several comments are in order. 
First,  both ferromagnetic and 120${\degree}$ magnon spectra are gapless  and the states 
retain a continuous symmetry on the classical level. Therefore, by the same 
Mermin-Wagner theorem, the ordering transition temperature $T_N$ is zero throughout their regions, 
as is shown in Figs.~\ref{fig_tn}(b) and \ref{fig_TN_d05}(b). 

Second, we have provided a clear example that a large frustration ratio can be highly misleading
as a guide to a quantum-disordered region in 2D.
In our case, quantum fluctuations are negligible and the 
ordered moment is nearly classical, but  $T_\text{\rm MF}/T_N$ can be  large, which 
naively would imply a proximity to a spin-liquid state \cite{Chen3}.

Third, because the degeneracy leading to  pseudo-Goldstone modes is accidental, 
quantum fluctuations will induce a gap in them via an order-by-disorder effect \cite{Rau_obd}.
We delegate the quantitative discussion of this effect to Appendix~\ref{app_hk_HF} which needs 
some further developments discussed in Sec.~\ref{sec_duality}.
Although the degeneracy will be lifted, the ordering temperature will remain suppressed in that region 
compared to the mean-field expectations. 

Lastly, while a detailed phenomenology of the pseudo-Goldstone modes and suppressed 
ordering temperature is achieved and demonstrated, it still leaves the question on the nature of the 
accidental degeneracy wide open. The system is in a well-ordered stripe phase, 
with the in-plane and out-of-plane angles of spins seemingly pinned by the energy minimization, with no obvious 
combinations of $\varphi$ and $\theta$ manifesting a continuous symmetry of the spin configuration 
in the crystallographic axes. In order to elucidate the nature of this symmetry, we need to consider a 
different set of axes used in the anisotropic bond-dependent models, which we 
discuss next.

\section{Cubic axes and generalized Kitaev-Heisenberg model}
\label{sec_cubic_axes}

\begin{figure}
\includegraphics[width=0.99\linewidth]{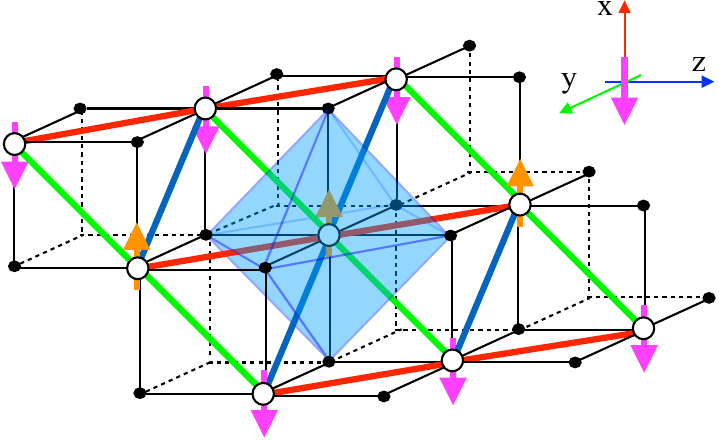}
\caption{Same as Fig.~\ref{fig_0}, side view. Octahedron of ligands is highlighted, cubic axes and bonds 
of the triangular lattice are shown. Spin ordering corresponds to the stripe-${\bf yz}$ phase for the 
parameters along the gapless $K$--$J$ line, see text.}
\label{fig_cubicaxes1}
\end{figure}

The underlying crystal structure of the triangular-lattice model considered in this work
is that of the 2D arrangement of the edge-sharing octahedra of ligands, such as O$^{2-}$, 
surrounding magnetic ions, such as Yb$^{3+}$ in the case of YbMgGaO$_4$, see Fig.~\ref{fig_cubicaxes1}. 
It is also a particular example of the  2D and 3D crystal structures built from the 
edge-sharing octahedra, such as the honeycomb and pyrochlore lattices, see Ref.~\cite{Rau18} for an overview.
  
Because of the crystal field effect  and since the superexchange processes between  magnetic ions 
are mediated by the ligands, this geometry precipitates bond-dependent anisotropic-exchange interactions
between magnetic moments of the ions with strong spin-orbit coupling \cite{Khal_ProgSupp,JK09,Winter_review,Rau18}. 
At the same time, this robust structure maintains high lattice symmetry of the resultant spin models,
such as the one discussed in Sec.~\ref{sec_classical_pd} for our model (\ref{HJpm}), 
which limits the number and the type  of allowed terms.

This lattice arrangement also makes natural the choice of axes that is different from the crystallographic ones that 
have been used in this work so far, the so-called cubic axes. 
The cubic axes are directly tied to the edges of the cubes, that is, to the bonds of the 
magnetic ions with ligands, see Fig.~\ref{fig_cubicaxes1}, as opposed to the crystallographic 
ones that  consider only  magnetic ions. 
Crucially, aside from this physical justification, 
some of the hidden symmetries of the model become apparent in this language.

One of the choices for the cubic axes is illustrated in Fig.~\ref{fig_cubicaxes1}, where we also outline the octahedron of 
the ligand sites and the ligand-to-magnetic-ion bonds that are forming  cubic shapes, 
with bonds and sites of the lattice being the same as in Fig.~\ref{fig_0}. 
Then, the transformation from the cubic to crystallographic reference frame, ${\bf S}_{\rm cryst}\!=\!
\hat{\mathbf{R}}_c{\bf S}_{\rm cubic}$, is given by 
\begin{align}
\hat{\mathbf{R}}_c=\left(
\begin{array}{ccc}
 0 & -\frac{1}{\sqrt{2}} & \frac{1}{\sqrt{2}} \\
 \sqrt{\frac{2}{3}} & -\frac{1}{\sqrt{6}} & -\frac{1}{\sqrt{6}} \\
 \frac{1}{\sqrt{3}} & \frac{1}{\sqrt{3}} & \frac{1}{\sqrt{3}} \\
\end{array} 
\right),
\label{eq_cubic_transform}
\end{align}

Next, the model \eqref{HJpm} in the  cubic axes can be rewritten as the extended Kitaev-Heisenberg ($K$--$J$) model
\begin{align}
\mathcal{H}=\sum_{\langle ij \rangle_\gamma}& \Big[
J_0 \mathbf{S}_i \cdot \mathbf{S}_j +K S^\gamma_i S^\gamma_j 
+\Gamma \left( S^\alpha_i S^\beta_j +S^\beta_i S^\alpha_j\right)\nonumber\\
+&\Gamma' \left( S^\gamma_i S^\alpha_j+S^\gamma_i S^\beta_j+S^\alpha_i S^\gamma_j
+S^\beta_i S^\gamma_j\right)\Big],
\label{H_JKGGp}
\end{align}
where we use conventions and notations borrowed from the much-studied extended $K$--$J$ model 
on the honeycomb lattice \cite{Winter_review}.
The diagonals of the faces of the  cubes in Fig.~\ref{fig_cubicaxes1} that 
connect magnetic ions form the triangular lattice with three different bonds denoted as  
$\{ \text{X,Y,Z}\}\equiv \{ \pm\bm{\delta}_1,\pm\bm{\delta}_2,\pm\bm{\delta}_3\}$. They are perpendicular to 
the corresponding cubic axes, i.e., the X-bond is perpendicular to the ${\rm x}$ axis, Y to ${\rm y}$, 
and Z to ${\rm z}$, respectively.
In the model (\ref{H_JKGGp}), these bonds are numerated as $\langle ij \rangle_\gamma$ with the triads of
$\{\alpha,\beta,\gamma\}$ being $\{{\rm y,z,x}\}$ on the X bond, $\{{\rm z,x,y}\}$ on the Y bond, 
and $\{{\rm x,y,z}\}$ on the Z bond, see Ref.~\cite{Rau18} for the   model in terms of  exchange matrices. 

The parameters of the extended $K$--$J$ model \eqref{H_JKGGp} are related to the parameters of the 
original model \eqref{HJpm} as 
\begin{align}
J_0&=\frac{1}{3}\left( 2J+\Delta J+2J_{\pm\pm}-\sqrt{2} J_{z\pm}\right),\nonumber\\
\label{eq_jkg_transform}
K&=-2J_{\pm\pm}+\sqrt{2}J_{z\pm},\\
\Gamma&=\frac{1}{3} \left( -J+\Delta J-4J_{\pm\pm}-\sqrt{2} J_{z\pm}\right),\nonumber\\
\Gamma'&=\frac{1}{6} \left( -2J+2\Delta J+4J_{\pm\pm}+\sqrt{2} J_{z\pm}\right).\nonumber
\end{align}
This relation was previously discussed in Refs.~\cite{Chaloupka,Rau18} with slightly different factor and axes conventions.

In Sec.~\ref{sec_classical_pd}~B, we have discussed  invariance of the model (\ref{HJpm}) 
to the simultaneous  change of sign of the  $J_{z\pm}$-term and a global 
$\pi$-rotation of the crystallographic axes about the $z$ axis as a justification to consider only $J_{z\pm}\!>\!0$.
In order to access  $J_{z\pm}\!<\!0$ in Eq.~(\ref{eq_jkg_transform}),  
rotation of the crystallographic axes $S^{x(y)} \!\rightarrow\! -S^{x(y)}$, 
leads to the following transformation in the 
$J_0 K\Gamma \Gamma'$ language  \cite{Chaloupka}
\begin{eqnarray}
\left( \begin{array}{c} 
J_0\\ 
K\\
\Gamma\\
\Gamma'
\end{array}\right)^\prime=\left( \begin{array}{cccc} 
1 & +4/9 &-4/9 &+4/9\\ 
0 & -1/3 &+4/3 &-4/3\\
0 & +4/9 &+5/9 &+4/9\\
0 & -2/9 &+2/9 &+7/9
\end{array}\right) \left( \begin{array}{c} 
J_0\\ 
K\\
\Gamma\\
\Gamma'
\end{array}\right),
\label{eq_negJzpm}
\end{eqnarray}
which returns the same extended Kitaev-Heisenberg model \eqref{H_JKGGp} and is, thus, self-dual.

\subsection{Degeneracies and Klein duality for $\Delta=1$}
\label{sec_duality}

\begin{figure}
\includegraphics[width=0.99\linewidth]{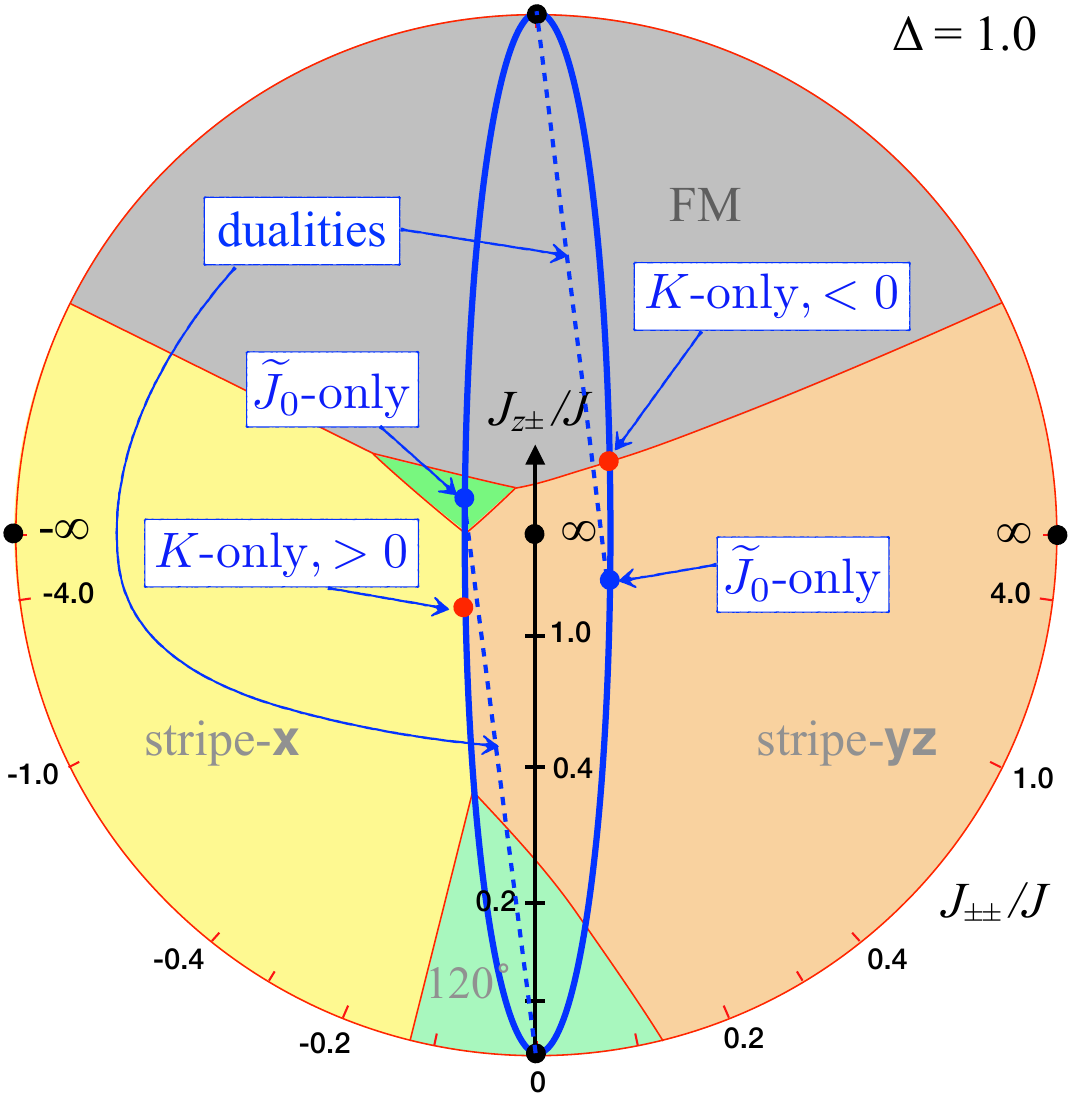}
\caption{Phase diagram of the model (\ref{HJpm}) for $\Delta\!=\!1$  as in Fig.~\ref{fig_pd_classical}.
The ellipse is the line of $J_{z\pm}\!=\!2\sqrt{2} |J_{\pm\pm}|$ for which model (\ref{HJpm}) corresponds to 
Kitaev-Heisenberg model (\ref{H_JKGGp}). Special points with higher symmetries and some Klein duality 
connections are highlighted, see text.}
\label{fig_duality}
\end{figure}

As was mentioned in Secs.~\ref{sec_lswt}~B and \ref{sec_avgs_tn}~B, in the $\Delta\!=\!1$ plane of the 
phase diagram in Fig.~\ref{fig_pd_classical} there is a special line
defined by $J_{z\pm}\!=\!2\sqrt{2}J_{\pm \pm}$, see Eq.~(\ref{eq_yzzero}), for which 
accidental degeneracy of the magnon spectrum in the stripe-${\bf yz}$ phase occurs.
One can immediately see from Eq.~(\ref{eq_jkg_transform}) that along this line 
two parameters of the model vanish, $\Gamma\!=\!\Gamma^\prime\!=\!0$,
reducing the model in the cubic axes (\ref{H_JKGGp}) to a  simpler and more symmetric 
Kitaev-Heisenberg model
\begin{align}
\mathcal{H}=\sum_{\langle ij \rangle_\gamma}
J_0 \mathbf{S}_i \cdot \mathbf{S}_j +K S^\gamma_i S^\gamma_j\, ,
\label{eq_HJKline}
\end{align}
with $J_0\!=\!J+2J_{\pm\pm}$ and $K\!=\!-6 J_{\pm\pm}$. 
This line of correspondence to the Kitaev-Heisenberg model is shown in the phase diagram in Fig.~\ref{fig_duality}
as an oval, with the left half of it obtained via a relation in (\ref{eq_negJzpm}) for $J_{z\pm}\!<\!0$.

Not only does this correspondence to the higher-symmetry model hint at the source of the enigmatic degeneracies, 
but it also provides a deeper insight into connections between different parts of the phase diagram, 
thanks to the prior works on the model (\ref{eq_HJKline}) on the honeycomb and triangular lattices
\cite{Khal_ProgSupp,Kimchi14,Ioannis,Trebst_tr}.

Contrary to their explicit anisotropic character, compass models often 
exhibit continuous symmetries in the classical limit~\cite{Nussinov,K1K2}. 
For instance, it is easy to see that for a classical ferromagnetic state, the Kitaev term in (\ref{eq_HJKline}) 
is invariant under the global spin rotation, thus, demonstrating an emergent $O(3)$ symmetry \cite{Ioannis,Trebst_tr}. 
This consideration is directly relevant to the accidental degeneracy in the stripe-${\bf yz}$ phase via 
the so-called Klein duality transformation~\cite{Khal_ProgSupp,Kimchi14,Ioannis,Trebst_tr}. 
This is a four-sublattice transformation, in which one spin, $\mathbf{S} (\mathbf{r})$, is left intact while 
spins connected to it via the X(Y,Z) bonds, $\mathbf{S}(\mathbf{r}+\bm{\delta}_\gamma)$, 
are rotated around the ${\rm x(y,z)}$ cubic axes by $\pi$. Crucially, this transformation leaves the 
$K$--$J$ Hamiltonian (\ref{H_JKGGp}) invariant, with the parameters redefined as
\begin{align}
\label{eq_jkdual}
\widetilde{J}_0=-J_0, \ \ \  \widetilde{K}=2J_0+K,
\end{align}
and in terms of $J$ and $J_{\pm\pm}$  as
\begin{align}
\widetilde{J}=-\frac{J}{3}-\frac{8J_{\pm\pm}}{3}, \ \ \ 
\widetilde{J}_{\pm\pm}=-\frac{J}{3}+\frac{J_{\pm\pm}}{3}.
\label{eq_jpmduality}
\end{align}
One can easily see in Fig.~\ref{fig_cubicaxes1} that the described spin 
transformation converts the  stripe-${\bf yz}$ state into the ferromagnetic one. Since the FM state of 
the Kitaev-Heisenberg model (\ref{H_JKGGp}) is invariant under  global spin rotation, the  stripe-${\bf yz}$ state
must also be invariant under a corresponding four-sublattice spin rotation, demonstrating accidental $O(3)$ symmetry
that naturally leads to the pseudo-Goldstone modes in the quasiclassical limit. 

In the crystallographic notations, the out-of-plane spin angle in the  stripe-${\bf yz}$ phase 
along the $K$--$J$ line is $\tan \theta\! =\!- 1/\sqrt{2}$, see Sec.~\ref{sec_lswt}~B, 
which is precisely along one of the cubic axes, see Fig.~\ref{fig_cubicaxes1}.  
Interestingly, as was shown for the  $K$--$J$ model on the honeycomb~\cite{Chaloupka2,Trebst_honey} 
and triangular lattices~\cite{Trebst_tr}, precisely this orientation is chosen by quantum fluctuations from a classically 
degenerate $O(3)$ manifold of states. In our case, this choice is made on the classical level by restricting ourselves
with the two-sublattice stripe state of the surrounding phase in Fig.~\ref{fig_duality}.

We underscore that the Klein duality is not restricted to the classical limit as the preceding discussion may
seem to suggest, but it rather implies an exact similarity of the ground state and excitations, up to the 
described spin rotations ~\cite{Khal_ProgSupp,Kimchi14}. 
This means that the entire section of the   $K$--$J$ line in Fig.~\ref{fig_duality}
belonging to the  stripe-${\bf yz}$ phase is dual to the entire section of the same line within the FM phase,
as can be confirmed from Eq.~(\ref{eq_jpmduality}), with the point marked as ``$K$-only'' being self-dual.
For instance, one of the implications of the Klein duality is the existence of a point in the stripe-${\bf yz}$ phase that is 
dual to the isotropic Heisenberg ferromagnetic point and thus must be free from any quantum fluctuations  \cite{JK10}.
Its coordinates, according to Eq.~\eqref{eq_jpmduality}, are $ \widetilde{J}\!=\!\widetilde{J}_{\pm\pm}\!>\!0$ 
and  $J_{z\pm}\!=\!2\sqrt{2} | J_{\pm \pm}|$,
and it is marked  as the ``$\widetilde{J}_0$-only'' point in Fig.~\ref{fig_duality} with the dashed line emphasizing 
the duality relation.

In practice, Klein duality means that  the calculations for quantities such as the N\'eel  temperature in stripe-${\bf yz}$ 
phase along the $K$--$J$ line can be performed in the ferromagnetic phase instead.
Since the $K$--$J$  model is $O(3)$ symmetric only on the classical level, of interest is 
the gap in the pseudo-Goldstone mode that is generated by the order-by-disorder effect. 
In Appendix~\ref{app_hk_HF}, we present calculations of the Hartree-Fock corrections to the magnon spectrum 
for the ferromagnetic Kitaev-Heisenberg model and use Klein duality to obtain the self-consistent 
transition temperature in the stripe-${\bf yz}$ phase. Although the fluctuation-induced gap $E_g$ in the magnon spectrum 
is small, the transition temperature is very sensitive to it, 
$T_N\!\propto\!T_{\rm MF}/\ln \left(T_{\rm MF}/E_g\right)$ \cite{Katanin}, 
leading to a finite $T_N$ except for the  $\widetilde{J}_0$-only point. The resultant transition temperatures are still 
significantly suppressed compared to the mean-field expectations (\ref{eq_tcmf}).

The implications for the rest of the phase diagram in Fig.~\ref{fig_duality} are the following.
A section of the $K$--$J$ line was previously identified as  a ``nematic'' phase \cite{Trebst_tr,Ioannis,Tohyama,Rau_tr}. 
In Fig.~\ref{fig_duality}, this section coincides with the  boundary between the stripe-${\bf x}$ and stripe-${\bf yz}$ 
phases and is unlikely to represent a separate phase on its own.   

Another insight is provided by the Klein duality into the dual 120${\degree}$ phase. 
The isotropic antiferromagnetic Heisenberg model has a Klein-dual point within the dual 120${\degree}$ phase,
marked as the second ``$\widetilde{J}_0$-only'' point in Fig.~\ref{fig_duality}  
with the coordinates  $\widetilde{J}\!=\!\widetilde{J}_{\pm\pm}\!<\!0$, see Eq.~\eqref{eq_jpmduality}. 
This relation also elucidates the nature of the dual 120${\degree}$ state. It is obtained from the three-sublattice 
120${\degree}$ state by the Klein rotations of spins around cubic axes resulting in the 
12-sublattice state. Since the cubic axes are not collinear with the crystallographic 
ones, the resultant state is, generally, noncoplanar, see Ref.~\cite{Khal_ProgSupp} for the 
projection of such a state onto the triangular-lattice plane. 
It is only when the plane of the initial 120${\degree}$ state is chosen to be parallel to one of the principal 
planes of the cubic coordinate system, the Klein rotations will keep spins of the dual 120${\degree}$ state in the same 
plane. This situation is sketched in Fig.~\ref{fig_pd_classical}.

The most important implication of the correspondence of the model (\ref{HJpm}) 
to the Kitaev-Heisenberg model (\ref{eq_HJKline}) along the line $J_{z\pm}\!=\!2\sqrt{2}J_{\pm \pm}$ in 
Fig.~\ref{fig_duality} is that it necessitates the existence of a spin-liquid region in the $S\!=\!1/2$ case 
that is Klein dual to the spin liquid found by us in Ref.~\cite{topography}. The confirmation of this 
is the subject of Sec.~\ref{sec_dmrg}.

\vspace{-0.2cm}

\subsection{Gapless modes at $\Delta\!<\!1$}

As was discussed in Secs.~\ref{sec_lswt}~B and \ref{sec_avgs_tn}~B,
the pseudo-Goldstone modes in the magnon spectrum of the stripe-${\bf yz}$ phase
persist for $\Delta\!<\!1$ along the lines defined by  
$J_{z\pm}\!=\left[ 4J_{\pm\pm}\!+\!J(1-\Delta)\right]/\sqrt{2}$; see Eq.~(\ref{eq_yzzero}). 
For the extended $K$--$J$ model  (\ref{H_JKGGp}) with  parameters
in Eq.~(\ref{eq_jkg_transform}), this condition  means that only one of the off-diagonal terms 
remains zero along these  lines  ($\Gamma\!=\!0$), while the other one does not ($\Gamma^\prime\!\neq\!0$),
leading to the $K$--$J$--$\Gamma^\prime$ model with $J_0\!=\!J\!+\!2J_{\pm\pm}$ and
$\Delta$-dependent $K$ and $\Gamma^\prime$
\begin{align}
 K=J(\Delta-1)-6 J_{\pm\pm}, \ \  \Gamma'=J(\Delta-1)/2.
\end{align}
Thus, while for $\Delta\!=\!1$ the accidental degeneracies are associated with the high symmetry of the
Kitaev-Heisenberg model, in this case their origin is more subtle.

For the $K$--$J$--$\Gamma^\prime$ model, Klein duality transformation does not leave the Hamiltonian
form invariant. While Heisenberg and Kitaev terms do preserve their structure with the change of
 \begin{align}
\widetilde{J}_0=-J_0, \ \ \ 
\widetilde{K}=2J_0+K,
\end{align}
the symmetric $\Gamma'$-term becomes antisymmetric. For example,  on the ${\rm X}$-bond 
\begin{align}
\Gamma'\left( S^{\rm x}_i S^{\rm y}_j+S^{\rm y}_i S^{\rm x}_j+S^{\rm x}_i S^{\rm z}_j+S^{\rm z}_i S^{\rm x}_j\right)
\end{align}
becomes
\begin{align}
\Gamma'\left( S^{\rm x}_i S^{\rm y}_j-S^{\rm y}_i S^{\rm x}_j+S^{\rm x}_i S^{\rm z}_j-S^{\rm z}_i S^{\rm x}_j\right).
\end{align}
However, for $\Delta\!<\!1$ and along the the $K$--$J$--$\Gamma^\prime$ lines 
the spin orientation in the stripe-${\bf yz}$ phase remains along one of the cubic axes.
Therefore, the $\Gamma^\prime$-term does not contribute to the classical energy and 
leaves the accidental degeneracy of the Kitaev-Heisenberg model intact. 

\vspace{-0.15cm}

\section{Quantum realm: dual spin liquids in the $S=1/2$ model}
\label{sec_dmrg}

There are several new aspects of the anisotropic-exchange model (\ref{HJpm}) that have been 
discussed in this work so far. These are the classification of all its single-${\bf Q}$ classical phases,
the identification of the instabilities of some of them to more complex multi-${\bf Q}$ states, 
and various quantum and thermal effects in the magnetically ordered phases,
see Secs.~\ref{sec_classical_pd}--\ref{sec_avgs_tn}.
There is also a fruitful connection with an extended Kitaev-Heisenberg model (\ref{H_JKGGp}) 
that uncovers hidden symmetries and relates different parts of the phase diagram 
to each other, see Sec.~\ref{sec_cubic_axes}. 
Here, we build upon these insights using DMRG for the $S\!=\!1/2$ model. 

\begin{figure}
\includegraphics[width=0.99\linewidth]{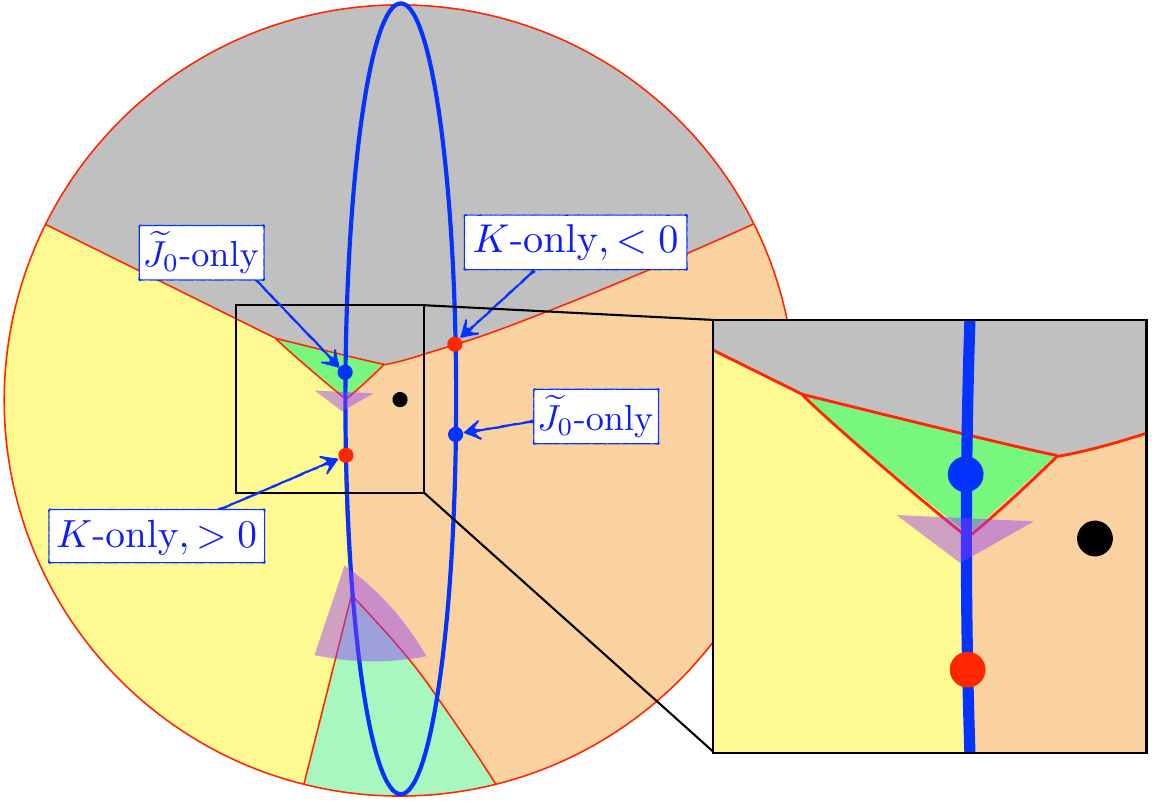}
\caption{Phase diagram for $\Delta\!=\!1.0$ from Fig.~\ref{fig_duality} with highlighted spin-liquid regions.
The lower SL region is from  Ref.~\cite{topography}.
The magnified region emphasizes the vicinity of the dual spin-liquid region that is studied here by DMRG. }
\label{fig_phd_inset}
\vskip -0.3cm
\end{figure}

In our prior work, Ref.~\cite{topography}, we have discovered a  spin-liquid (SL) region
of the 3D phase diagram of the model (\ref{HJpm})  using DMRG, with the sketch of 
its base shown in Fig.~\ref{fig_phd_inset} 
as a  triangle enveloping the tricritical point of the 120$\degree$ and the stripe phases. 
According to  Ref.~\cite{topography}, the SL phase 
occupies a  distorted cone shape with the base at $\Delta\!=\!1.0$, the widest dimensions 
$J_{z\pm}\!\simeq\![0.27,0.45]J$ and $J_{\pm\pm}\!\simeq\![-0.17,0.1]J$, and  the tip of the cone 
protruding along the $XXZ$ axis down to $\Delta\!\agt\!0.7$. 

It is important to note that for $\Delta\!=\!1.0$, the SL area in Fig.~\ref{fig_phd_inset} includes a 
segment of the line that corresponds to the pure Kitaev-Heisenberg model 
(\ref{eq_HJKline}), discussed in Sec.~\ref{sec_duality}.
Thus, in a retrospect, our discovery is also a discovery of a spin liquid in the 
$K$--$J$ model on the triangular lattice, without the benefit of having an exact Kitaev-like solution 
in this geometry. We note that this finding is in  disagreement with the previous numerical studies of this 
model \cite{Trebst_tr,Tohyama}, which, we believe, have missed the SL phase
due to finite-size effects \cite{Trebst_tr} or unfavorable boundary conditions \cite{Tohyama}. 

\begin{figure*}
\includegraphics[width=0.99\linewidth]{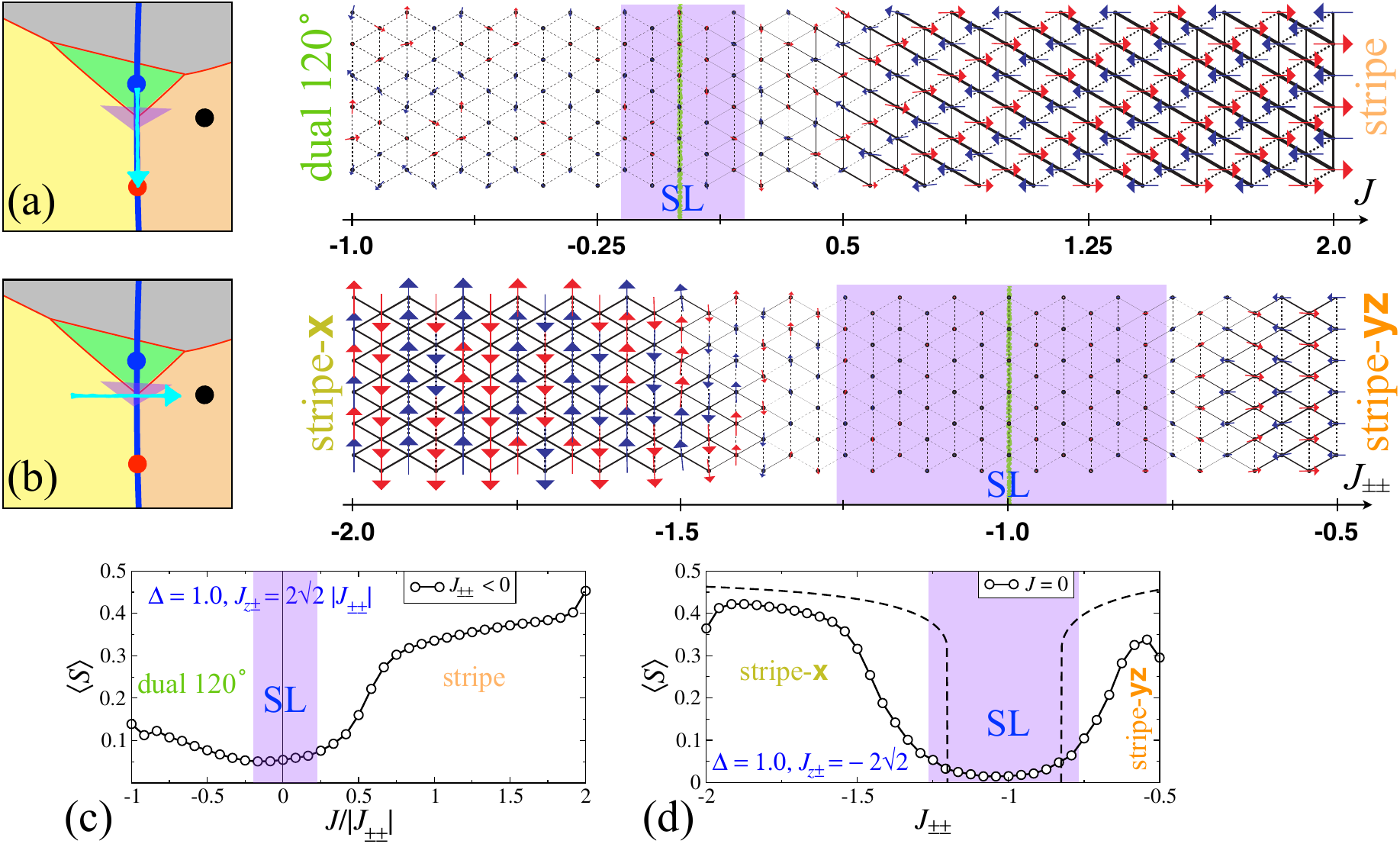}
\vskip -0.1cm
\caption{Long-cylinder DMRG scans (a) along the $K$--$J$ line from the dual 120$\degree$ to the 
$K$-only point, and (b) normal to it at the classical tricritical point of the dual 120$\degree$ and the stripe phases,
$J\!=\!0$ and $J_{z\pm}\!=\!-2\sqrt{2}J_{\pm\pm}$. Sketches of the phases  
indicate the direction and extent of each scan, with (c) and (d) showing $\langle S\rangle$ along the scans.
Shaded areas indicate putative dual spin-liquid regions. Dashed lines in (d) are the LSWT results for 
$\langle S\rangle$, Eq.~(\ref{eq_avgs}), in the stripe phases.}
\label{fig_SL_DMRG1}
\vskip -0.3cm
\end{figure*}

Crucially, the Klein duality along the line of a correspondence to the Kitaev-Heisenberg model 
necessitates the existence of another spin liquid region, sketched in Fig.~\ref{fig_phd_inset}.
Thus, in our present DMRG study, we investigate the previously unexplored parts of the phase diagram
in order to confirm the existence of this dual SL phase.
As in Ref.~\cite{topography}, we use several complementary 
approaches: the long-cylinder 1D  ``scans'' with one parameter varied along the length of the $6\!\times\!36$ cylinder
to explore different phases \cite{ZhuWhite,us},  the shorter cylinders with fixed parameters 
[$6\!\times\!20$, ``non-scans''], as well as the  intensity maps of the structure factor, ${\cal S}({\bf q})$, 
and correlation lengths; see also Appendix~\ref{app_DMRG}. 
We use different boundary conditions and ranges of the varied parameter to exclude unwanted proximity effects.

For the DMRG calculations in the $6\!\times\!36$ cylinders,  we typically perform $20\!-\!24$ sweeps and  keep up to 
$m\!=\!1600\!-\!2000$ states  depending on the complexity of the Hamiltonian  with truncation error $<\!10^{-5}$. 
For the $6\!\times \!20$ cylinders,  the protocol is 24 sweeps and up to $m\!=\!2000$ states with 
truncation error $<\!10^{-6}$. In the real-space images of cylinders below, the size of the arrows represents the 
projection of  local spin in the lattice plane and the color is used to indicate the sign of the out-of-plane tilt 
angle of the spins. The thickness of the nearest-neighbor bonds is proportional to the magnitude of the 
$\langle {\bf S}_i {\bf S}_j\rangle$ correlation, with the (solid) dashed  lines representing (anti)ferromagnetic sign of it.

\vspace{-0.15cm}

\subsection{Dual spin-liquid region}

Figure~\ref{fig_SL_DMRG1} summarizes  the results from the two long-cylinder scans through the putative
dual SL region, first along the $K$--$J$ line and second normal to it at the tricritical point of the 
dual 120$\degree$ and the stripe phases. The 
portions of the phase diagram shown on the left of the cylinder images in Figs.~\ref{fig_SL_DMRG1}(a), (b)  
indicate the direction and extent of each scan, and  Figs.~\ref{fig_SL_DMRG1}(c), (d) show the 
magnitude of the ordered moment $\langle S\rangle$ along the scans.

In Fig.~\ref{fig_SL_DMRG1}(a), the scan covers the entire stretch 
from the dual 120$\degree$ point, that is, the point that is Klein dual to the isotropic Heisenberg 
antiferromagnetic 120$\degree$ point,  to the self-dual $K$-only point, $K\!>\!0$, see also Fig.~\ref{fig_duality}. 
In the notations of the Kitaev-Heisenberg model (\ref{eq_HJKline}), this scan is from 
$\{J_0,K\}\!=\!\{-1,2\}/\sqrt{5}$ to $\{J_0,K\}\!=\!\{0,1\}$, with the normalization of $J_0^2+K^2\!=\!1$. 
Using duality transformation (\ref{eq_jkdual}), one can see that the 
dual 120$\degree$ point  indeed corresponds to the $\widetilde{J}_0$-only model.
In the notations of the original anisotropic-exchange model (\ref{HJpm}), the scan is performed along the 
$J_{z\pm}\!=\!2\sqrt{2}|J_{\pm\pm}|$ line with $\Delta\!=\!1.0$ by varying $J/|J_{\pm\pm}|$ from 
$-1$ to $2$, as is indicated in   Figs.~\ref{fig_SL_DMRG1}(a) and  \ref{fig_SL_DMRG1}(c).

We note that no special boundary conditions are applied in any of the long-cylinder scans, yet 
the dual 120$\degree$ order appears naturally on the left end of the cylinder in Fig.~\ref{fig_SL_DMRG1}(a) 
and is clearly discernible. Nearly half of this scan is occupied by what is clearly identifiable as a stripe 
state, which is intermediate between the stripe-${\bf x}$ and stripe-${\bf yz}$ states 
in that the spins are neither parallel nor
perpendicular to any of the bonds. It is also tilted out of the lattice plane similar to the stripe-${\bf yz}$ state.
As was pointed out above, see Sec.~\ref{sec_duality}, this section of the $K$--$J$ line, 
which is exposed here in a wider parameter space, corresponds to the boundary between two stripe phases and 
should not represent a separate phase in the quantum limit. In Refs.~\cite{Trebst_tr,Tohyama},  this boundary is 
referred to as the nematic phase even in the quantum limit, while Ref.~\cite{Li_CSL} called it a spin liquid. 
We believe that both terms must have been used in error, see also Appendix~\ref{app_DMRG} for a DMRG scan 
across the right end of the scan in Fig.~\ref{fig_SL_DMRG1}(a), the $K$-only point.

\begin{figure*}
\includegraphics[width=0.99\linewidth]{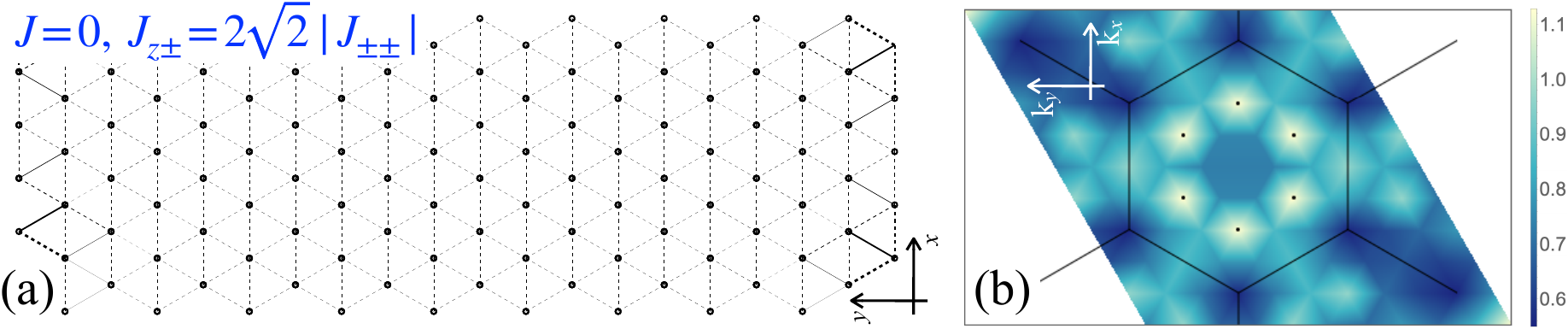}
\vskip -0.1cm
\caption{(a) $6\!\times\!20$ DMRG cluster for $J\!=\!0$ and $J_{z\pm}\!=\!-2\sqrt{2}J_{\pm\pm}$.
(b) Intensity map of ${\cal S}({\bf q})$, Eq.~(\ref{eq_Sq}), from  (a), see text.}
\label{fig_SL_DMRG2}
\vskip -0.3cm
\end{figure*}

Despite a significant range of the scan in  Fig.~\ref{fig_SL_DMRG1}(a), there is a clear 
suppression of order  between the ordered phases. The shaded regions in  Figs.~\ref{fig_SL_DMRG1}(a) and (c)
indicate the range that is obtained from the Klein duality of the extent of the ``original'' spin liquid  
along the $K$--$J$ line in Fig.~\ref{fig_phd_inset}, $J/|J_{\pm\pm}|\!\approx\![-0.22,0.23]$. It  agrees with the 
minimum in the ordered moment and we refer to it as the dual SL region.
Similar to the SL phase of Ref.~\cite{topography}, which comes from melting of  the 120${\degree}$ phase, 
the dual spin liquid appears to be born out of the dual 120${\degree}$ phase. 

The second scan through the dual SL region is shown in  Fig.~\ref{fig_SL_DMRG1}(b), with the 
vertical line within the shaded region marking its intersect with the first scan. It is also a 
tricritical point of the classical dual 120$\degree$ and the stripe phases. 
The scan in  Fig.~\ref{fig_SL_DMRG1}(b) is not on the $K$--$J$ line, so the natural notations 
here are of the anisotropic-exchange model (\ref{HJpm}), in which this scan corresponds to the $J\!=\!0$ line, the 
line where the $XXZ$ part of the model is absent and the only two variables are  $J_{z\pm}$ and $J_{\pm\pm}$.
Keeping $J_{z\pm}\!=\!2\sqrt{2}$ in order to match  the first scan at $J_{\pm\pm}\!=\!-1$,  we vary 
$J_{\pm\pm}$ from $-2$ to $-0.5$.  The  stripe-${\bf x}$ and stripe-${\bf yz}$ phases are clearly seen in the scan, 
separated by a rather wide shaded region, $J_{\pm\pm}\!\approx\![-1.26,-0.76]$, 
where magnetic order is suppressed. In Fig.~\ref{fig_SL_DMRG1}(d), the dashed lines  show 
the ordered moment as given by the $1/S$ calculations in the stripe phases, indicating that 
magnon instability boundaries are not unlike the boundaries of the putative dual SL phase.

As is in the previous studies \cite{topography}, the long-cylinder scans are only a part of the evidence for the spin liquid.
In order to confirm the SL, we perform DMRG non-scans on  $6\!\times\!20$ clusters with fixed parameters.  
Figure~\ref{fig_SL_DMRG2}(a) shows one of them for the point that is at the core of the suggested dual SL region,
at $J\!=\!0$ and $J_{z\pm}\!=\!-2\sqrt{2}J_{\pm\pm}$, a tricritical point of the classical phases.
Since it is on the $K$--$J$ line, its coordinates in the $K$--$J$ language correspond to $\{J_0,K\}\!=\!\{-1,3\}/\sqrt{10}$.

The  cluster presented in Fig.~\ref{fig_SL_DMRG2}(a) shows no discernible traces of any magnetic order and has 
weak nearest-neighbor $\langle {\bf S}_i {\bf S}_j\rangle$ correlations of ferromagnetic sign. 
The boundary conditions  are open with one site removed on each side of the cylinder to avoid spinon localization, 
common to the $Z_2$ spin-liquid states \cite{ZhuWhite,topography}
and indicative of them. Without the sites removed, a weak pattern similar to the dual  120$\degree$ structure 
appears at the boundaries (not shown) and decays exponentially toward the center of the cluster,
all supportive of the SL state. This behavior is in clear contrast with the results of a similar analysis of the 
dual 120$\degree$ ($\widetilde{J}_0$-only) point  offered in Appendix~\ref{app_DMRG}.
There, the long-cylinder scans also suggest a possible SL state, but 
the $6\times 20$ non-scan cluster verification demonstrates a robust 12-sublattice order
with a power-law decay from the boundary, as expected.

A more comprehensive insight into the spin-spin correlations and into the nature of the SL state is given by 
the static structure factor observable in experiments,
\begin{equation}
\label{eq_Sq}
{\cal S}({\bf q})\!=\!\frac{1}{N}\sum_{\alpha\beta,ij}\left(\delta_{\alpha\beta}-\frac{q_\alpha q_\beta}{q^2}\right)
\big\langle S_i^\alpha S_j^\beta\big\rangle e^{i{\bf q}({\bf R}_i-{\bf R}_j)}. 
\end{equation}
We obtain it from the Fourier transform
of the real-space spin-spin correlation function $\langle S_i^\alpha S_j^\beta\rangle$,
determined from the DMRG ground-state wave function with all $|i-j|$ distances that are available in the cylinder.  

In Fig.~\ref{fig_SL_DMRG2}(b), we present an intensity map of ${\cal S}({\bf q})$ from the cluster 
in Fig.~\ref{fig_SL_DMRG2}(a). It shows clear maxima at the $\mathbf{Q}_{\rm d120{\degree}}\!=\!(2\pi/3,0)$
and equivalent  points, associated with the dual 120${\degree}$ order, see Sec.~\ref{sec_classical_pd}~B.
This is in accord with the ``original'' spin liquid of Ref.~\cite{topography} having broad maxima in ${\cal S}({\bf q})$
at the $\mathbf{Q}_{\rm 120{\degree}}$.
These results also suggest that the dual spin liquid can be seen as a result of a ``melting'' of the dual 120${\degree}$  
phase in the same way that the spin liquid of Ref.~\cite{topography} can be seen as a molten 120${\degree}$ phase, both
maintaining the shape of the structure factor similar to that of the parent ordered states.

Altogether, we have confirmed the existence of the second SL region in the phase diagram of the anisotropic-exchange 
model, with  properties that are in accord with the expectations based on the Klein duality relation along the 
$K$--$J$ line. This confirmation of the dual spin liquid strengthens our case for both SL regions.

Two additional notes are in order. In Sec.~\ref{sec_swt_inst}~B, we remark on a similarity 
of the spin-liquid region in the quantum $S\!=\!1/2$ model from Ref.~\cite{topography}
with the quasiclassical region of instability of the 120$\degree$ phase to a multi-${\bf Q}$ long-range spiral, 
or a $Z_2$-vortex-like state, suggesting a possible connection between the two.
This similarity may imply that the discussed spin liquids should originate from the melting of the $Z_2$-vortex and dual 
$Z_2$-vortex states 
rather than their more simple 120${\degree}$ and dual 120${\degree}$ counterparts. However, 
we reiterate here that while the long-range distortions of the ordered states found in DMRG clusters are possible,
there are no traces of the expected shifts of the peaks  from the commensurate ordering vectors in the SL structure factors.
There is also no detectable residual chirality in the spin-liquid wave functions that may be expected to survive from 
the $Z_2$-vortex states, see Ref.~\cite{topography}. Thus, the relation between the quasiclassical and quantum ground states 
of the problem in this region  deserves further investigation.

Lastly, a superficial observation is that both the ``original'' and the dual spin liquids seem to be centered at the 
tricritical points of the classical  single-${\bf Q}$ ordered phases. In the $K$--$J$ notations
they correspond to the $K\!=\!J_0\!>\!0$ and $K\!=\!-3J_0\!>\!0$ points, respectively. 
Interestingly, in the anisotropic-exchange notations, the dual tricritical point corresponds to
the model (\ref{HJpm}) with no $XXZ$ terms and only anisotropic $J_{z\pm}$ and $J_{\pm\pm}$ terms. 
While we are unable to draw any useful insight from this observation 
or identify any quasi-classical degeneracy that can be affiliated 
with this model, the chance that this point can have a special solution may be an intriguing possibility.

\subsection{Phase diagram of the $K$--$J$ $S=1/2$ model}

\begin{figure}
\includegraphics[width=0.99\linewidth]{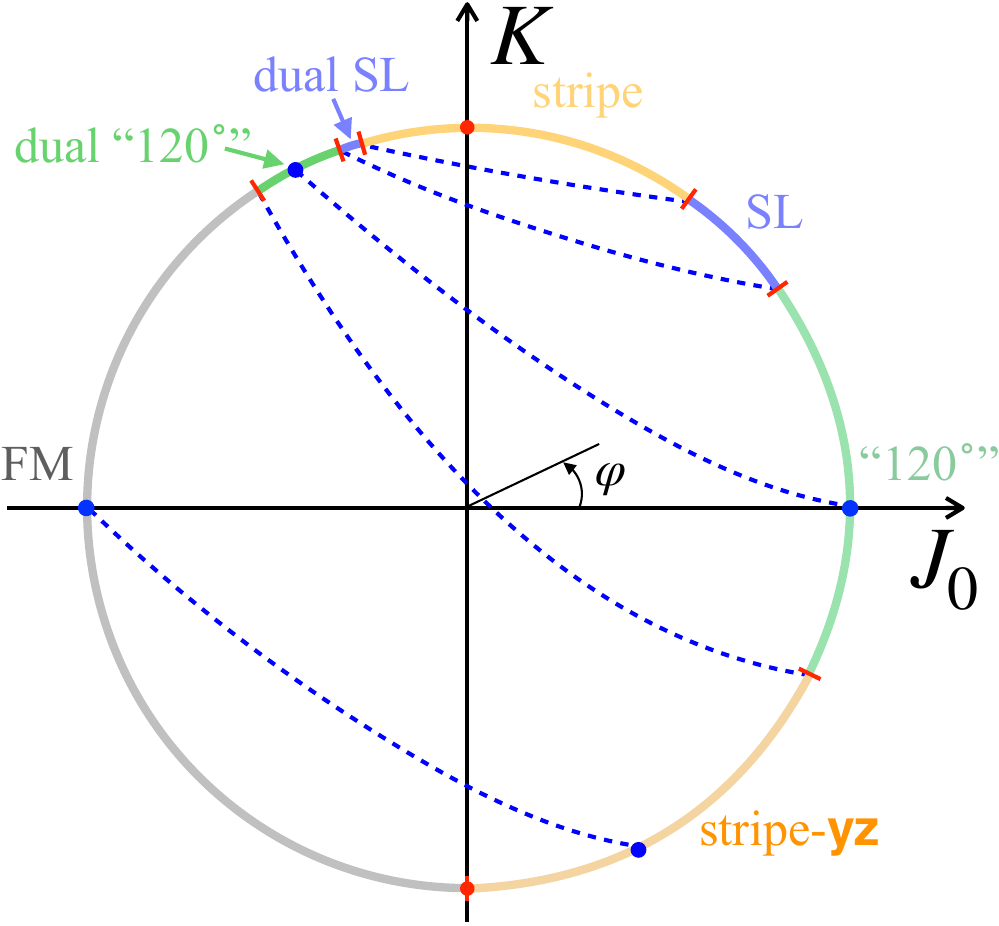}
\caption{The 1D phase diagram of the quantum $S\!=\!1/2$ $K$--$J$ model (\ref{eq_HJKline}) 
using DMRG results from Sec.~\ref{sec_dmrg}~A and Ref.~\cite{topography}.  
Pure Heisenberg, pure Kitaev and their Klein-dual points are marked and the duality 
relation between them and SL and ``120${\degree}$'' phases is emphasized by dashed lines.}
\label{fig_phd_dmrg}
\vskip -0.3cm
\end{figure}

Although the full parameter space of the nearest-neighbor anisotropic-exchange model (\ref{HJpm}),
or, equivalently, of the extended $K$--$J$--$\Gamma$--$\Gamma^\prime$ model (\ref{H_JKGGp})
on the triangular lattice is three dimensional, their relation to a simpler and more symmetric $K$--$J$ 
model (\ref{eq_HJKline}) along a one-dimensional line has proven to be very informative.
Thus, we conclude this Section by summarizing our DMRG results  for the quantum $S\!=\!1/2$ 
$K$--$J$ model (\ref{eq_HJKline}) in the form of its 1D phase diagram, shown in Fig.~\ref{fig_phd_dmrg},
where we use the standard parametrization, $J_0\!=\!\cos \varphi$ and $K\!=\!\sin \varphi$, 
and the positive (counterclockwise) $\varphi$ direction corresponds to  the negative  (clockwise) direction
along the ellipse in our 2D phase diagram in Fig.~\ref{fig_duality}, 
see Eq.~\eqref{eq_HJKline}. 

We fully agree with the previous studies of the triangular-lattice $K$--$J$ model 
\cite{Ioannis,Rau_tr,Trebst_tr, Tohyama,Li_CSL,Avella} on the location and  boundaries 
of the FM and stripe-${\bf yz}$ phases. Note that 
we deviate in the notations for what we refer to  as the ``120${\degree}$'' and dual ``120${\degree}$'' 
phases. They were quasiclassically identified as the multi-${\bf Q}$ distortions of such 120${\degree}$ orders 
and were called the  $Z_2$ vortex crystal and its dual phases \cite{Ioannis,Trebst_tr}.
These regions are still designated as the $Z_2$ vortex and the dual $Z_2$ vortex states in the $S\!=\!1/2$ 
case  \cite{Trebst_tr,Tohyama}, although the evidence for the persistence of such distortions in this limit is slim.
We note that some deviations from the precise spiral orders  near the isotropic limit of the model may have 
been detected in our prior DMRG work, Ref.~\cite{topography}.
While we do not conclusively confirm or rule them out, it seems that these 
deviations from the ideal orders, even if exist,  play only a minor role in the energetics of the phases in 
the quantum limit, see Sec.~\ref{sec_swt_inst}. 
Hence, we simply use quotation marks for the ``120${\degree}$'' in referring to them.

We disagree with the previous works on the presence of a nematic phase in the quantum limit 
\cite{Trebst_tr,Tohyama}. 
While it may exist as a fully classical curiosity, quantum fluctuations select a stripe state 
that is intermediate between the stripe-${\bf x}$ and stripe-${\bf yz}$ phases,
see Sec.~\ref{sec_duality} and Appendix~\ref{app_DMRG}. Denoting this sector  a spin liquid \cite{Li_CSL} 
must have been in error. In Fig.~\ref{fig_phd_dmrg}, we refer to it as a ``stripe'' phase 
to distinguish from the stripe-${\bf yz}$ phase.

Our most important findings are the two regions of the spin-liquid phase. They occur in a proximity of, and,
arguably, as a result of a melting of their respective parent ``120${\degree}$'' phases. This is also evidenced
by the structure factor discussed  in Sec.~\ref{sec_dmrg}~A and in Ref.~\cite{topography}. 
The connection of our ``SL'' region to a SL phase of the fully isotropic $J_1$--$J_2$ Heisenberg model
was established in Ref.~\cite{topography}, and was referred to as a SL isomorphism. 
We have found that the spin-spin correlations  are very similar 
in them and that there is a path in the 4D phase diagram that provides a continuous link between the two. 
This connection further strengthens the argument for our ``molten 120${\degree}$'' phase scenario.

The respective SL regions in Fig.~\ref{fig_phd_dmrg} are as follows. For the region marked ``SL'', 
$K/J_0\!\approx\! [0.71,1.40]$ and  $\varphi/\pi\!\approx\![0.20,0.30]$. While this may or may not be 
a coincidence, the $K\!=\!J_0$ point is included in this range. 
By the  Klein-duality transformation, the region referred to as ``dual SL'' occurs. Its boundaries agree well 
with the DMRG presented in Sec.~\ref{sec_dmrg}~A and are $K/J_0 \!=\! [-3.4,-2.71]$
and $\varphi/\pi\!\approx\![0.59,0.61]$. As was discussed above, it is centered around $K\!=\!-3J_0$ point,
which translates to the vanishing point for the $XXZ$ part of anisotropic-exchange model (\ref{HJpm})
with only anisotropic $J_{\pm\pm}$ and $J_{z\pm}$ terms present. 

Lastly, the mean-field Schwinger-boson study \cite{Punk} has discussed various spin liquids 
in the triangular-lattice $K$--$J$ model. The SL region in Fig.~\ref{fig_phd_dmrg} 
is affiliated with their ``SL2'' state, but that latter state has the structure factor that is very different from 
${\cal S}({\bf q})$ found by DMRG \cite{topography} and thus can be ruled out.
We also reiterate that the spin liquids that we identify 
are not consistent with the ``open spinon Fermi-surface'' SL state 
proposed for YMGO \cite{Chen2}. 

\section{Summary}
\label{sec_conc}

In this work, we have provided an extensive, if not exhaustive, overview of the phase diagram
of the nearest-neighbor triangular-lattice anisotropic-exchange model, which is relevant to
a growing family of the rare-earth-based magnets and other materials with strong spin-orbit interactions. 
We have explored ordered phases, identified the principal ones that occupy a majority 
of the parameter space, and obtained explicit expressions for their non-trivial spin-wave excitation spectra.
We have also identified and characterized transitions of some of the ordered phases to more complex 
multi-${\bf Q}$ states  and demonstrated the effectiveness of the analysis of such transitions with the 
help of  magnon instabilities. 

In the studies of the quantum and finite-temperature effects in the well-ordered phases, a 
number of accidental degeneracies that lead to emergent continuous symmetries 
of the classical states  have been discussed, and the effect of order-by-disorder on them has 
been analyzed. Another systematic and enigmatic  accidental degeneracy has been found in the nearly classical 
stripe phase, leading to a strongly suppressed ordering temperature that can be falsely attributed to a 
proximity to an exotic state.
This degeneracy  has been connected to the correspondence of the original model to
an extended Kitaev-Heisenberg model, in which the degeneracies have a more natural explanation.

This connection has been particularly fruitful for uncovering hidden symmetries and in relating 
different parts of the phase diagram  to each other via the Klein-duality transformation. 
In a rather spectacular manifestation of the correspondence to the Kitaev-Heisenberg model,
a new region of the spin-liquid state  that is Klein dual to the spin liquid found by us  in a prior work has 
been confirmed for the $S\!=\!1/2$ model using an unbiased DMRG approach. 
Both the original and the dual spin liquids occur in  proximity 
to their parent ordered phases that are also dual to each other. This finding strengthens our case for both 
spin-liquid regions in the phase diagram of the quantum model.
As a corollary, we have also provided a one-dimensional phase diagram of the quantum 
Kitaev-Heisenberg model on the triangular lattice that updates and corrects previous results.

In conclusion, the present work, together with our prior works in the phase diagram of the 
anisotropic-exchange model on a triangular lattice,  creates a foundation for the studies of the 
large group of materials with anisotropic exchanges, clears the path to a consistent interpretation of 
the current and future experiments, and gives important new insights into fundamental properties of 
quantum magnets with spin-orbit-generated low-energy Hamiltonians.

\begin{acknowledgments}

We are indebted to Natalia Perkins for setting us straight on the path to dualities, for 
persistence regarding the $Z_2$ vortex state, and for numerous other fruitful discussions.
We would like to thank Ioannis Rousochatzakis for forgiveness and for a number of useful comments, 
Itamar Kimchi for private communication concerning the correspondence of the anisotropic-exchange 
to the generalized Kitaev-Heisenberg model, and Gang Chen for communications and overall tolerance. 
We are also thankful to George Jackeli 
for pointing out the method used in Eq.~\eqref{eq_uk} and for several useful conversations. 
We are forever grateful to Martin Mourigal for nudging us into the anisotropic triangular-lattice realm.
This work was supported by the U.S. Department of Energy,
Office of Science, Basic Energy Sciences under Award No. DE-FG02-04ER46174 (P. A. M. and A. L. C.)
and by the NSF through grant DMR-1812558 (Z. Z. and S. R. W.).
\end{acknowledgments}

\appendix


\section{LSWT details}
\label{app_stryz}

The spin Hamiltonian in the local  axes and without the anharmonic terms is 
\begin{align}
\mathcal{H}=&\sum_{\langle ij\rangle}\Big(\widetilde{J}_{ij}^{xx} \widetilde{S}^x_i \widetilde{S}^x_j 
+\widetilde{J}_{ij}^{yy} \widetilde{S}^y_i \widetilde{S}^y_j 
+\widetilde{J}_{ij}^{zz} \widetilde{S}^z_i \widetilde{S}^z_j \nonumber\\
&\phantom{\sum_{\langle ij\rangle}}+\widetilde{J}_{ij}^{xy} \widetilde{S}^x_i \widetilde{S}^y_j +
\widetilde{J}_{ij}^{yx} \widetilde{S}^y_i \widetilde{S}^x_j\Big) \,.
\label{app_HlocalSWT_general}
\end{align}
\emph{Stripe-${\bf yz}$ phase.}---%
For the stripe-${\bf yz}$ state, elements of the rotated exchange matrix $\widetilde{\bm J}_{ij}$ are
\begin{align}
\widetilde{J}_{ij}^{xx}=&\left(J-2J_{\pm\pm}\cos\varphi_\alpha\right)\sin\theta_i \sin\theta_j+
\Delta J\cos\theta_i \cos\theta_j\nonumber\\
\label{App_stryz} &
-J_{z\pm}\left(\cos\theta_i\sin\theta_j+\sin\theta_i\cos\theta_j\right)\cos\varphi_\alpha\,,\nonumber\\
\widetilde{J}_{ij}^{yy}=&J+2J_{\pm\pm}\cos\varphi_\alpha\,,\\
\widetilde{J}_{ij}^{zz}=&\left(J-2J_{\pm\pm}\cos\varphi_\alpha\right)\cos\theta_i \cos\theta_j+
\Delta J\sin\theta_i \sin\theta_j\nonumber\\
&+J_{z\pm}\left(\cos\theta_i\sin\theta_j+\sin\theta_i\cos\theta_j\right)\cos\varphi_\alpha\,,\nonumber\\
\widetilde{J}_{ij}^{xy}=&2J_{\pm\pm}\sin\theta_i\sin\varphi_\alpha-
J_{z\pm}\cos\theta_i\sin\varphi_\alpha\,,\ \ \widetilde{J}_{ij}^{yx}=\widetilde{J}_{ji}^{xy},\nonumber
\end{align}
where  $\theta_{i(j)}$ for the two sublattices are defined by the choice of ${\bf Q}\!=\!M^\prime$, 
$\theta_A\!=\!\theta$ and $\theta_B\!=\!\theta+\pi$, see Sec.~\ref{sec_lswt}.B.
After some algebra, (\ref{App_stryz}) yields 
the following expressions for the matrix elements of the LSWT matrices in
Eq.~(\ref{ABstripex})
\begin{align}
A_{\bf k}&=2J\Delta+ 2\big(J\left(1-\Delta\right)
+4J_{\pm \pm}\big) \cos^2\theta -4J_{z\pm} \sin 2 \theta\nonumber\\
&+\Big(2J+\big(J(\Delta-1)+2J_{\pm \pm}\big) \cos^2\theta 
-J_{z\pm} \sin 2 \theta \Big)c_1,\nonumber\\
B_{\bf k}&=\Big(\big(J(1-\Delta)+J_{\pm \pm}\big)\cos^2\theta-2J_{\pm \pm} 
-J_{z\pm} \sin2\theta/2 \Big)\nonumber\\
& \times\big(c_2 +c_3\big) 
-i\sqrt{3}\Big(2J_{\pm\pm}\sin\theta-J_{z\pm} \cos \theta\Big) \big(c_2 -c_3 \big) ,\nonumber\\
C_{\bf k}&=-\Big(2J +\big(J(\Delta-1) -J_{\pm\pm}\big) \cos^2 \theta 
+J_{z\pm}\sin \theta \cos \theta\Big)\nonumber\\
&  \quad\quad \times\big(c_2+c_3\big),\\
D_{\bf k}&=-\Big(\big(J \left(1-\Delta\right)  -2J_{\pm\pm} \big)\cos^2 \theta +4J_{\pm\pm} 
 \nonumber\\
& \phantom{=-\Big(\big(J \left(1-\Delta\right)  -2J_{\pm\pm} \big)\cos^2 \theta\ }+J_{z\pm} \sin 2\theta\Big) c_1 ,
 \nonumber
\end{align}
where $c_\alpha\!=\!\cos {\bf k}\bm{\delta}_\alpha$.

\begin{figure}
\includegraphics[width=0.99\linewidth]{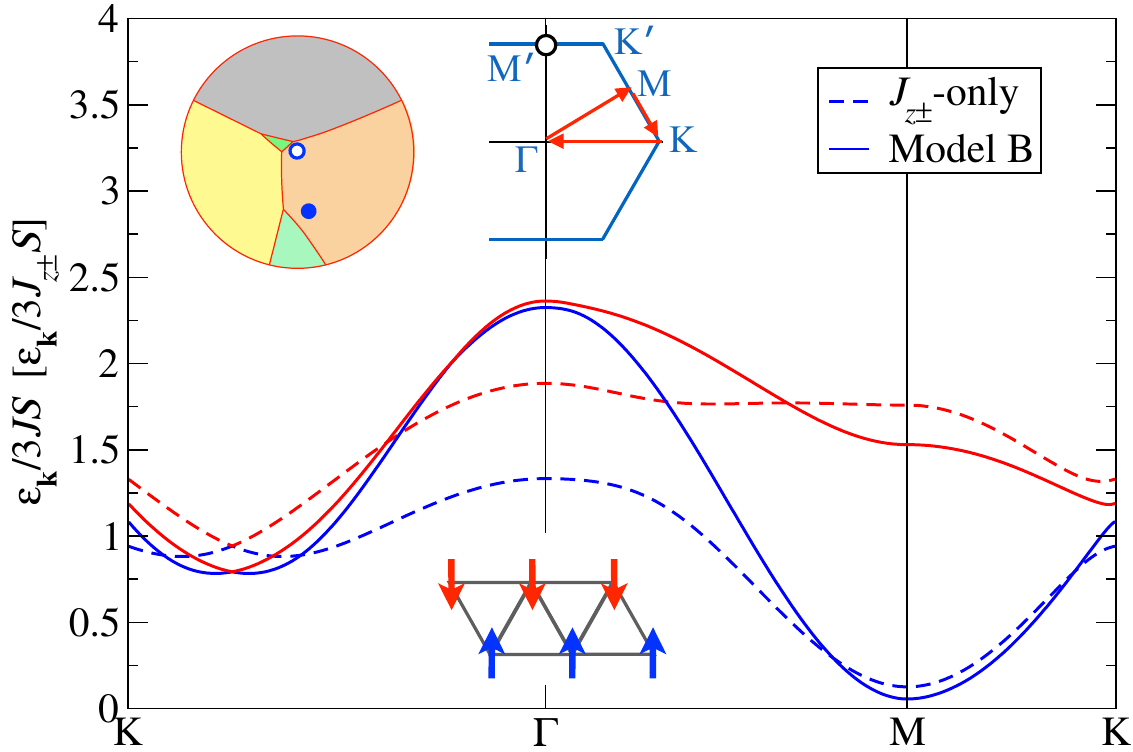}
\caption{Magnon energies $\varepsilon_{1,2{\bf k}}$ from Eq.~(\ref{E12stripe}) vs ${\bf k}$ along the 
$\Gamma MK\Gamma$ path for the model (\ref{HJpm}) with only $J_{z\pm}$ term present (dashed lines) and for the Model B, 
$J\!>\!0$, $\Delta\!=\!0.76$, $J_{\pm\pm}\!=\!0.26J$, and $J_{z\pm}\!=\!0.45J$ (solid lines). Both are  
within the stripe-${\bf yz}$ phase, and the ordering vector is at $M^\prime$ as indicated. 
Left inset:  the 2D phase diagram with the points indicating the chosen set of parameters.}
\label{fig_app_lswtxyz}
\end{figure}

\begin{figure}
\includegraphics[width=0.99\linewidth]{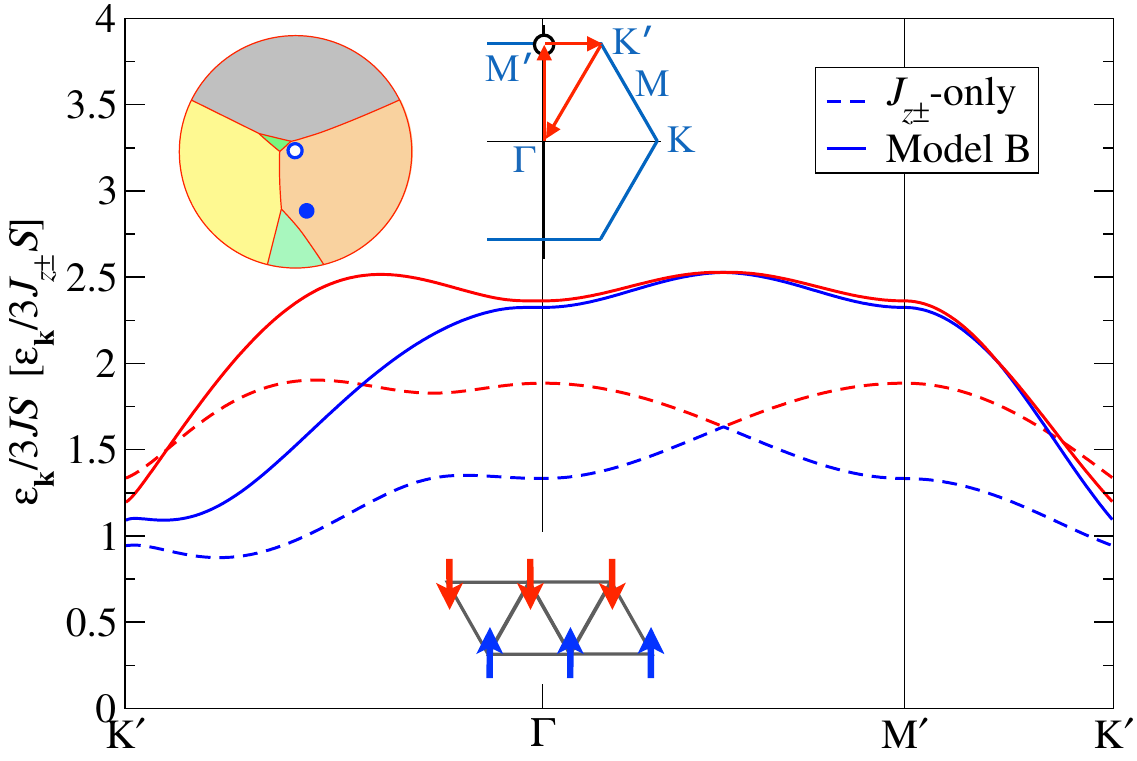}
\caption{Same as Fig.~\ref{fig_app_lswtxyz} along the $\Gamma M^\prime K^\prime\Gamma$ path.}
\label{fig_app_lswtxyz2}
\end{figure}

In our Figures \ref{fig_app_lswtxyz} and \ref{fig_app_lswtxyz2} we present the magnon spectra 
for two points within the stripe-${\bf yz}$
phase. The motivation for their choice is the following. The $J_{z\pm}$-term was initially suggested as 
the main source of frustration
\cite{Chen3}, so the model with the rest of the terms absent, the $J_{z\pm}$-only model, is of interest. 
The parameter set referred to as ``Model B,'' $\Delta\!=\!0.76$, $J_{\pm\pm}\!=\!0.26J$, and $J_{z\pm}\!=\!0.45J$,  
was discussed \cite{Ruegg,Chen4,MM2} as potentially relevant to YbMgGaO$_4$ based on a restricted 
fit of neutron-scattering results in high field and on a matching of the observed static structure factor 
by a semiclassical simulation of the spin-spin correlations \cite{Ruegg,Chen4}. 

Fig.~\ref{fig_app_lswtxyz} shows $\varepsilon_{1,2{\bf k}}$ from Eq.~(\ref{E12stripe}) vs ${\bf k}$ along the 
$\Gamma MK\Gamma$ path and Fig.~\ref{fig_app_lswtxyz2} along the $\Gamma M^\prime K^\prime\Gamma$ 
path, respectively, where $M^\prime$ is the ordering vector. 
The two models are away from the accidental degeneracy surface (\ref{eq_yzzero}) discussed in Sec.~\ref{sec_lswt}~B, but still
demonstrate a nearly gapless mode at the $M$ point. This yields an ordering temperature 
that is suppressed compared to the mean-field value, which naively would suggest a proximity to a spin-liquid state.  
However, we have confirmed that no signs of strong quantum fluctuations are present in either of the cases. 
For $S=1/2$, LSWT for the $J_{z\pm}$-only model gives $\langle S \rangle\!=\!0.4869$, supported by the DMRG result 
$\langle S \rangle\!=\!0.4795$.  Model B is also nearly classical with LSWT $\langle S \rangle\!=\!0.4763$ 
and DMRG giving $\langle S \rangle\!=\!0.4694$.

\begin{figure}
\includegraphics[width=0.99\linewidth]{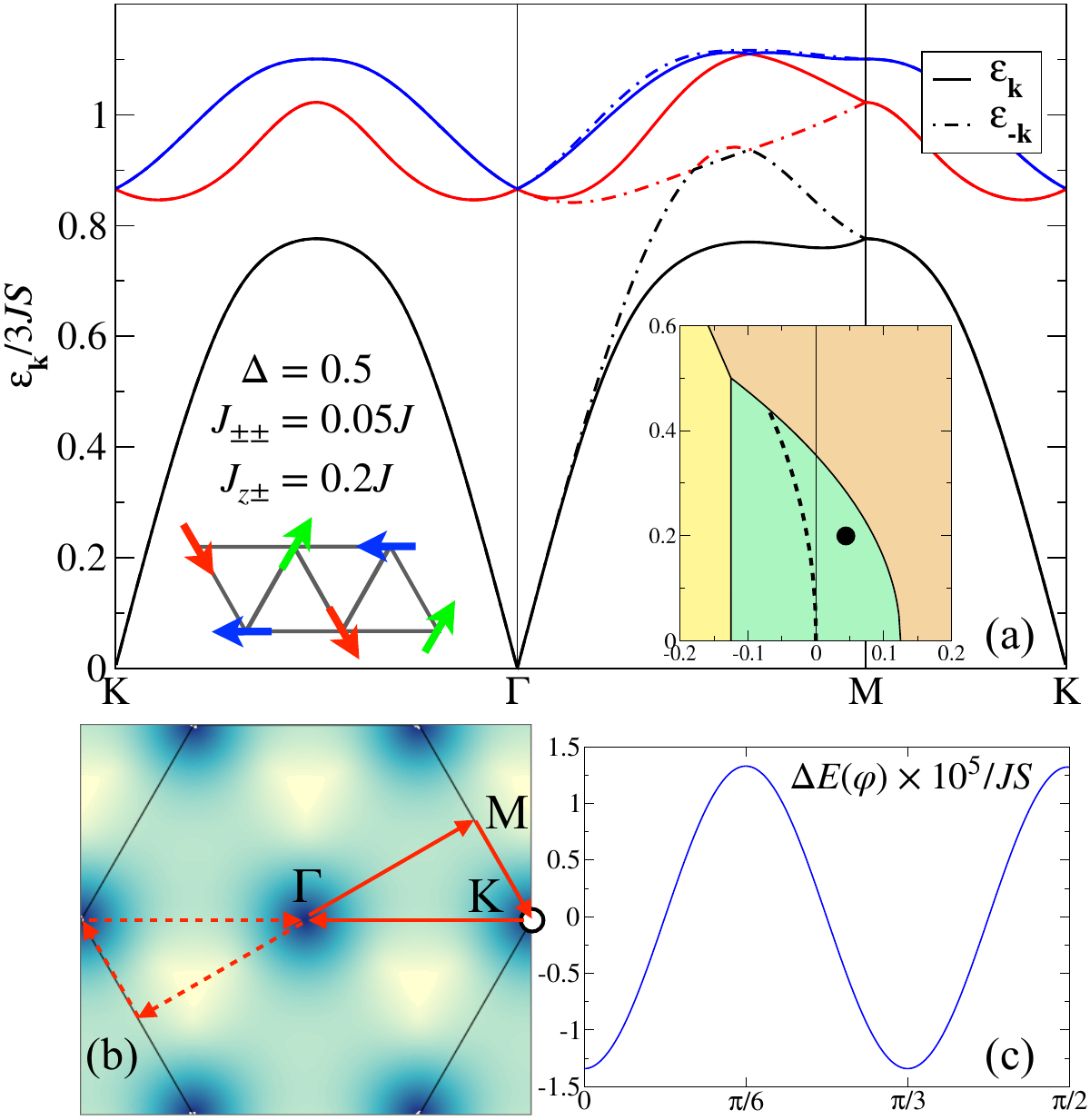}
\caption{Same as Fig.~\ref{fig_120}. (a) Magnon energies 
along the two reciprocal ${\bf k}$-contours (solid and dashed) in (b) for the parameters shown in the graph. 
(b)  Intensity plot of the lower branch. (c) The $\varphi$-dependent part of the zero-point energy 
selecting the structure sketched in (a) with $\varphi_0=\pi n /3$.}
\label{fig_app_120}
\end{figure}

\emph{120$\,{\degree}$ phase.}---%
In Fig.~\ref{fig_app_120}, we provide another example of the nonreciprocal spectrum in the 120${\degree}$ phase, 
discussed in Sec.~\ref{sec_lswt}~C. The choice of parameters is  to the right of the dashed line in the inset
of Fig.~\ref{fig_app_120}(a),
which corresponds to quantum order-by-disorder selecting a state with spins along the bonds,  
$\varphi\!=\!\pi n /3$.

\emph{FM phase.}---%
For the ferromagnetic state, elements of the rotated exchange matrix $\widetilde{\bm J}_{ij}$ are
\begin{align}
\widetilde{J}_{\alpha}^{xx}=& \big(J+2J_{\pm\pm}\cos\left(\varphi_\alpha+2\varphi\right)\big)\sin^2\theta+
\Delta J\cos^2\theta\nonumber\\
& -J_{z\pm}\sin2\theta\, \sin\left(\varphi-\varphi_\alpha\right),\nonumber\\
\widetilde{J}_{\alpha}^{yy}=&J-2J_{\pm\pm}\cos\left(\varphi_\alpha+2\varphi\right),\\
\widetilde{J}_{\alpha}^{zz}=&\big[J+2J_{\pm\pm}\cos\left(\varphi_\alpha+2\varphi\right)\big]\cos^2\theta+
\Delta J\sin^2\theta\nonumber\\
& +J_{z\pm}\sin2\theta\, \sin\left(\varphi-\varphi_\alpha\right),\nonumber\\
\widetilde{J}_{\alpha}^{xy}=& -2J_{\pm\pm}\sin\theta \sin\left(\varphi_\alpha+2\varphi\right)
-J_{z\pm}\cos\theta \cos\left(\varphi-\varphi_\alpha\right),\nonumber
\end{align}
and $\widetilde{J}_{\alpha}^{yx}\!=\!\widetilde{J}_{\alpha}^{xy}$, 
where  $\varphi$ and  $\theta$ are the global angles of the ordered magnetic moment, see Sec.~\ref{sec_lswt}.D.

\begin{figure}
\includegraphics[width=0.99\linewidth]{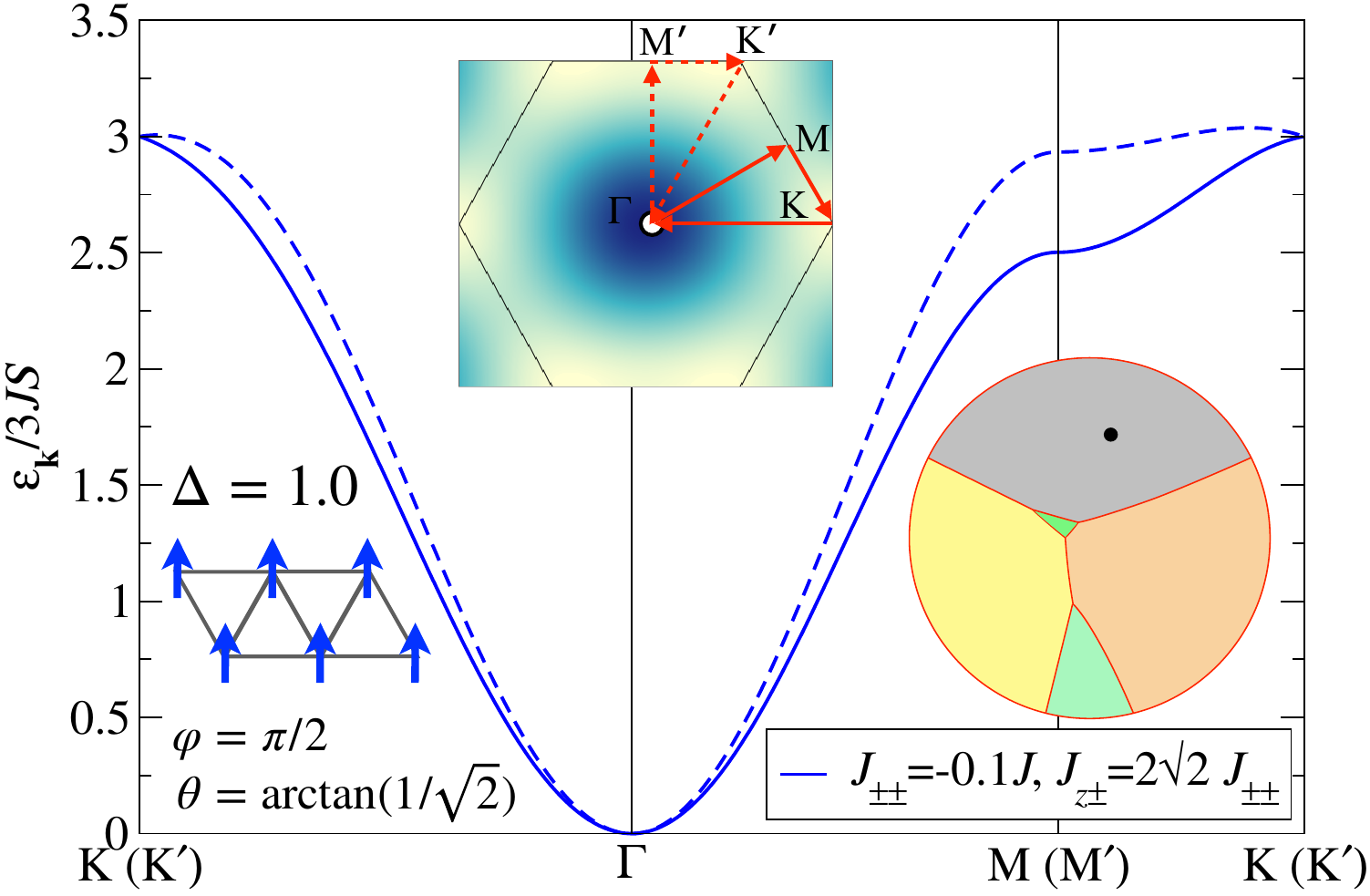}
\vskip -0.1cm
\caption{Magnon energies along the two contours shown in the upper inset  
for a  point in the FM phase (right inset) that belongs to the line  corresponding 
to the $K$--$J$ model:  $J\!<\!0$,  $\Delta\!=\!1.0$, $J_{\pm\pm}\!=\!-0.1J$, and 
$J_{z\pm}\!=\!2\sqrt{2}J_{\pm\pm}$, see text. Upper inset shows the intensity plot of the magnon dispersion.
A sketch depicts the orientation selected  by the order-by-disorder mechanism with the chosen angles
$\varphi\!=\!\pi/2$ and $\theta\!=\!\arctan \left(1/\sqrt{2}\right)$.}
\label{fig_app_fm}
\vskip -0.5cm
\end{figure}

In Fig.~\ref{fig_app_fm}, we  demonstrate the magnon spectrum in the FM phase  
that is discussed in Sec.~\ref{sec_lswt}~D. Here it is shown for the  $\Delta\!=\!1$
plane of the phase diagram and also for the parameters that fall on the line corresponding 
to the $K$--$J$ model, see Sec.~\ref{sec_cubic_axes}, for which
order-by-disorder fluctuations select the cubic axes as a preferred direction for the ordered moment.
The Goldstone mode is clearly quadratic in ${\bf k}$, while the rest of the features discussed 
in Sec.~\ref{sec_lswt}~D are preserved. 

\section{Gap and $T_N$ on the $K$--$J$ line}
\label{app_hk_HF}

Since the degeneracy leading to the pseudo-Goldstone modes along the $K$--$J$ line in Fig.~\ref{fig_duality} 
is accidental, quantum fluctuations induce a gap and make the transition temperature $T_N$ (\ref{eq_tn}) finite in 2D.
Generally, calculations of such order-by-disorder  gaps can be rather involved, see, e.g., Ref.~\cite{Rau_obd}.
In the considered case of the stripe-${\bf yz}$ state in the  $K$--$J$
model, the problem is simplified by the absence of the cubic anharmonicities, because the 
spin orientation is along one of the cubic axes. The other simplification is the Klein duality to the 
ferromagnetic state, for which  calculations are straightforward. 

The LSWT  spectrum in the  ferromagnetic phase of the Kitaev-Heisenberg model (\ref{eq_HJKline})
is $\varepsilon_\mathbf{k}\!=\!\sqrt{A_\mathbf{k}^2-B_\mathbf{k}^2}$, where
\begin{align}
A_\mathbf{k}&=-6J_0 S \left(1-\gamma_\mathbf{k} \right)-KS\left( 2-c_2-c_3\right),\nonumber\\
B_\mathbf{k}&=-KS\left( c_2-c_3\right),
\label{eq_hk_e}
\end{align}
$c_\alpha\!=\!\cos \mathbf{k} \bm{\delta}_\alpha$, and $\gamma_\mathbf{k}$ is from Eq.~(\ref{gammaFM}), 
see Ref.~\cite{Avella}.

\begin{figure}
\includegraphics[width=0.99\linewidth]{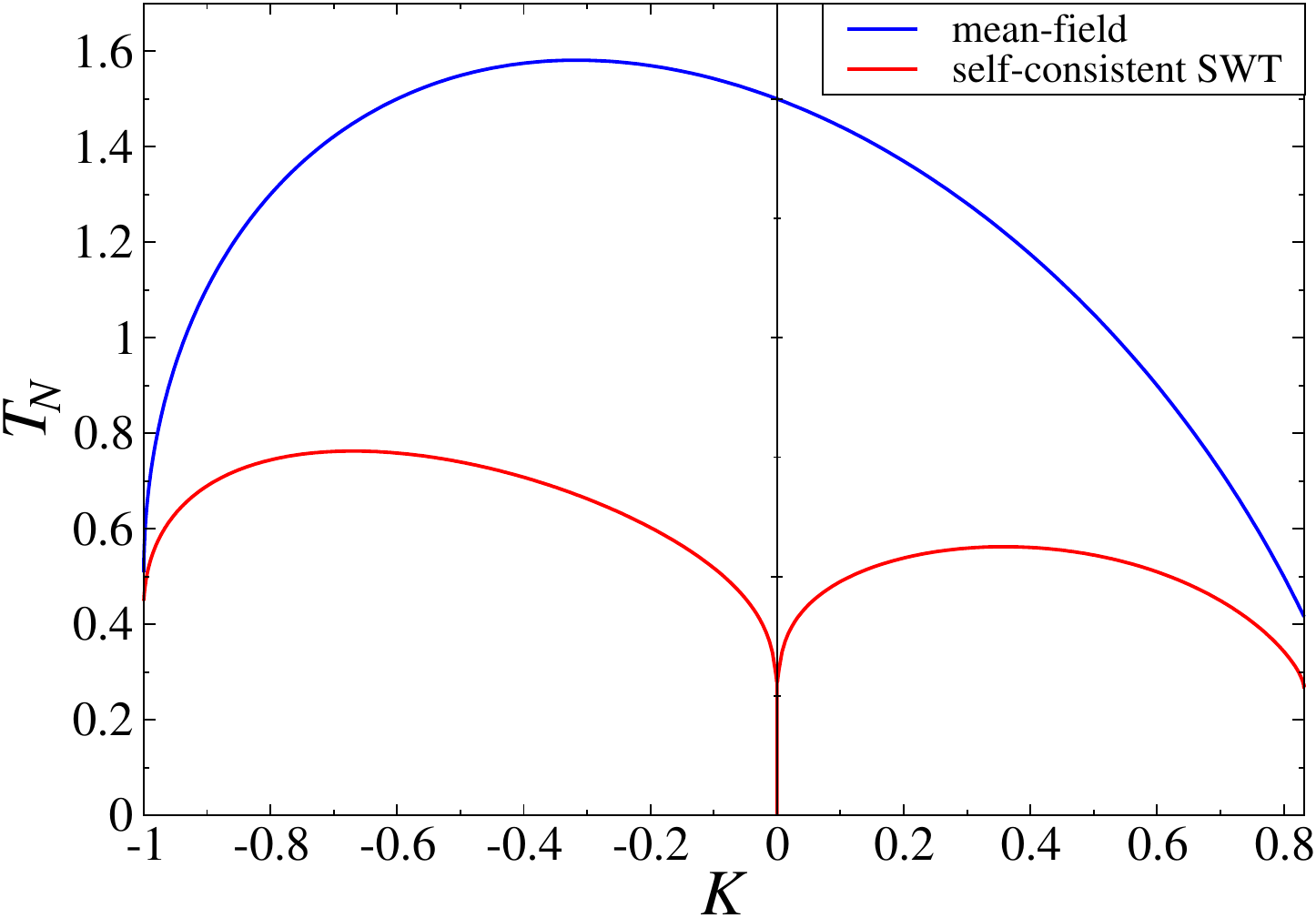}
\vskip -0.1cm
\caption{Ordering temperature for the FM phase of the $K$--$J$ model calculated using Eq.~(\ref{app_eq_tn}), 
lower curve, and the mean-field result, Eq.~\eqref{eq_tcmf}, upper curve. 
$T_N$ and $K$ are in units of $\left(J_0^2+K^2\right)^{1/2}$ and $J_0\!<\!0$. }
\label{fig_JK_TN}
\vskip -0.3cm
\end{figure}

The Hartree-Fock  corrections from the quartic terms in the SWT Hamiltonian 
to $A_\mathbf{k}$ and $B_\mathbf{k}$ are 
\begin{align}
\delta A_\mathbf{k}&\!=\!2J_0 \big[ 3n\left(1\!-\!\gamma_\mathbf{k}\right)\!+\!m_1
\left( c_1 \!-\!1\right)\!+\!m_2 \left(c_2\!+\!c_3\!-\!2\right)\big]\nonumber\\
\label{eq_HFAB}
&  +K\big[ n\left(2-c_2-c_3\right)+2(m_1 c_1 -m_2-\Delta_2)\big],\\
\delta B_\mathbf{k}&=\left(-2J_0\Delta_2+K n\right)\left(c_2-c_3 \right),\nonumber
\end{align}
with  $n\!=\!\langle a^\dagger_i a^{\phantom \dagger}_i \rangle$, 
$m_\alpha\!=\!\langle a^\dagger_i a^{\phantom \dagger}_{i+\bm{\delta}_\alpha} \rangle$, and 
$\Delta_\alpha\!=\!\langle a_i a_{i+\bm{\delta}_\alpha} \rangle$,
\begin{align}
&n\!=\!\frac{1}{2N}\sum_\mathbf{k} \left(\frac{A_{\bf k}}{\varepsilon_\mathbf{k}}-1\right), \ \ \ \ 
m_\alpha\!=\!\frac{1}{2N}\sum_\mathbf{k} \frac{c_\alpha A_{\bf k}}{\varepsilon_\mathbf{k}}, \nonumber \\
&\Delta_\alpha\!=\!\frac{1}{2N}\sum_\mathbf{k}  \frac{c_\alpha B_{\bf k}}{\varepsilon_\mathbf{k}},
\end{align}
where  $m_2\!=\!m_3$, $\Delta_2\!=\!-\Delta_3$, and $\Delta_1\!=\!0$. 

Within the $1/S$-approximation, the spectrum becomes
\begin{align}
\widetilde{\varepsilon}_\mathbf{k}=\sqrt{\left(A_\mathbf{k}+\delta A_\mathbf{k} \right)^2-\left(B_\mathbf{k}
+\delta B_\mathbf{k} \right)^2}.
\end{align}

However, the most important effect of the 
quantum fluctuations is the gap at ${\bf k}\!=\!{\bf 0}$,
\begin{align}
E_g=|\delta A_{\bf 0}|=2|K\left( m_1 -m_2-\Delta_2\right)|,
\end{align}
which we have also verified by the recently developed method of Ref.~\cite{Rau_obd}.

Since the LSWT spectrum is parabolic, one can approximate the renormalized spectrum as 
\begin{align}
\widetilde{\varepsilon}_\mathbf{k}=E_g+\varepsilon_\mathbf{k},
\label{eq_renEk}
\end{align}
which should suffice for the regularization of the divergences leading to the vanishing ordering temperature
due to Bose factors in Eq.~\eqref{eq_app_rpa_s}.
Thus, our self-consistent transition temperature in the Kitaev-Heisenberg model is given by a 
simplified version of Eq.~(\ref{eq_tn}) with the renormalized spectrum (\ref{eq_renEk})
\begin{align}
\frac{1}{T_N}=\frac{1}{N}\sum_{\mathbf{k}} \frac{1+2v_{\mathbf{k}}^2}{\widetilde{\varepsilon}_{\mathbf{k}}}.
\label{app_eq_tn}
\end{align}
The results of using (\ref{app_eq_tn}) for the FM phase of the $K$--$J$ model
are presented in Fig.~\ref{fig_JK_TN},  where we denote the transition temperature  as $T_N$ having in 
mind Klein duality to the stripe-${\bf yz}$ part of the $K$--$J$ line in Fig.~\ref{fig_duality}.  
Both $T_N$ and $K$ are in units of $\left(J_0^2+K^2\right)^{1/2}$ and $J_0\!<\!0$. 

Although the  gap $E_g$ is small, the transition temperature depends singularly on it, 
$T_N\!\propto\!T_{\rm MF}/\ln \left(T_{\rm MF}/E_g\right)$ \cite{Katanin}, 
leading to a finite and sizable $T_N$ with an exception to the phase boundaries and to the 
isotropic, $O(3)$ symmetric $K\!=\!0$ point (``$J_0$-only'' point), where $T_N$ remains 
zero. The resultant transition temperatures are still  significantly suppressed 
compared to the mean-field results of Eq.~\eqref{eq_tcmf}.

\begin{figure*}
\includegraphics[width=0.99\linewidth]{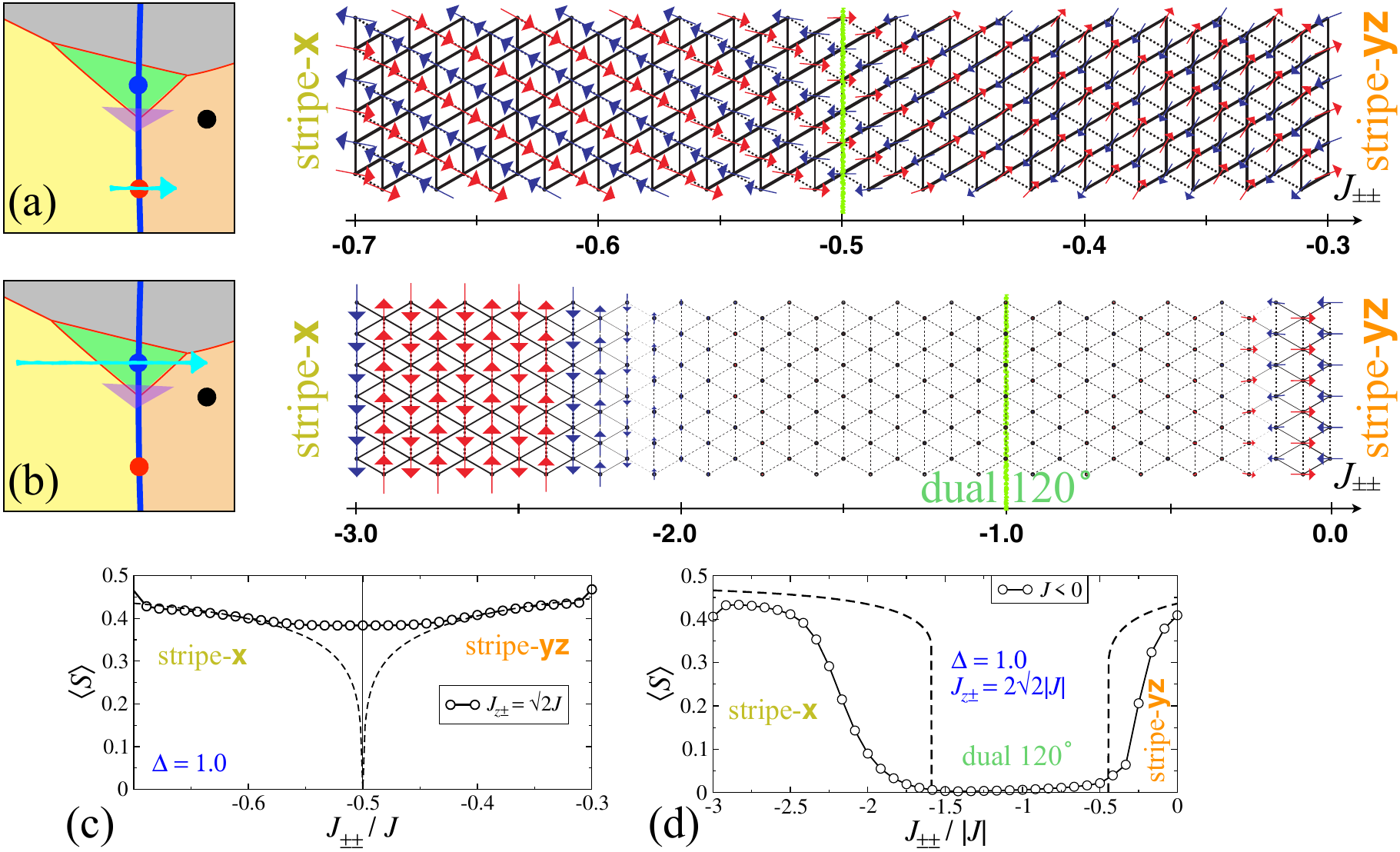}
\caption{Long-cylinder DMRG scans (a) across the $K$-only point ($\Delta\!=\!1.0$, 
$J_{\pm\pm}\!=\!-0.5J$, and $J_{z\pm}\!=\!\sqrt{2} J$) vs $J_{\pm\pm}/J$ from $-0.7$ to $-0.3$, and (b) 
across the dual 120$\degree$ point ($\Delta\!=\!1.0$, $J\!<\!0$,
$J_{\pm\pm}\!=\!J$, and $J_{z\pm}\!=\!2\sqrt{2} |J|$) vs $J_{\pm\pm}/|J|$ from $-3.0$ to $0$. 
Both scans go from the stripe-${\bf x}$ to the stripe-${\bf yz}$ phase and are normal to the $K$--$J$ line in 
Fig.~\ref{fig_duality}, with the vertical lines in the clusters showing the intersect with it. 
Sketches of the phases  
indicate the direction and extent of each scan, with (c) and (d) showing $\langle S\rangle$ along the scans.
Dashed lines in (c) and (d) are the LSWT results for $\langle S\rangle$, Eq.~(\ref{eq_avgs}), in the stripe phases.}
\label{fig_app_DMRG1}
\end{figure*}

\section{DMRG details}
\label{app_DMRG}

Figure~\ref{fig_app_DMRG1} summarizes  results from the two long-cylinder scans, one 
across the $K$-only point separating the stripe-${\bf x}$ and stripe-${\bf yz}$ phases, and the other 
through the dual 120$\degree$ region, also connecting the stripe phases. Both scans are  
perpendicular to the $K$--$J$ line in Fig.~\ref{fig_duality} and the vertical lines in the clusters 
show the point of intersect with it. The cutouts
of the phase diagram shown on the left of the cylinder images in  Figs.~\ref{fig_app_DMRG1}(a) and 
\ref{fig_app_DMRG1}(b) indicate the direction and extent of each scan, and  Fig.~\ref{fig_app_DMRG1}(c) and 
\ref{fig_app_DMRG1}(d) show the 
magnitude of the ordered moment $\langle S\rangle$ along the scans.

\begin{figure}[t]
\includegraphics[width=0.99\linewidth]{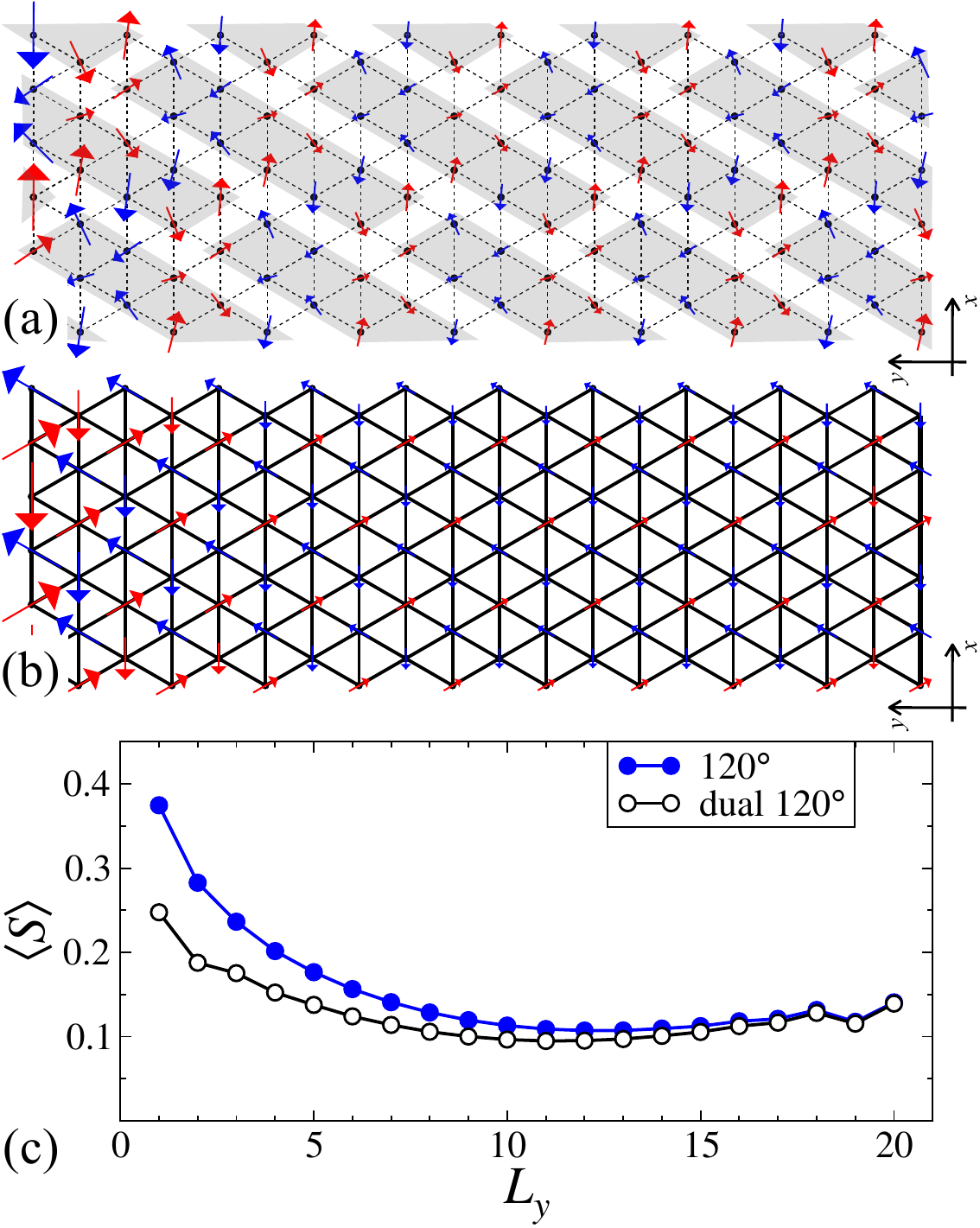}
\caption{DMRG non-scan calculation in the $6\times 20$ cluster for the (a) 
dual 120$\degree$ point,  and (b) 120$\degree$ point for comparison. 
In (a) the 12-sublattice structure is highlighted. (c) Ordered moment $\langle S \rangle$ along the length 
of the cluster.}
\label{fig_app_sdecay}
\end{figure}

In Fig.~\ref{fig_app_DMRG1}(a), a scan is performed by varying $J_{\pm\pm}/J$ from $-0.7$ to $-0.3$ 
for fixed $\Delta\!=\!1.0$
and  $J_{z\pm}\!=\!\sqrt{2}J$, so that  at the $K$--$J$ line $J_{z\pm}\!=\!-2\sqrt{2}J_{\pm\pm}$ as elsewhere.
The scan shows a clear crossover from  the stripe-${\bf x}$ to  stripe-${\bf yz}$
phase. The spins on the left edge of the cylinder in the  stripe-${\bf x}$ phase 
are slightly tilted off the lattice plane and continuously deform into the stripe-${\bf yz}$ order on the right edge 
with only a small variation
of the ordered moment near $\langle S\rangle\!\approx\!0.4$, see Fig.~\ref{fig_app_DMRG1}(c). 
There is no indication of a nematic \cite{Trebst_tr,Tohyama} or a spin-liquid state \cite{Li_CSL} at the $K$--$J$ line,
nor there is a sign of a proximity to any.

The dashed lines in Fig.~\ref{fig_app_DMRG1}(c) show the ordered moment as given by the $1/S$ calculations in the 
neighboring stripe phases, indicating a direct  transition.   
We have also performed a DMRG non-scan $6\times 20$ cluster calculation at the 
 $K$-only point (not shown)
with the results very similar to the central part of the scan in Fig.~\ref{fig_app_DMRG1}(a),
showing a robust order in the form of a stripe phase with the same ordering vector at the $M$-point
and spins oriented in between  the stripe-${\bf x}$ and stripe-${\bf yz}$ orders.
Given the smoothness of the crossover in Fig.~\ref{fig_app_DMRG1}(a) and that the ordering vector does not change, it is 
not clear to us whether the two stripe phases remain distinct in the presence of quantum fluctuations.

In Fig.~\ref{fig_app_DMRG1}(b), a scan is performed across the dual 120$\degree$ point 
by varying $J_{\pm\pm}/|J|$ from $-3.0$ to 0.0 for fixed $\Delta\!=\!1.0$
and  $J_{z\pm}\!=\!2\sqrt{2}|J|$ ($J\!<\!0$ here).
There are well-ordered stripe-${\bf x}$ and stripe-${\bf yz}$ phases in the two ends of the cylinder, separated by 
a region where the order is suppressed, see also Fig.~\ref{fig_app_DMRG1}(d). This may seem surprising as one 
expects the dual 120$\degree$ ordered phase in this region. One of the obvious reasons for a suppression of the  order 
is a very large gradient of parameters in this scan, combined with a large unit cell of the 12-sublattice structure.
In Fig.~\ref{fig_app_DMRG1}(d), the dashed lines  show 
the ordered moment as given by the $1/S$ calculations in the stripe phases, showing that for $S\!=\!1/2$ 
the dual 120$\degree$ state expands and claims some of the stripe regions, similar to the 
expansion of the ``original'' 120$\degree$  state, see Ref.~\cite{topography}.

To verify that the intermediate phase is not a spin liquid, we have 
performed a DMRG non-scan calculation in the $6\times 20$ cluster for the $\widetilde{J}_0$-only, dual 120$\degree$ point 
($\Delta\!=\!1.0$, $J\!<\!0$, $J_{\pm\pm}\!=\!J$, and $J_{z\pm}\!=\!2\sqrt{2} |J|$)
with the results shown in Fig.~\ref{fig_app_sdecay}(a).
Here, the dual 120$\degree$ boundary conditions are applied at the left edge only, with a robust 12-sublattice 
order parameter showing a power-law decay with a finite asymptote, characteristic of the well-ordered phase, 
see Fig.~\ref{fig_app_sdecay}(c).

The observed ordered moment of the dual 120${\degree}$ state in this cluster is about $\langle S \rangle\! \approx\! 0.09$.
Since the Klein duality implies that all observables between dual points should be the same, this value may seem 
to contradict the value of the ordered moment at the Heisenberg 120${\degree}$ point, 
$\langle S \rangle\! \approx\! 0.2$ \cite{White07}. 
As we verify in Fig.~\ref{fig_app_sdecay}(b), the smaller value of the ordered magnetic moment is due to 
the mixed boundary conditions and the resultant 
aspect ratio of the cluster being far from being ``optimal,'' according to Ref.~\cite{White07}. 
In Figs.~\ref{fig_app_sdecay}(b) and (c) we demonstrate that the 120${\degree}$ order on the same cluster 
behaves very similarly and is in  quantitative agreement with the Klein-duality expectations.


\end{document}